\documentclass[twoside,twocolumn,9pt]{article}
\usepackage{extsizes}
\usepackage[super,sort&compress,comma]{natbib}
\usepackage[version=3]{mhchem}
\usepackage[left=1.5cm, right=1.5cm, top=1.785cm, bottom=2.0cm]{geometry}
\usepackage{balance}
\usepackage{mathptmx}
\usepackage{sectsty}
\usepackage{graphicx}
\usepackage{lastpage}
\usepackage[format=plain,justification=justified,singlelinecheck=false,font={stretch=1.125,small,sf},labelfont=bf,labelsep=space]{caption}
\usepackage{float}
\usepackage[hidelinks,draft]{hyperref}
\usepackage{fancyhdr}
\usepackage{fnpos}
\usepackage[english]{babel}
\addto{\captionsenglish}{%
  
}
\usepackage{array}
\usepackage{droidsans}
\usepackage{charter}
\usepackage[T1]{fontenc}
\usepackage[usenames,dvipsnames]{xcolor}
\usepackage{setspace}
\usepackage[compact]{titlesec}
\usepackage{xspace}

\usepackage{subcaption}
\usepackage{braket}
\usepackage{multirow}
\usepackage{bm}
\usepackage[floats=float]{chemscheme}

\usepackage{epstopdf}

\definecolor{cream}{RGB}{222,217,201}

\newcommand{\angstrom}{\mbox{\normalfont\AA}\xspace}

\begin{document}

\pagestyle{fancy}
\thispagestyle{plain}
\fancypagestyle{plain}{
\renewcommand{\headrulewidth}{0pt}
}

\makeFNbottom
\makeatletter
\renewcommand\LARGE{\@setfontsize\LARGE{15pt}{17}}
\renewcommand\Large{\@setfontsize\Large{12pt}{14}}
\renewcommand\large{\@setfontsize\large{10pt}{12}}
\renewcommand\footnotesize{\@setfontsize\footnotze{7pt}{10}}
\makeatother

\renewcommand{\thefootnote}{\fnsymbol{footnote}}
\renewcommand\footnoterule{\vspace*{1pt}%
\color{cream}\hrule width 3.5in height 0.4pt \color{black}\vspace*{5pt}}
\setcounter{secnumdepth}{5}

\makeatletter
\renewcommand\@biblabel[1]{#1}
\renewcommand\@makefntext[1]%
{\noindent\makebox[0pt][r]{\@thefnmark\,}#1}
\makeatother
\renewcommand{\figurename}{\small{Fig.}~}
\sectionfont{\sffamily\Large}
\subsectionfont{\normalsize}
\subsubsectionfont{\bf}
\setstretch{1.125} 
\setlength{\skip\footins}{0.8cm}
\setlength{\footnotesep}{0.25cm}
\setlength{\jot}{10pt}
\titlespacing*{\section}{0pt}{4pt}{4pt}
\titlespacing*{\subsection}{0pt}{15pt}{1pt}

\fancyfoot{}
\fancyfoot[LO,RE]{\vspace{-7.1pt}\includegraphics[height=9pt]{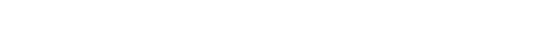}}
\fancyfoot[CO]{\vspace{-7.1pt}\hspace{11.9cm}\includegraphics{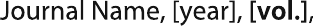}}
\fancyfoot[CE]{\vspace{-7.2pt}\hspace{-13.2cm}\includegraphics{head_foot/RF}}
\fancyfoot[RO]{\footnotesize{\sffamily{1--\pageref{LastPage} ~\textbar  \hspace{2pt}\thepage}}}
\fancyfoot[LE]{\footnotesize{\sffamily{\thepage~\textbar\hspace{4.65cm} 1--\pageref{LastPage}}}}
\fancyhead{}
\renewcommand{\headrulewidth}{0pt}
\renewcommand{\footrulewidth}{0pt}
\setlength{\arrayrulewidth}{1pt}
\setlength{\columnsep}{6.5mm}
\setlength\bibsep{1pt}

\makeatletter
\newlength{\figrulesep}
\setlength{\figrulesep}{0.5\textfloatsep}

\newcommand{\topfigrule}{\vspace*{-1pt}%
\noindent{\color{cream}\rule[-\figrulesep]{\columnwidth}{1.5pt}} }

\newcommand{\botfigrule}{\vspace*{-2pt}%
\noindent{\color{cream}\rule[\figrulesep]{\columnwidth}{1.5pt}} }

\newcommand{\dblfigrule}{\vspace*{-1pt}%
\noindent{\color{cream}\rule[-\figrulesep]{\textwidth}{1.5pt}} }

\makeatother

\twocolumn[
  \begin{@twocolumnfalse}
{\includegraphics[height=30pt]{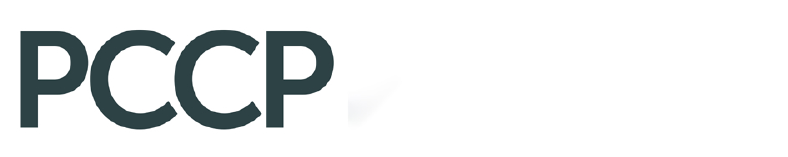}\hfill\raisebox{0pt}[0pt][0pt]{\includegraphics[height=55pt]{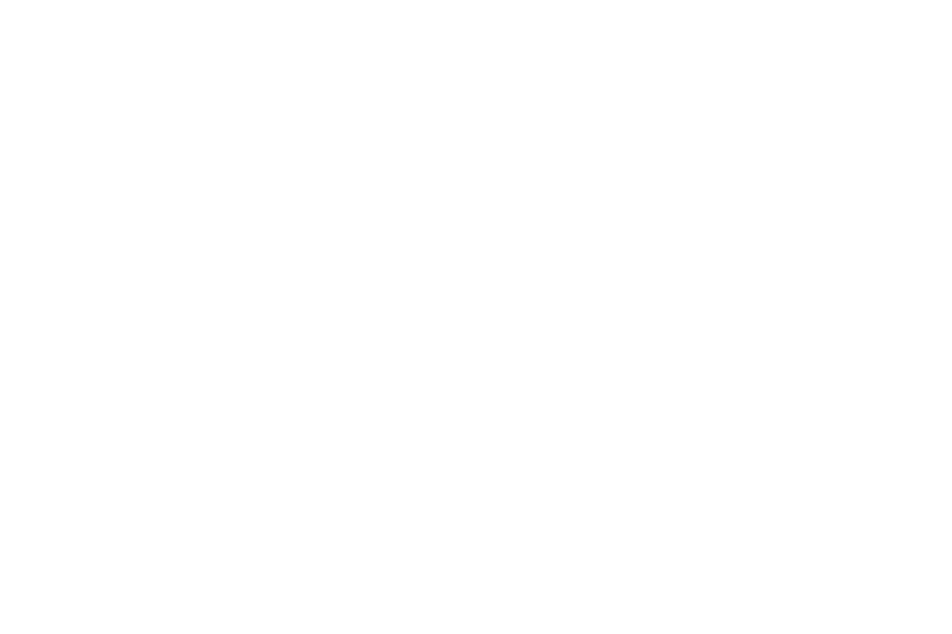}}\\[1ex]
\includegraphics[width=18.5cm]{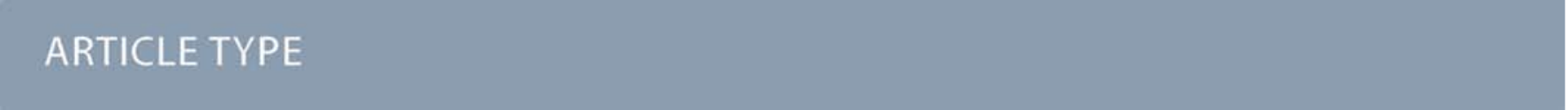}}\par
\vspace{1em}
\sffamily
\begin{tabular}{m{4.5cm} p{13.5cm} }

\includegraphics{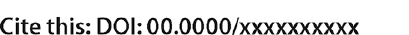} & \noindent\LARGE{\textbf{Simulating X-ray Photoelectron Spectra With Strong Electron Correlation Using Multireference Algebraic Diagrammatic Construction Theory}} \\
\vspace{0.3cm} & \vspace{0.3cm} \\

 & \noindent\large{Carlos E. V. de Moura,\textit{$^{a}$} and Alexander Yu. Sokolov$^{\ast}$\textit{$^{a}$}} \\

\includegraphics{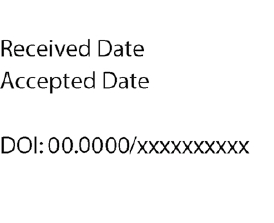} & \noindent\normalsize{
We present a new theoretical approach for the simulations of X-ray photoelectron spectra of strongly correlated molecular systems that combines multireference algebraic diagrammatic construction theory (MR-ADC) [\textit{J. Chem. Phys., 2018,} \textbf{149}, 204113] with core-valence separation (CVS) technique.
The resulting CVS-MR-ADC approach has a low computational cost while overcoming many challenges of the conventional multireference theories associated with the calculations of excitations from inner-shell and core molecular orbitals. 
Our results demonstrate that the CVS-MR-ADC methods are as accurate as single-reference ADC approximations for predicting core ionization energies of weakly-correlated molecules, but are more accurate and reliable for systems with multireference character, such as stretched nitrogen molecule, ozone, and isomers of benzyne diradical. 
We also highlight the importance of multireference effects for the description of core-hole screening  that determines the relative spacing and order of peaks in the XPS spectra of strongly correlated systems.
} \\

\end{tabular}

 \end{@twocolumnfalse} \vspace{0.6cm}

  ]

\renewcommand*\rmdefault{bch}\normalfont\upshape
\rmfamily
\section*{}
\vspace{-1cm}


\footnotetext{\textit{$^{a}$~Department of Chemistry and Biochemistry, The Ohio State University, Columbus, Ohio, 43210, USA. E-mail: vieirademoura.2@osu.edu, sokolov.8@osu.edu}}

\footnotetext{\dag~Electronic Supplementary Information (ESI) available: active spaces used in the CVS-MR-ADC calculations, atomic coordinates, Mulliken atomic charges, core ionization energies, and spectroscopic factors for the ozone and benzyne molecules. See DOI: 10.1039/d1cp05476g/}



\section{Introduction}

Recently, X-ray spectroscopies have become widely used tools to investigate the electronic structure and dynamics of molecules and materials.\cite{Lin.2017,Chergui.2017,Kraus.2018,Norman.2018}
The increase in popularity of X-ray techniques is in part due to the growing availability and accessibility of X-ray radiation sources,\cite{Pellegrini.2016,Geloni.2017,Young.2018,Piancastelli.2019, Seres.2004,Popmintchev.2010,Li.2017,Barreau.2020} as well as the ability of X-ray spectroscopies to probe the excited states of core electrons that are sensitive to local electronic structure and geometric environment.  
Among many methods that employ X-ray radiation, X-ray photoelectron spectroscopy (XPS) is by far the most commonly used.\cite{Greczynski.2019,Stevie.2020}
Phenomenologically, XPS is based on the photoelectric effect, measuring the kinetic energy of electrons ionized from the core atomic orbitals.
Although, the XPS technique is most widely employed to study the surfaces of solids,\cite{briggs2003surface,watts2019introduction} it has been also applied to investigate chemical systems in liquid\cite{Winter.2007,Nishizawa.2010,Thurmer.2021} and gas phases.\cite{Banna.2006,Kraus.2014,Tao.2018}

Along with the experimental advances in X-ray photoionization techniques, reliable interpretation of the XPS spectra requires simulations of core-ionized states using accurate electronic structure methods.
However, computations of core-excited states are very challenging as they require simulating electronic transitions with energies much higher than the ionization threshold, using large uncontracted basis sets, and incorporating orbital relaxation, electron correlation, and relativistic effects.\cite{Norman.2018,Bokarev.2020,Kasper.2020,Besley.2020,Besley.2021,Rankine.2021}
For this reason, most of the currently available methods for the simulations of XPS introduce simplifications based on the assumption that the ground-state electronic structure is well described using a single (reference) electronic configuration.
These single-reference methods developed using a variety of theoretical approaches\cite{Salpeter:1951p1232,Hedin:1965p796,vanSchilfgaarde:2006p226402,vanSetten:2013p232,Stener:2003p115,Besley:2010p12024,Peng:2015p4146,Barth:1980p149,Nooijen:1995p6735,Sen:2013p2625,Dutta:2014p3656,Coriani:2015p181103,Nascimento:2017p2951,Liu.2019,Barth:1985p867,Wenzel.2014,Schirmer.2001,Thiel.2003} can be used to simulate the XPS spectra of weakly-correlated molecules and materials, but may be unreliable for chemical systems that exhibit strong electron correlation.

Incorporating strong correlation into the simulations of XPS requires a multiconfigurational description of the ground-state electronic structure.\cite{Lischka.2018,Park.2020,Khedkar.2021} 
However, most of the available multireference approaches are not designed to compute core-level excited states as they can only simulate electronic excitations in the frontier (so-called active) molecular orbitals with strongly correlated electrons.
This problem has been partially addressed by treating the core orbitals as active and placing restrictions on orbital occupations in the self-consistent calculations of multiconfigurational ground-state wavefunctions,\cite{Rocha.2011,Rocha.2011vjw,Moura.2013,Corral.201794l,Battacharya.2021,Agren:1993p45,Josefsson.2012,Pinjari:2016p477,Guo:2016p3250} at a cost of introducing approximations that are difficult to control.
Alternatively, core-level excited states can be calculated using linear-response\cite{Yeager:1979p77,Graham:1991p2884,Yeager:1992p133,Nichols:1998p293,HelmichParis:2019p174121,Paris.2021,Kohn:2019p041106} or equation-of-motion\cite{Datta:2012p204107,Maganas.2019} multireference approaches that incorporate excitations from non-active molecular orbitals.
However, these methods may produce unphysical (complex) excitation energies due to the non-Hermitian nature of underlying equations. 

In this work, we present a new approach for the XPS simulations of strongly correlated systems that combines multireference algebraic diagrammatic construction theory (MR-ADC)\cite{Sokolov.2018,Chatterjee.2019,Chatterjee.2020,Mazin.2021} with core-valence separation technique (CVS).\cite{Cederbaum.1980,Barth.1981} 
The MR-ADC approach is naturally suited for the simulations of core-level excitations combining several attractive features: (i) low computational cost (similar to that of multireference perturbation theories),\cite{Finley:1998p299,Andersson:1990p5483,Andersson:1992p1218,Angeli.2001b,Angeli.2001,Angeli.2004} (ii) Hermitian equations, and (iii) ability to calculate excitations from all molecular orbitals, including inner-shell and core.
To access the high-energy core-ionized states, we employ the CVS technique that has been successfully used for the simulations of X-ray spectra using a variety of theoretical methods.\cite{Coriani.2015c1e,Vidal.2019,Vidal.2019b,Vidal.2020,Thielen.2021,Koppel.1997,Barth.1999,Trofimov.2000,Schirmer.2001,Thiel.2003,Wenzel.2014,Wenzel.2015,Liu.2019,Zheng.2019,Paris.2021,Peng.2019,Garner.2020,Seidu.2019}
We first investigate the accuracy of MR-ADC approximations for the calculations of K-edge core ionization energies in small weakly-correlated molecules (\autoref{subsec:CVS-benchmark}) and along the dissociation pathway of molecular nitrogen (\autoref{subsec:dissociation-n2}).
We then apply the MR-ADC methods to simulate the XPS spectra of ozone (\autoref{sec:ozone}) and three isomers of benzyne diradicals (\autoref{sec:benzynes}) that exhibit significant multireference character in their ground electronic states. 

\section{Theory}\label{sec:Theory}

\subsection{Multireference Algebraic Diagrammatic Construction Theory for Photoelectron Spectra}

\begin{figure}[t!]
	\centering
	\includegraphics[width=\columnwidth]{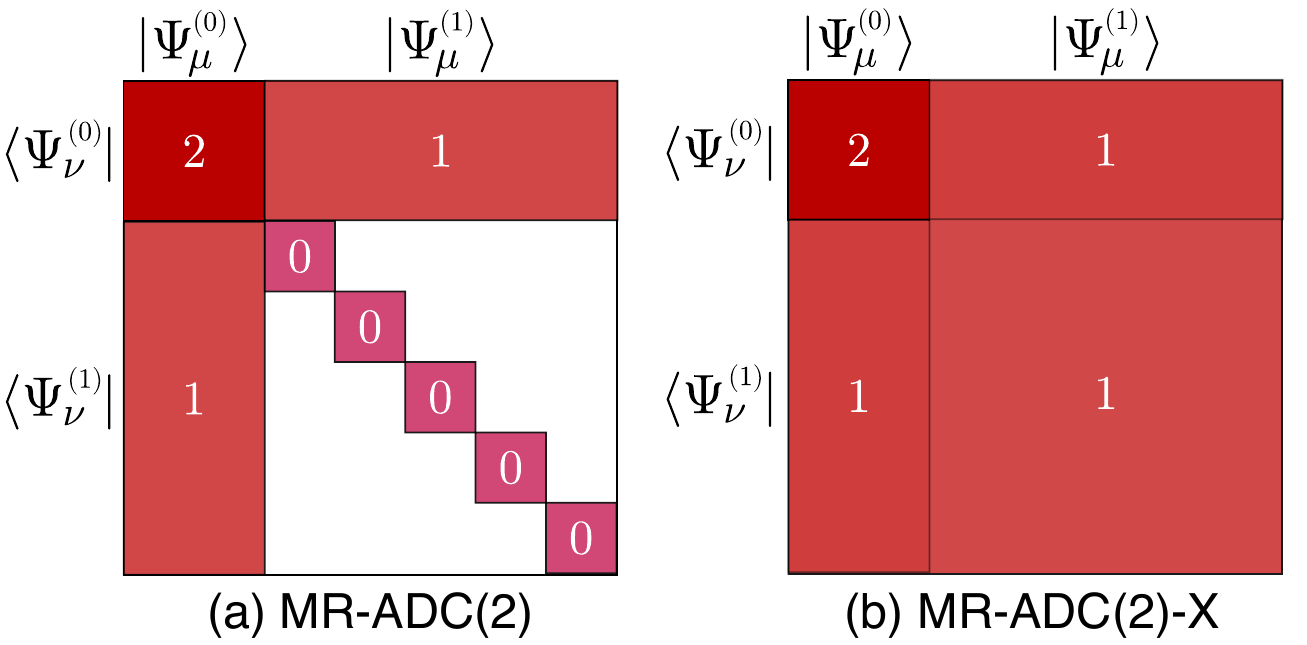}
	\caption{Schematic representation of the effective Hamiltonian matrix $\mathbf{M}$ for the (a) MR-ADC(2) and (b) MR-ADC(2)-X approximations. Zeroth- and first-order excitations are labeled as $\ket{\Psi_\mu^{(0)}}$ and $\ket{\Psi_\mu^{(1)}}$, respectively. Nonzero matrix blocks are highlighted in color. Numbers represent the perturbation order to which the effective Hamiltonian is evaluated in each matrix block. }
	\label{fgr:mr-adc-matrix-compare}
\end{figure}

We begin with a brief introduction to the MR-ADC theory for ionization energies and photoelectron spectra.
For more details, we refer the readers to our previous publications.\cite{Sokolov.2018,Chatterjee.2019,Chatterjee.2020,Mazin.2021}
Ionization of a many-electron system irradiated with light of frequency $\omega$ is described by the backward component of the one-particle Green's function,\cite{dickhoff2008many,fetter2012quantum} $G_{pq}(\omega)$:
\begin{equation}
	\label{eq:greens_function}
	G_{pq}(\omega) = \langle \Psi | a_{p}^{\dag} (\omega - H - E)^{-1} a_{q} | \Psi \rangle
\end{equation}
where $\ket{\Psi}$ and $E$ are the ground-state eigenfunction and eigenvalue of the electronic Hamiltonian $H$, respectively.
The operators $a_{p}^{\dag}$  and $a_{q}$ are the particle creation and annihilation operators from the second quantization formalism of quantum mechanics.\cite{Dirac.1927,feynman1998statistical}
The Green's function \eqref{eq:greens_function} contains the information about energies and probabilities of all one-electron ionization transitions in the photoelectron spectrum of the system.

\begin{figure*}[t!]
	\centering
	\includegraphics[width=0.9\textwidth]{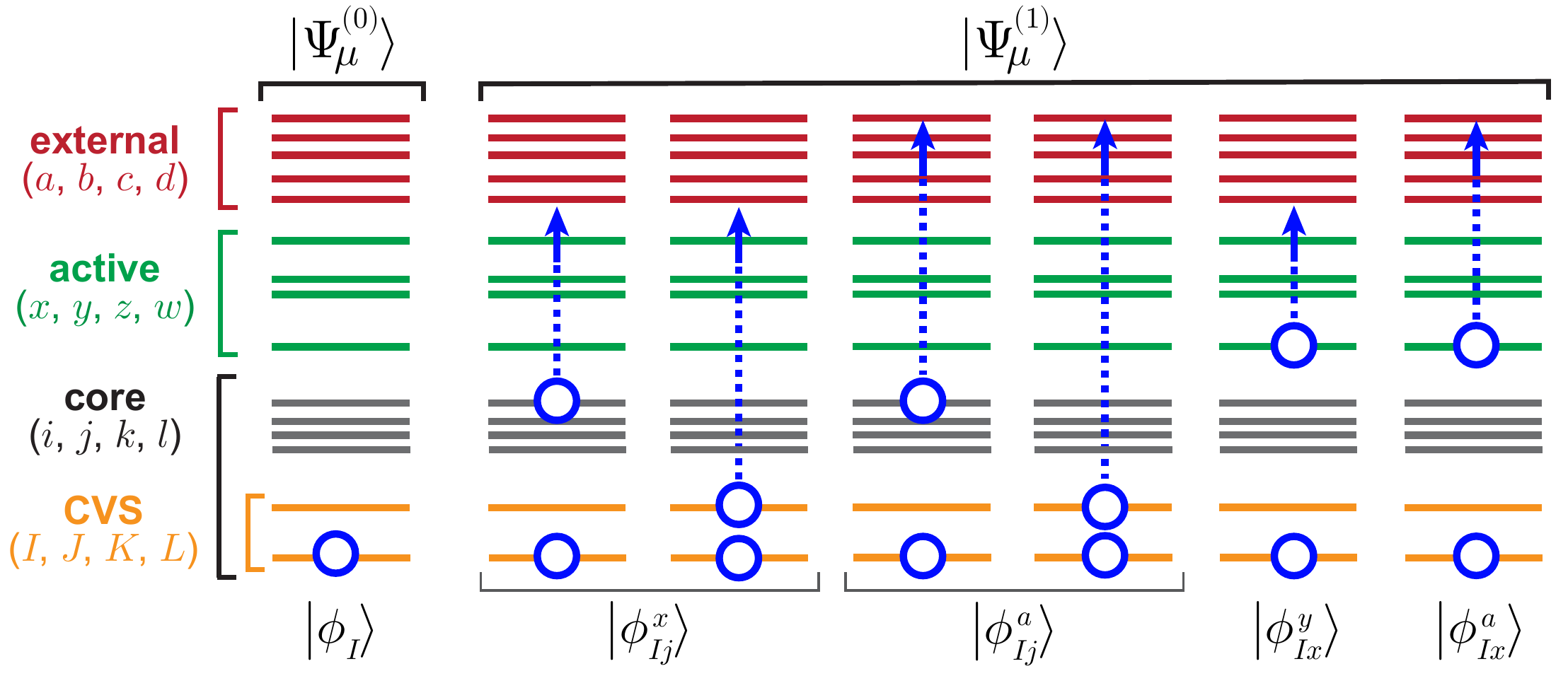}
	\caption{Schematic representation of electronic excitations included in the CVS-MR-ADC method. A circle with a dashed line and an arrow indicate a single excitation. Empty circles represent the ionized orbitals.}
	\label{fgr:excitations-cvs-ip}
\end{figure*}

MR-ADC provides a computationally efficient approach to calculate the ionization energies and transition probabilities by approximating the Green's function of a strongly correlated system using multireference perturbation theory.
Here, the ground-state wavefunction $\ket{\Psi}$ is assumed to be well-approximated by a multiconfigurational reference wavefunction $| \Psi_0 \rangle = \sum_K C_K | \psi_K \rangle$, which is obtained from a complete active-space configuration interaction (CASCI) or self-consistent field (CASSCF) calculation\cite{Werner.1980,Werner.1981,Knowles.1985} including contributions from Slater determinants $| \psi_K \rangle$ with all possible occupations in the active (frontier) molecular orbitals (\autoref{fgr:excitations-cvs-ip}).
The remaining orbitals are classified as core (doubly occupied) and external (unoccupied).
Further, the Hamiltonian $H$ is separated into the zeroth-order $H^{(0)}$ and perturbation $V$ contributions.
In MR-ADC, $H^{(0)}$ is chosen to be the Dyall Hamiltonian,\cite{Dyall.1995,Angeli.2001b,Angeli.2001,Angeli.2004} which includes all active-space terms of the full electronic Hamiltonian $H$.
This choice ensures that MR-ADC is exact when all orbitals are included in the active space (i.e.\@ full configuration interaction) or is equivalent to the single-reference ADC theory\cite{Schirmer:1982p2395,Schirmer:1983p1237} when there are no active orbitals.\cite{Sokolov.2018}

Expanding the Green's function in the perturbative series and truncating the expansion at the \textit{n}th order
\begin{equation}
	\label{eq:greens_function_approx}
	\mathbf{G}(\omega) \approx \mathbf{G}^{(0)}(\omega) + \mathbf{G}^{(1)}(\omega) + \dots + \mathbf{G}^{(n)}(\omega)
\end{equation}
defines the MR-ADC(n) approximation.
The MR-ADC(n) propagator $\mathbf{G}(\omega)$ in \autoref{eq:greens_function_approx} is expressed as follows:
\begin{equation}
	\mathbf{G}(\omega) = \mathbf{T} (\omega \mathbf{S} - \mathbf{M})^{-1} \mathbf{T}^{\dag}
\end{equation}
where $\mathbf{T}$ is the effective transition moments matrix, $\mathbf{S}$ is the overlap matrix, and $\mathbf{M}$ is the effective Hamiltonian (or so-called Liouvillian) matrix, all calculated to the chosen \textit{n}th perturbation order.
The MR-ADC(n) ionization energies are computed as the eigenvalues of the $\mathbf{M}$ matrix from a generalized eigenvalue problem:
\begin{equation}
\label{eq:mradc-eigenvalue-problem}
	\mathbf{M} \mathbf{Y} = \mathbf{S} \mathbf{Y} \boldsymbol{\Omega}
\end{equation}
\autoref{eq:mradc-eigenvalue-problem} is solved using a multiroot implementation of the Davidson algorithm\cite{Davidson.1975,Liu.1978} that calculates the eigenvalues ($\boldsymbol{\Omega}$) and eigenvectors ($\mathbf{Y}$) for a specified number of lowest-energy transitions in the photoelectron spectrum.

Once the eigenvalue problem is solved, the spectral information is obtained by computing the matrix of spectroscopic amplitudes $\mathbf{X}$:
\begin{equation}
	\mathbf{X} = \mathbf{T} \mathbf{S}^{-1/2} \mathbf{Y}
\end{equation}
The elements of $\mathbf{X}$ can be used to compute the spectroscopic factors\cite{Chatterjee.2019,Banerjee.2021}
\begin{equation}
	P_{\alpha} = \sum_p |X_{p \alpha}|^2
\end{equation}
that provide information about the intensity of a photoelectron transition with ionization energy $\Omega_\alpha$.
The $\mathbf{X}$ amplitudes can be also used to calculate the photoelectron spectral function (so-called density of states)
\begin{equation}
	\label{eq:dos}
	A(\omega) = - \frac{1}{\pi} \mathrm{Im}[\mathrm{Tr} \; \mathbf{G}(\omega)]
\end{equation}
where the MR-ADC(n) Green's function $\mathbf{G}(\omega)$ is computed as:
\begin{equation}
	\mathbf{G}(\omega) = \mathbf{X}(\omega \mathbf{1} - \boldsymbol{\Omega})^{-1} \mathbf{X}^{\dag}
\end{equation}
This straightforward access to the spectroscopic properties is an important advantage of MR-ADC over traditional multireference perturbation 
methods.\cite{Finley:1998p299,Andersson:1990p5483,Andersson:1992p1218,Angeli.2001b,Angeli.2001,Angeli.2004}

In this work, we will employ two MR-ADC methods: the strict second-order approximation (MR-ADC(2)) and its extended version (MR-ADC(2)-X).\cite{Chatterjee.2019,Chatterjee.2020}
\autoref{fgr:mr-adc-matrix-compare} shows the perturbative structure of effective Hamiltonian matrix $\mathbf{M}$ for both methods where each matrix element corresponds to a pair of excitations.
The zeroth-order excitations (denoted as $\ket{\Psi_\mu^{(0)}}$) describe removal of one electron from core or active molecular orbitals.
The first-order excitations ($\ket{\Psi_\mu^{(1)}}$) describe detachment of an electron accompanied by a one-electron excitation between core, active, or external orbitals.
The main difference between the MR-ADC(2) and MR-ADC(2)-X methods is in the matrix block corresponding to the first-order excitations ($\langle \Psi_\nu^{(1)} |$ -- $| \Psi_\mu^{(1)} \rangle$), which is described at zeroth order in MR-ADC(2) and up to the first order in MR-ADC(2)-X (\autoref{fgr:mr-adc-matrix-compare}).
The higher-order treatment of the $\langle \Psi_\nu^{(1)} |$ -- $| \Psi_\mu^{(1)} \rangle$ block in MR-ADC(2)-X significantly improves the description of orbital relaxation effects for singly-ionized states and provides a better description of the satellite transitions, which involve an ionization and a one-electron excitation simultaneously.
In addition, the MR-ADC(2)-X method incorporates higher-order contributions to the effective transition moments matrix $\mathbf{T}$.\cite{Chatterjee.2019,Chatterjee.2020}

\subsection{MR-ADC With Core-Valence Separation for X-ray Photoelectron Spectra}

An important feature of MR-ADC is the ability to simulate electronic excitations involving all electrons and molecular orbitals of the system, in contrast to conventional multireference methods that can only simulate excitations in active orbitals.
This feature is particularly useful for simulating the electronic transitions in X-ray absorption or photoelectron spectra, which originate from doubly occupied core orbitals that are not strongly correlated and should not be included in the active space.
Similar to single-reference ADC or equation-of-motion coupled cluster theories, MR-ADC describes all single and double excitations from core molecular orbitals starting from its second-order approximation (MR-ADC(2)).\cite{Chatterjee.2019}
However, these core-level excitations are buried deep inside the MR-ADC eigenvalue spectrum and are difficult to access using the standard iterative diagonalization algorithms, such as the Davidson method used for solving \autoref{eq:mradc-eigenvalue-problem}.

A common approach to avoid this problem is to introduce the core-valence separation (CVS) approximation, originally proposed by Cederbaum et al. in 1980.\cite{Cederbaum.1980,Barth.1981}
In the CVS method, the excitation energies and properties of core- and valence-excited states are computed separately from each other by neglecting small couplings between the core- and valence-excited electronic configurations.
This decoupling is justified on the basis of large energetic separation between core and valence orbitals and the highly localized nature of the former.
The CVS approach has been widely used to compute core-level excitations with a variety of excited-state methods, such as configuration interaction,\cite{Asmuruf.2008} coupled cluster theory,\cite{Coriani.2015c1e,Vidal.2019,Vidal.2019b,Vidal.2020,Thielen.2021} single-reference ADC theory,\cite{Koppel.1997,Barth.1999,Trofimov.2000,Schirmer.2001,Thiel.2003,Wenzel.2014,Wenzel.2015,Liu.2019,Zheng.2019} linear-response CASSCF,\cite{Paris.2021} linear-response density cumulant theory,\cite{Peng.2019} second-order excited-state perturbation theory,\cite{Garner.2020}, and density functional theory combined with multireference configuration interaction.\cite{Seidu.2019}

To introduce the CVS approximation in the MR-ADC framework for ionized states, we first select several core orbitals of the reference wavefunction $\ket{\Psi_0}$ to form a set of ``CVS'' orbitals, which includes all lowest-energy molecular orbitals up to and including the orbital probed in the X-ray photoelectron spectrum to be simulated.
The core ionization energies and spectroscopic properties are then computed by solving the MR-ADC equations in the basis of zeroth- and first-order excitations ($\ket{\Psi_\mu^{(0)}}$ and $\ket{\Psi_\mu^{(1)}}$) that involve at least one CVS orbital, as shown schematically in \autoref{fgr:excitations-cvs-ip}.
We note that several variants of the CVS approximation have been proposed in the past that differ in the treatment of double excitations and frozen-core approximation.\cite{Barth.1999,Koppel.1997,Trofimov.2000,Wenzel.2014,Wenzel.2015,Coriani.2015c1e,Vidal.2019,Peng.2019,Liu.2019,Zheng.2019}
The CVS scheme used in this work does not introduce the frozen-core approximation, incorporates double excitations involving two CVS orbitals, and is equivalent to the one employed in 2015 by Coriani and Koch.\cite{Coriani.2015c1e}

Introducing the CVS approximation greatly reduces the size of matrix $\mathbf{M}$ (\autoref{fgr:mr-adc-matrix-compare}) diagonalized in \autoref{eq:mradc-eigenvalue-problem}, leading to significant computational savings.
An important contribution to the reduction in computational cost originates from neglecting the excitations involving only active orbitals, which are described by computing the CASCI wavefunctions of ionized system in the MR-ADC implementation for photoelectron spectra.\cite{Chatterjee.2019}
As a result, the CVS-approximated MR-ADC method does not require calculating the excited-state CASCI wavefunctions and transition reduced density matrices, so that only the reference (ground-state) wavefunction ($\ket{\Psi_0}$) and reduced density matrices are necessary for the calculations.

\section{Computational Details}\label{sec:ComputationalDetails}

We combined the CVS approximation with the second-order MR-ADC(2) and extended second-order MR-ADC(2)-X methods for electron ionization.\cite{Chatterjee.2019,Chatterjee.2020}
The resulting CVS-MR-ADC(2) and CVS-MR-ADC(2)-X methods were implemented in \textsc{Prism}, a standalone Python program for spectroscopic simulations of multireference systems.
To obtain the one- and two-electron integrals and the reference CASSCF wavefunctions, \textsc{Prism} was interfaced with the \textsc{Pyscf} software package.\cite{Sun.2020}
The CVS-MR-ADC core ionization energies and X-ray photoelectron spectra were compared to the energies and spectra calculated using the single-reference CVS-SR-ADC(2), CVS-SR-ADC(2)-X, and CVS-SR-ADC(3) methods.
These methods were implemented in a local version of \textsc{Pyscf} by modifying the implementation of non-Dyson single-reference ADC\cite{Banerjee.2019,Banerjee.2021} using the same CVS approach as employed in CVS-MR-ADC. Additional calculations using CVS-EOM-CCSD were performed using ORCA.\cite{Neese.2012,Neese.2020}

The MR-ADC calculations require specifying parameters for removing linearly-dependent excitations in the solution of the MR-ADC equations.\cite{Sokolov.2018}
As in our previous work,\cite{Chatterjee.2020,Chatterjee.2019,Mazin.2021} we employ the $\eta_d = 10^{-10}$ parameter to remove linearly-dependent double excitations.
For the single excitations, we use a modified truncation approach where the singles amplitudes are damped by a sigmoidal function
\begin{equation}
	\tilde{t}_{\mu} = t_{\mu} \left(1 + \exp\left(\frac{10}{\beta} (\log(\eta_s) - \log(s_\mu))\right)\right)^{-1}
	\label{eq:sigmoidal-damping}
\end{equation}
where $\tilde{t}_{\mu}$ is the damped singles amplitude, $s_\mu$ is the overlap metric eigenvalue corresponding to the amplitude $t_{\mu}$, $\eta_s$ is the singles truncation parameter, and $\beta$ is the damping strength parameter.
For small $\beta$ ($\sim 10^{-6}$) this approach is equivalent to the truncation scheme used in our previous work.\cite{Chatterjee.2019,Chatterjee.2020}
Using larger $\beta$ ensures that the computed ionization energies are smooth functions of the internuclear coordinates.
We employ $\eta_s = 10^{-5}$ and $\beta = 4$ in all MR-ADC calculations reported in this work.

The CVS-MR-ADC methods were first benchmarked against the CVS-EOM-CCSDT results reported by Liu et al.\cite{Liu.2019} for the K-edge core ionization energies of 16 closed-shell molecules: \ce{CO}, \ce{N2}, \ce{HF}, \ce{F2}, \ce{HCN}, \ce{CO2}, \ce{N2O}, \ce{H2O}, \ce{NH3}, \ce{CH2O}, \ce{CH4}, \ce{CH3CN}, \ce{CH3NC}, \ce{C2H2}, \ce{C2H4}, and \ce{CH3OH}.
These calculations employed the same equilibrium geometries, basis set (cc-pCVTZ-X2C),\cite{CFOUR.page} and scalar relativistic corrections (X2C)\cite{Dyall.2001.4,Liu.2009} as used by Liu et al.\cite{Liu.2019}.
To study the active space dependence of the CVS-MR-ADC results, for each molecule calculations were performed for two different active spaces, which we abbreviate as CAS[Small] and CAS[Large].
Detailed information about the composition of these active spaces can be found in the Electronic Supplementary Information (ESI).\dag

Next, we performed the CVS-SR-ADC and CVS-MR-ADC calculations of ionization energies and potential energy curves for the K-edge core-ionized state of molecular nitrogen (\ce{N2}) along its dissociation pathway.
We used the cc-pCVTZ basis set for all methods.
The CVS-MR-ADC calculations employed the CASSCF reference wavefunction with 10 electrons in 8 active orbitals (10e,8o).
The CVS-SR-ADC potential energies of the \ce{N2} core-ionized state were computed by adding the total ground-state second-order M\"oller--Plesset (MP2)\cite{Moller.1934} energy and the CVS-SR-ADC core ionization energies at each geometry. 
To calculate the \ce{N2} core-ionized potential energy curve using CVS-MR-ADC, we combined the ground-state energy computed using partially-contracted second-order N-electron valence perturbation theory (NEVPT2)\cite{Angeli.2001,Angeli.2001b} and the CVS-MR-ADC core ionization energies. 
The MP2 and NEVPT2 energies were computed using the CVS-SR-ADC and CVS-MR-ADC implementations, respectively.

Finally, we used the CVS-MR-ADC methods to calculate the X-ray photoelectron spectra of ozone molecule (\ce{O3}) and three benzyne singlet diradicals (\textit{ortho}-, \textit{meta}-, and \textit{para}-benzynes).
All multiconfigurational calculations of ozone employed the CASSCF reference wavefunction with 12 electrons in 9 active orbitals (12e,9o).
For the benzyne diradicals the (8e,8o) CASSCF reference wavefunction was used.
The non-relativistic CVS-MR-ADC calculations of \ce{O3} employed the cc-pCVTZ basis set,\cite{Dunning.1989.1,Kendall.1992,Woon.1995} while the recontracted cc-pCVTZ-X2C basis set\cite{CFOUR.page} was used when scalar relativistic effects were included using the X2C method.\cite{Dyall.2001.4,Liu.2009}
All benzyne calculations employed the cc-pCVDZ-X2C basis set along with the X2C treatment of scalar relativistic effects.
The equilibrium geometries of all four molecules were computed using the CASSCF method implemented in the MOLPRO package.\cite{Werner.2012,MOLPRO,Werner.2020} 
The \ce{O3} and benzyne optimized geometries and active spaces are reported in the ESI.\dag \ 
The ADC photoelectron spectra of all molecules were simulated in the sudden approximation by plotting the density of states (\autoref{eq:dos}) calculated by adding a small imaginary broadening to the core ionization energies.
The sudden approximation assumes the decoupling of a photoelectron from the ionized system and allows to avoid the explicit treatment of free electron wavefunction.
In addition, the calculated photoelectron intensities neglect vibronic effects.

Together with the CVS-MR-ADC results, we report core ionization energies of \ce{N2} and \ce{O3} computed using multireference configuration interaction with single and double excitations (MRCISD)\cite{Buenker.1974,Shepard.1992} that employed occupation restricted multiple active space (ORMAS) reference wavefunction.\cite{Ivanic.2003d2e,Ivanic.2003}
For \ce{N2}, the ORMAS calculations were performed using the two-step self-consistent procedure proposed by Rocha,\cite{Rocha.2011} in which the active space was split into two subsets of orbitals.
The first subset contained the core 1s orbitals of nitrogen atoms.
The second subset included the same active orbitals as in the (10e,8o) active space used in the CVS-MR-ADC calculations.
Core-ionized states were computed by restricting the occupation of the core subspace to have one electron less than that in the ground state.
A similar procedure has been successfully used in the calculations of core-ionized potential energy curves of diatomic molecules.\cite{Moura.2013,Corral.201794l,Battacharya.2021}
For \ce{O3}, the two-step self-consistent optimization of the ORMAS wavefunction leads to root-flipping problems due to two excited states of interest having the same symmetry.
For this reason, we performed the \ce{O3} ORMAS calculations using the Hartree--Fock orbitals. 
As for \ce{N2}, the active space consisted of two subsets: one including the oxygen 1s orbitals and another one incorporating the orbitals from the (12e,9o) active space used for CVS-MR-ADC.
All ORMAS and MRCISD calculations were performed using GAMESS\cite{GAMESS.2020} and a standalone Python script.\cite{isGAMESS}

\section{Results and Discussion}
\label{sec:Results}

\subsection{Benchmarking the Accuracy of CVS-MR-ADC for Weakly-Correlated Molecules}
\label{subsec:CVS-benchmark}

\begin{table*}[t!]
\small
  \caption{K-edge core ionization energies (eV) computed using the CVS-SR-ADC and CVS-MR-ADC methods with the cc-pCVTZ-X2C basis set and the X2C scalar relativistic corrections.
  	Asterisk indicates ionization in the 1s orbital of an atom.
  	Also shown are the reference core ionization energies from CVS-EOM-CCSDT\cite{Liu.2019} and experiment,\cite{Jolly.1984,Beach.1984} as well as mean absolute errors ($\Delta_{\mathrm{MAE}}$) and standard deviations ($\Delta_{\mathrm{STD}}$) relative to the CVS-EOM-CCSDT results.}
  \label{tbl:cvs-benchmark}
  \begin{tabular*}{0.98\textwidth}{@{\extracolsep{\fill}}llllllllll}
    \hline
    \multirow{2}{*}{molecule} & \multirow{2}{*}{SR-ADC(2)} & \multirow{2}{*}{SR-ADC(2)-X} & \multirow{2}{*}{SR-ADC(3)} & \multicolumn{2}{c}{MR-ADC(2)} & \multicolumn{2}{c}{MR-ADC(2)-X} & \multirow{2}{*}{EOM-CCSDT} & \multirow{2}{*}{Experiment} \\
     & & & & CAS[Small] & CAS[Large] & CAS[Small] & CAS[Large] & & \\
    \hline
\ce{\textbf{C}^*2H4}    & 292.30 & 290.25 & 293.55 & 292.28 & 293.31 & 290.69 & 290.59 & 290.85 & 290.82 \\
\ce{\textbf{C}^*H4}     & 292.18 & 290.32 & 293.20 & 292.24 & 293.73 & 290.29 & 291.42 & 290.86 & 290.91 \\
\ce{\textbf{C}^*2H2}    & 292.47 & 290.61 & 294.05 & 292.84 & 293.62 & 290.60 & 290.77 & 291.36 & 291.14 \\
\ce{CH3N\textbf{C}^*}   & 294.73 & 292.42 & 294.69 & 294.56 & 294.83 & 291.96 & 292.37 & 292.35 & 292.37 \\
\ce{\textbf{C}^*H3OH}   & 294.03 & 291.98 & 294.78 & 294.40 & 294.88 & 292.02 & 292.58 & 292.47 & 292.43 \\
\ce{CH3\textbf{C}^*N}   & 294.10 & 292.23 & 295.14 & 294.23 & 293.98 & 292.06 & 292.19 & 292.81 & 292.45 \\
\ce{\textbf{C}^*H3CN}   & 294.25 & 292.40 & 295.19 & 294.29 & 294.81 & 292.28 & 292.77 & 292.90 & 292.98 \\
\ce{\textbf{C}^*H3NC}   & 294.83 & 292.94 & 296.12 & 295.12 & 294.94 & 293.11 & 293.00 & 293.41 & 293.35 \\
\ce{H\textbf{C}^*N}     & 294.88 & 292.91 & 295.66 & 295.00 & 295.51 & 292.82 & 293.60 & 293.59 & 293.40 \\
\ce{\textbf{C}^*H2O}    & 296.58 & 294.30 & 296.56 & 296.14 & 297.18 & 294.23 & 294.79 & 294.62 & 294.47 \\
\ce{\textbf{C}^*O}      & 298.58 & 296.39 & 297.80 & 298.10 & 298.31 & 295.63 & 296.37 & 296.47 & 296.21 \\
\ce{\textbf{C}^*O2}     & 300.53 & 298.35 & 299.48 & 299.76 & 299.62 & 297.12 & 297.27 & 298.03 & 297.69 \\
\ce{\textbf{N}^*H3}     & 405.73 & 404.64 & 409.18 & 405.69 & 407.46 & 404.86 & 405.47 & 405.55 & 405.52 \\
\ce{CH3C\textbf{N}^*}   & 405.91 & 404.73 & 410.22 & 407.25 & 408.10 & 404.87 & 405.58 & 405.71 & 405.64 \\
\ce{HC\textbf{N}^*}     & 406.88 & 405.74 & 411.10 & 408.06 & 409.71 & 405.85 & 406.95 & 406.88 & 406.78 \\
\ce{CH3\textbf{N}^*C}   & 406.23 & 405.45 & 411.61 & 407.48 & 409.50 & 405.36 & 407.12 & 407.02 & 406.67 \\
\ce{\textbf{N}^*NO}     & 409.58 & 408.30 & 412.57 & 411.16 & 411.19 & 408.60 & 408.07 & 408.92 & 408.71 \\
\ce{\textbf{N}2^*}      & 410.31 & 408.93 & 413.20 & 411.91 & 412.14 & 409.71 & 409.01 & 410.03 & 409.98 \\
\ce{N\textbf{N}^*O}     & 413.98 & 412.77 & 416.11 & 413.85 & 415.77 & 412.01 & 412.80 & 413.15 & 412.59 \\
\ce{CH3\textbf{O}^*H }  & 538.00 & 537.83 & 544.62 & 538.65 & 539.16 & 538.39 & 538.19 & 539.00 & 539.11 \\
\ce{CH2\textbf{O}^*}    & 538.18 & 538.09 & 545.83 & 539.34 & 543.02 & 539.44 & 540.10 & 539.44 & 539.48 \\
\ce{H2\textbf{O}^*}     & 538.68 & 538.53 & 544.74 & 538.99 & 541.26 & 538.53 & 540.18 & 539.79 & 539.90 \\
\ce{C\textbf{O}2^*}     & 540.22 & 540.38 & 547.73 & 541.17 & 543.09 & 539.98 & 540.02 & 541.40 & 541.28 \\
\ce{NN\textbf{O}^*}     & 539.95 & 540.33 & 548.86 & 540.87 & 544.16 & 540.07 & 540.93 & 541.63 & 541.42 \\
\ce{C\textbf{O}^*}      & 540.95 & 541.07 & 549.06 & 542.99 & 545.67 & 541.03 & 543.05 & 542.57 & 542.55 \\
\ce{H\textbf{F}^*}      & 691.99 & 692.75 & 699.99 & 693.13 & 697.10 & 692.86 & 695.13 & 694.22 & 694.23 \\
\ce{\textbf{F}2^*}      & 695.33 & 695.07 & 702.59 & 695.00 & 698.96 & 695.22 & 697.02 & 696.72 & 696.69 \\
$\Delta_{\mathrm{MAE}}$ & 1.26   & 0.84   & 3.77   & 1.22   & 2.20   & 0.82   & 0.44 \\
$\Delta_{\mathrm{STD}}$ & 1.40   & 0.51   & 1.75   & 1.10   & 0.68   & 0.47   & 0.55 \\
  \hline
  \end{tabular*}
\end{table*}

\begin{figure}[t!]
	\centering
	\includegraphics[width=\columnwidth]{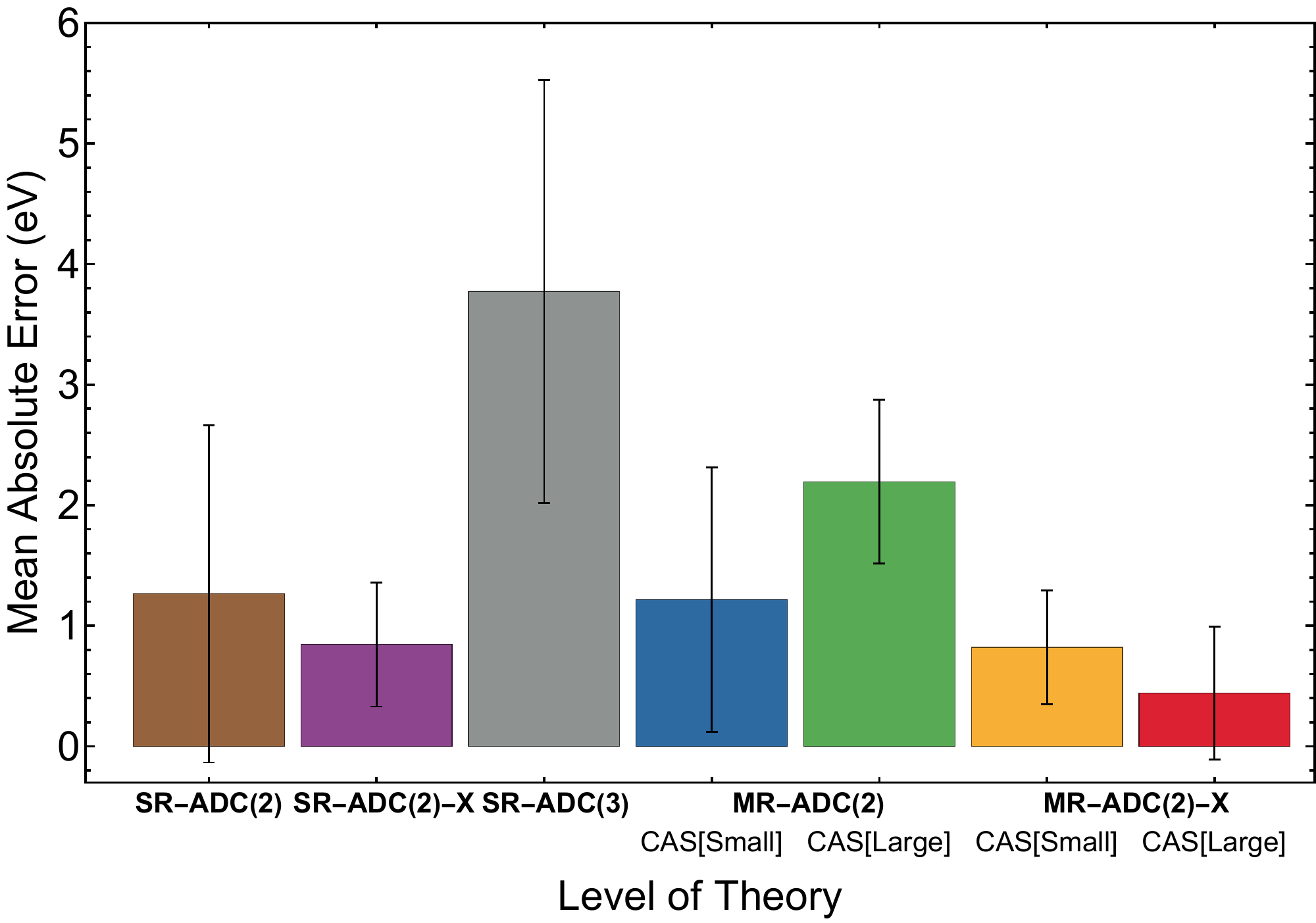}
	\caption{Mean absolute errors ($\Delta_{\mathrm{MAE}}$) in the K-edge core ionization energies of weakly-correlated molecules computed by the CVS-SR-ADC and CVS-MR-ADC methods, relative to CVS-EOM-CCSDT results.\cite{Liu.2019} Error bars show the corresponding standard deviation of errors ($\Delta_{\mathrm{STD}}$). All calculations used the cc-pCVTZ-X2C basis set and the X2C description of scalar relativistic effects.}
	\label{fgr:mean-absolute-error-plot}
\end{figure}

\begin{figure}[t!]
	\centering
	\includegraphics[width=\columnwidth]{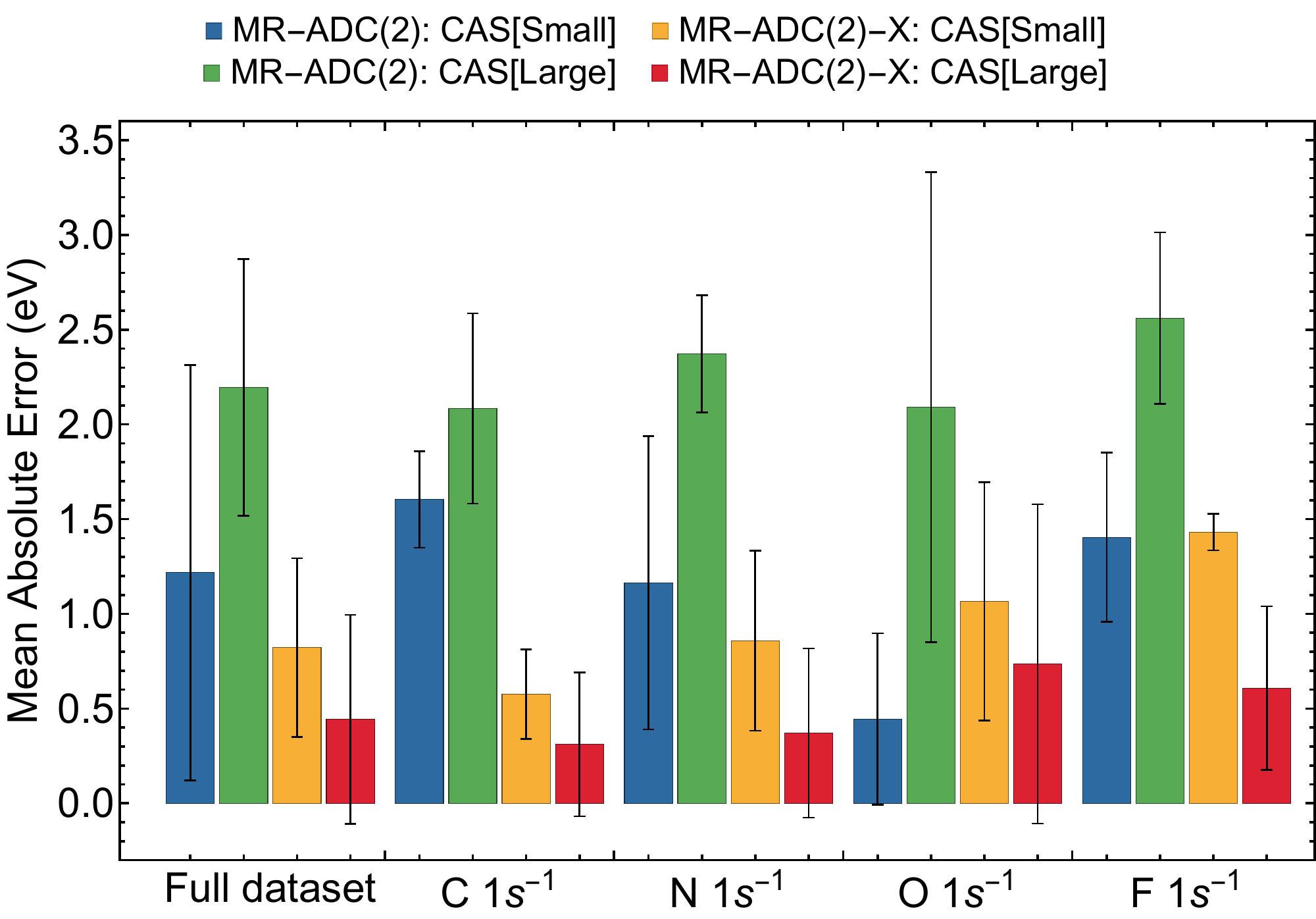}
	\caption{Mean absolute errors ($\Delta_{\mathrm{MAE}}$) in core ionization energies for different K-edges of weakly-correlated molecules computed using the CVS-MR-ADC methods, relative to the CVS-EOM-CCSDT results.\cite{Liu.2019} Error bars show the corresponding standard deviation of errors ($\Delta_{\mathrm{STD}}$). All calculations used the cc-pCVTZ-X2C basis set and the X2C description of scalar relativistic effects.}
	\label{fgr:mean-absolute-error-edge-plot}
\end{figure}

Before we apply CVS-MR-ADC to multireference systems, we first benchmark its accuracy for molecules that do not exhibit strong correlation effects.
Our goal here is to assess the CVS-MR-ADC accuracy in capturing the weak (dynamic) electron correlation in core-ionized electronic states and to compare it to the accuracy of well-established single-reference ADC methods (CVS-SR-ADC). 
To accomplish this, we performed the CVS-MR-ADC calculations for the benchmark set of Liu and co-workers\cite{Liu.2019} that contains highly-accurate (CVS-EOM-CCSDT) reference data for 27 K-edge ionization transitions in 16 weakly-correlated small molecules.
The CVS-MR-ADC core ionization energies are reported in \autoref{tbl:cvs-benchmark}, along with the results from CVS-SR-ADC, the reference CVS-EOM-CCSDT energies computed by Liu et al.,\cite{Liu.2019} and experimental data for each transition.\cite{Jolly.1984,Beach.1984}
All calculations were performed using the cc-pCVTZ-X2C basis set and include the X2C scalar relativistic corrections.
\autoref{tbl:cvs-benchmark} also shows the mean absolute errors ($\Delta_{\mathrm{MAE}}$) and standard deviations of errors ($\Delta_{\mathrm{STD}}$) for each method relative to CVS-EOM-CCSDT.

In contrast to CVS-SR-ADC, the CVS-MR-ADC calculations require choosing an active space. 
(When no orbitals are included in the active space, CVS-SR-ADC and CVS-MR-ADC are equivalent). 
While in multireference systems the active space is usually chosen to include the frontier orbitals with strongly correlated electrons, different choices of active space are possible for weakly-correlated molecules, such as the molecules in our benchmark set.
In this case, including the orbitals in active space can be used to incorporate some orbital relaxation and dynamic correlation effects that are not captured by the low-order single-reference methods (e.g., improved description of double and higher excitations).
To study the active-space dependence of CVS-MR-ADC results in molecules that do not exhibit strong correlation, we employed two types of active spaces, which we denote as CAS[Small] and CAS[Large].
The CAS[Large] active space was designed to include most of the valence orbitals, except for some occupied and virtual orbitals with large negative and positive orbital energies, respectively.
For all molecules, CAS[Small] included the highest-occupied and lowest-unoccupied molecular orbitals (HOMO and LUMO).
In addition, for molecules with $\pi$-bonds, all $\pi$-bonding and antibonding orbitals were included in CAS[Small].
For molecules without $\pi$-bonds, CAS[Small] also included an antibonding counterpart of HOMO.

The performance of all ADC methods measured by $\Delta_{\mathrm{MAE}}$ and $\Delta_{\mathrm{STD}}$ is illustrated in \autoref{fgr:mean-absolute-error-plot}.
At each level of perturbation theory, the CVS-MR-ADC results are within the standard deviation of CVS-SR-ADC results, indicating that the performance of both methods for weakly-correlated molecules in our benchmark set is similar.
The best agreement with the reference core ionization energies from CVS-EOM-CCSDT is demonstrated by the extended second-order approximations (CVS-SR- and CVS-MR-ADC(2)-X) that show $\Delta_{\mathrm{MAE}}$ < 1 eV and $\Delta_{\mathrm{STD}}$ $\sim$ 0.5 eV.
The second-order methods (CVS-SR- and CVS-MR-ADC(2)) exhibit intermediate accuracy with $\Delta_{\mathrm{MAE}}$ ranging from 1.22 to 2.20 eV and $\Delta_{\mathrm{STD}}$ within the 0.7 -- 1.4 eV range.
It is important to note that the computed errors in core ionization energies originate from a careful balance of errors in the ground and core-ionized electronic state energies and that increasing the level of theory may affect this balance and worsen the performance. 
This is demonstrated by the CVS-SR-ADC(3) approximation that shows the largest errors ($\Delta_{\mathrm{MAE}}$ = 3.77 eV, $\Delta_{\mathrm{STD}}$ = 1.75 eV) among all levels of ADC theory.
The worse performance of CVS-SR-ADC(3) relative to CVS-SR-ADC(2)-X has been also observed by Wenzel et al.\cite{Wenzel.2015} in the simulations of X-ray absorption spectra, indicating that the third-order approximation is not well-balanced for the calculations of core-excited states.

\autoref{fgr:mean-absolute-error-plot} demonstrates that the errors of CVS-MR-ADC methods exhibit different dependence on the active space. 
As discussed above, for weakly-correlated molecules this dependence originates primarily from the differences in description of dynamic correlation effects and orbital relaxation, i.e.\@ choice of reference (CASSCF) orbitals. 
The CVS-MR-ADC(2) method shows a significant variation in the results upon enlarging the active space from CAS[Small] to CAS[Large], leading to an increase in $\Delta_{\mathrm{MAE}}$ by $\sim$ 1 eV.
The active space dependence is significantly weakened in the CVS-MR-ADC(2)-X method that incorporates a higher-level description of orbital relaxation effects leading to a much smaller change in $\Delta_{\mathrm{MAE}}$ (0.38 eV) as a result of increasing the active space size.
Since CVS-MR-ADC(2) and CVS-MR-ADC(2)-X provide a similar description of dynamic correlation effects, the stronger active-space dependence of CVS-MR-ADC(2) can be attributed to the well-known sensitivity of the second-order perturbation theories (such as MR-ADC(2)) to the choice of reference orbitals.\cite{Nakayama.1998,Andrzejak.2011} 
Increasing the active space size, shifts the balance of error cancellation in the CVS-MR-ADC(2) results to higher $\Delta_{\mathrm{MAE}}$ (still within $\Delta_{\mathrm{STD}}$ of CVS-SR-ADC(2), \autoref{fgr:mean-absolute-error-plot}), while lowering $\Delta_{\mathrm{MAE}}$ for CVS-MR-ADC(2)-X. 

To analyze the performance of CVS-MR-ADC methods for different K-edges, we computed $\Delta_{\mathrm{MAE}}$ and $\Delta_{\mathrm{STD}}$ in the C, N, O, and F K-edge core ionization energies shown in \autoref{fgr:mean-absolute-error-edge-plot}.
Using CAS[Small], the CVS-MR-ADC results show some variation in core ionization energies with changes in $\Delta_{\mathrm{MAE}}$ of up to $1.2$ eV between K-edges of different elements.
Increasing the active space size weakens the K-edge dependence of $\Delta_{\mathrm{MAE}}$ to $\sim$ 0.5 eV, giving rise to a more consistent performance of CVS-MR-ADC across the C, N, O, and F K-edges.

Overall, our benchmark results indicate that the second-order CVS-MR-ADC methods provide accurate predictions of the K-edge core ionization energies for small weakly-correlated molecules. 
While both CVS-MR-ADC(2) and CVS-MR-ADC(2)-X have a similar computational cost ($\mathcal{O}(N^5)$ scaling with the basis set size $N$), CVS-MR-ADC(2)-X is more accurate and exhibits a weaker active-space dependence.
Although for this benchmark study the performance of CVS-MR-ADC and CVS-SR-ADC is similar, the multireference theory is expected to be more accurate and reliable for molecules that exhibit strong electron correlation.
In the following, we demonstrate this by computing the core ionization energies of \ce{N2} along bond dissociation (\autoref{subsec:dissociation-n2}) and the X-ray photoelectron spectra of ozone (\autoref{sec:ozone}) and benzyne (\autoref{sec:benzynes}) singlet diradicals with multireference ground-state electronic structure.

\subsection{Core Ionization of Molecular Nitrogen Along Dissociation Pathway}
\label{subsec:dissociation-n2}

\begin{figure}[t!]
	\centering
	\includegraphics[width=\columnwidth]{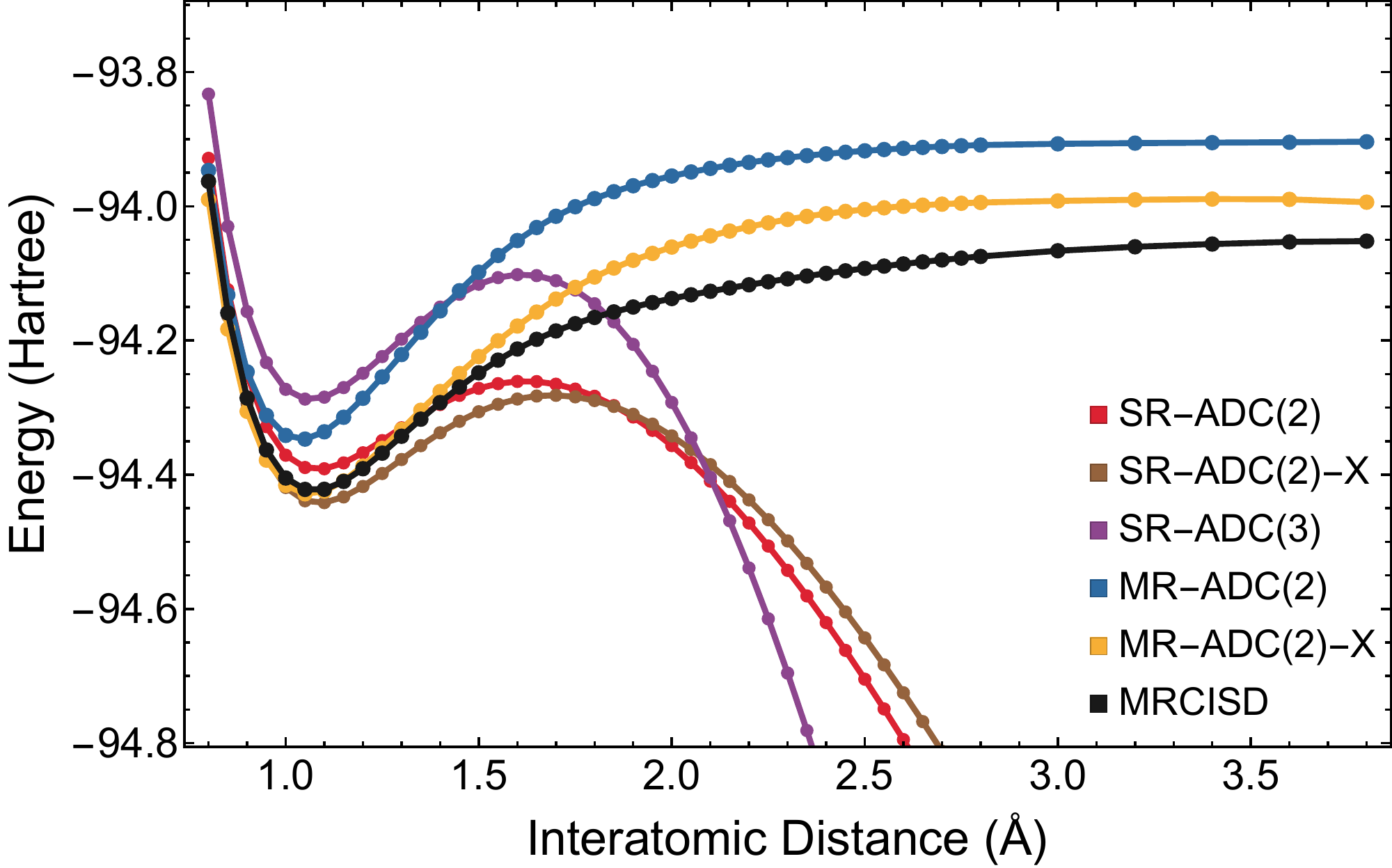}
	\caption{Potential energy curves for the K-edge core-ionized excited state of molecular nitrogen computed using the CVS-SR-ADC, CVS-MR-ADC, and MRCISD methods with the cc-pCVTZ basis set. Multiconfigurational calculations were performed using a CASSCF(10e,8o) reference wavefunction.}
	\label{fgr:potential-energy-plot}
\end{figure}

\begin{figure}[t!]
	\centering
	\includegraphics[width=\columnwidth]{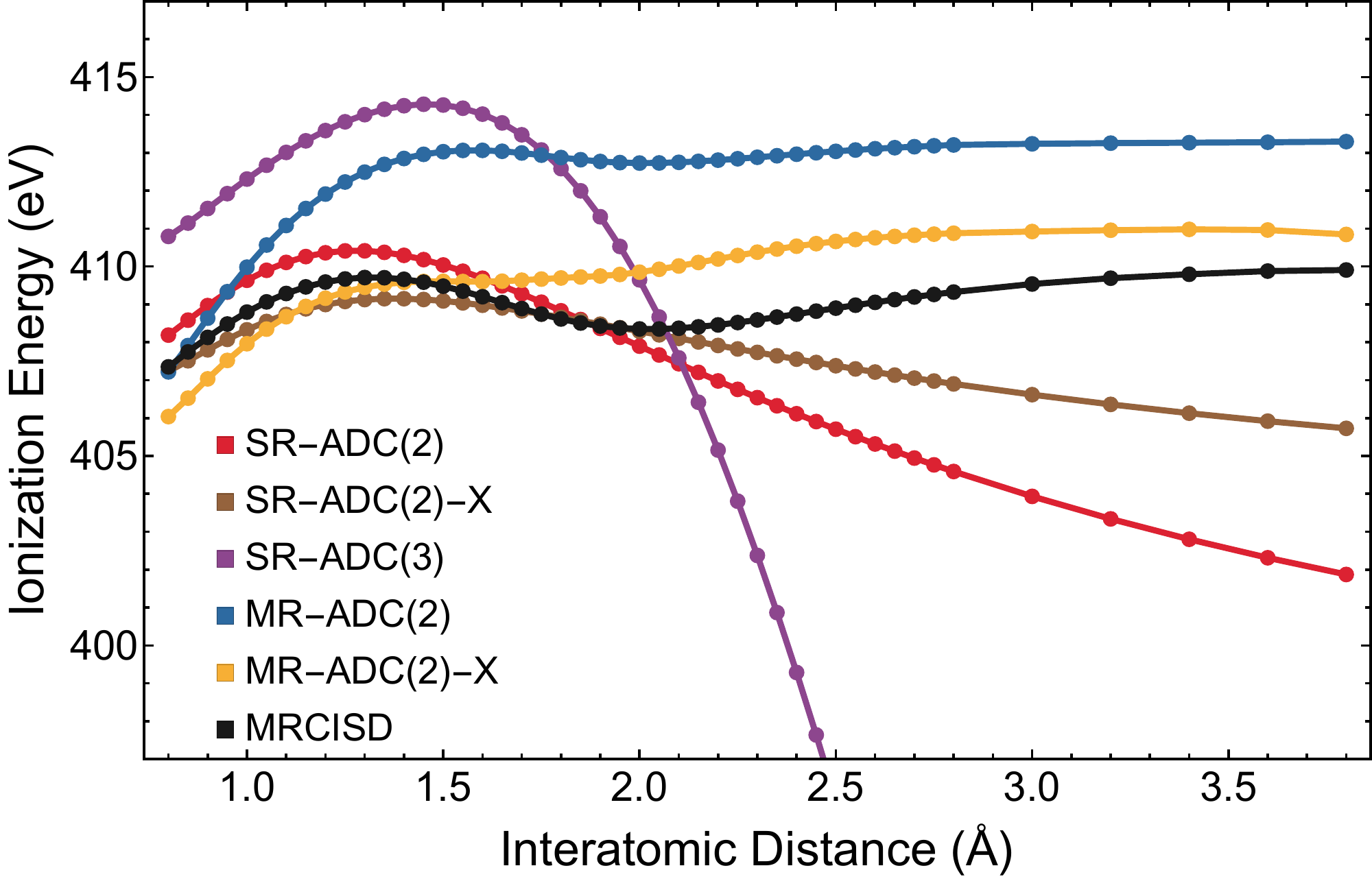}
	\caption{K-edge core ionization energy along the dissociation pathway of molecular nitrogen computed using the CVS-SR-ADC, CVS-MR-ADC, and MRCISD methods with the cc-pCVTZ basis set. Multiconfigurational calculations were performed using a CASSCF(10e,8o) reference wavefunction.}
	\label{fgr:ionization-potential-plot}
\end{figure}

To assess the performance of CVS-MR-ADC for multireference systems, we begin by computing the core ionization energies and the potential energy curves of the first core-ionized state along dissociation of the nitrogen molecule (\ce{N2}).\cite{Moura.2013,Corral.201794l,Battacharya.2021}
Calculations of core ionization energies along bond dissociation pathways find important applications in the interpretation of time-resolved XPS experiments, which can probe molecules at non-equilibrium geometries where multireference effects are significant.

\autoref{fgr:potential-energy-plot} shows the potential energy curves (PEC's) for the lowest-energy K-edge core-ionized state of \ce{N2} computed using the CVS-SR-ADC and CVS-MR-ADC methods along with the reference results from MRCISD.
At short internuclear distances (r(N$-$N) $\le$ 1.3 \angstrom), PEC computed using CVS-MR-ADC(2)-X overlaps with the reference PEC from MRCISD showing the best performance out of all ADC methods.
CVS-SR-ADC(2) and CVS-MR-ADC(2) produce similar energies for r(N$-$N) $\le$ 1.10 \angstrom), but their PEC's deviate from each other at longer distances where the CVS-MR-ADC(2) curve is more parallel to the PEC from MRCISD.
Among all ADC methods, CVS-SR-ADC(3) shows the largest error in the computed total energy of core-ionized state at short distances (r(N$-$N) $\le$ 1.5 \angstrom).
Upon increasing the internuclear separation, the PEC's computed using all three CVS-SR-ADC methods show an unphysical barrier at $\sim$ 1.6 -- 1.7 \angstrom and diverge away from the MRCISD PEC at even longer distances. 
The CVS-MR-ADC methods produce qualitatively correct potential energy curves with CVS-MR-ADC(2)-X demonstrating the best agreement with MRCISD at all distances.

The core ionization energies of \ce{N2} computed using the ADC and MRCISD methods are plotted in \autoref{fgr:ionization-potential-plot} as a function of the N$-$N distance.
As for the total energies, the worst agreement with MRCISD is shown by CVS-SR-ADC(3), which produces significant errors ($>$ 5 eV) in core ionization energy at shorter bond lengths (r(N$-$N) $\le$ 1.6 \angstrom) and a diverging curve at longer distances. 
CVS-SR-ADC(2) and CVS-MR-ADC(2) show similar results near equilibrium, but their ionization energy curves separate at longer distances.
The CVS-MR-ADC(2) curve shows the qualitative features of the MRCISD curve with an inflection point at $\sim$ 1.9 \angstrom (same point is at $\sim$ 1.6 \angstrom for MRCISD) and a flat dissociation region for r(N$-$N) $\ge$ 3 \angstrom.
In contrast, the ionization energy computed using CVS-SR-ADC(2) continues to change significantly well past 3 \angstrom.
The CVS-SR-ADC(2)-X and CVS-MR-ADC(2)-X methods demonstrate the best agreement with MRCISD.
Although the ionization energies computed using both methods are within 2 eV of the MRCISD results for r(N$-$N) $\le$ 2 \angstrom, the CVS-MR-ADC(2)-X curve is more parallel to the MRCISD curve at longer distances showing an inflection point at 1.6 \angstrom and a flat dissociation region. 
The CVS-SR-ADC(2)-X curve continues to decrease in energy past 3 \angstrom, although at a slower pace compared to CVS-SR-ADC(2).
Single-point calculations at 4 and 5 \angstrom in the dissociation region reveal significant ($\sim$ 0.55 eV) changes in the CVS-SR-ADC(2)-X core ionization energy due to its inability to properly treat multireference effects.
At 5 \angstrom, the CVS-SR-ADC(2)-X error in ionization energy ($-5.26$ eV) significantly exceeds that of CVS-MR-ADC(2)-X (0.62 eV), relative to MRCISD. 

\subsection{Core-Ionized States of Ozone}
\label{sec:ozone}

\begin{figure*}[t!]
	\centering
	\includegraphics[width=1.0\textwidth]{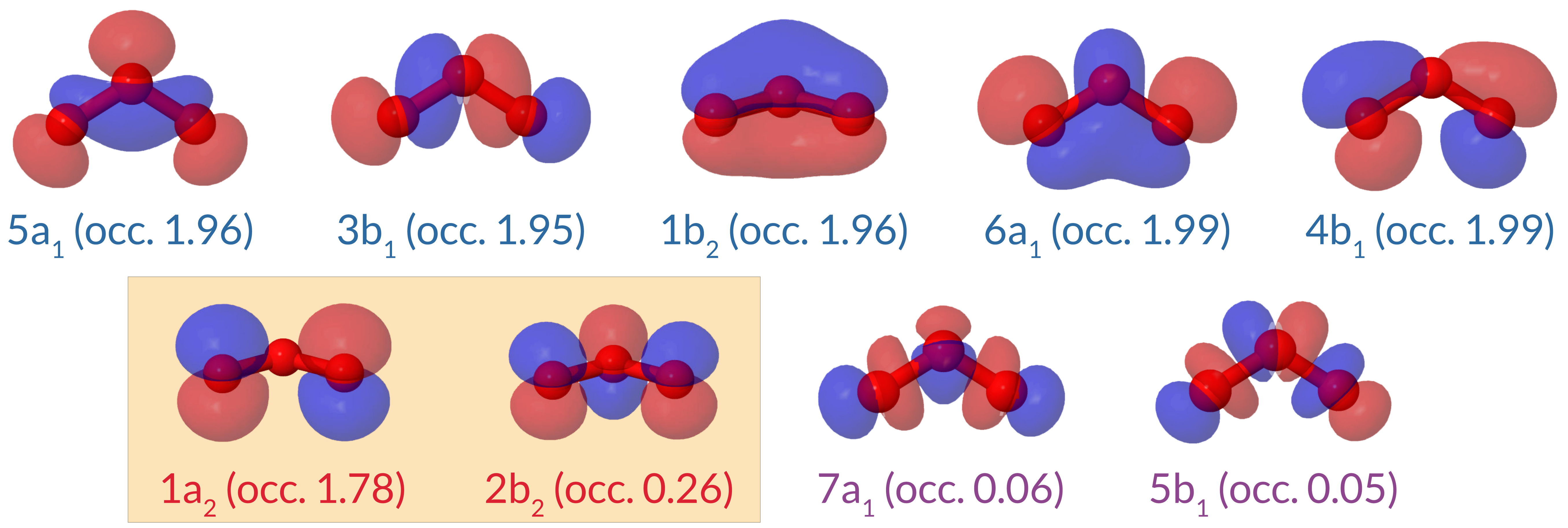}
	\caption{Natural orbitals and occupations of the ozone molecule computed using CASSCF with the (12e,9o) active space (cc-pCVTZ basis set). Two frontier orbitals describing the ozone diradical character are highlighted.}
	\vspace{1\parsep}
	\label{fgr:natorbs-ozone}
\end{figure*}

\begin{table*}[t!]
	\small
	\caption{Oxygen K-edge core ionization energies (eV) of ozone. 
		\ce{O_C} and \ce{O_T} stand for the central and terminal oxygen atoms, respectively, and the difference between ionization energies of these sites is presented as $\Delta_{\mathrm{CT}}$. 	
		All multireference methods used the CASSCF(12e,9o) reference wavefunction. 
		Core ionization energies were computed using the cc-pCVTZ basis set.
		Also shown are the X2C scalar relativistic corrections computed using the cc-pCVTZ-X2C basis set (in parentheses) and the experimental results from Ref.\@ \citenum{Banna.1977}.}
	\label{tbl:cvs-ozone}
	\begin{tabular*}{0.98\textwidth}{@{\extracolsep{\fill}}lllllllllllllll}
		\hline
		ionization             & SR-ADC(2) & SR-ADC(2)-X & SR-ADC(3) & EOM-CCSD & MR-ADC(2) & MR-ADC(2)-X & MRCISD & Experiment &  \\ \hline
		\ce{O_T} ($1a_1^{-1}$) & 540.64    & 540.49      & 548.17    & 544.15   & 543.47    & 540.62      & 545.92 & 541.5      &  \\
		                       & (+0.38)   & (+0.38)     & (+0.39)   &          & (+0.38)   & (+0.38)     &        &            &  \\
		\ce{O_T} ($1b_2^{-1}$) & 541.64    & 540.49      & 548.17    & 544.16   & 543.47    & 540.63      & 545.92 &            &  \\
		                       & (+0.38)   & (+0.38)     & (+0.39)   &          & (+0.38)   & (+0.38)     &        &            &  \\
		\ce{O_C} ($2a_1^{-1}$) & 546.46    & 546.09      & 551.37    & 549.23   & 548.11    & 545.06      & 550.31 & 546.2      &  \\
		                       & (+0.37)   & (+0.37)     & (+0.38)   &          & (+0.38)   & (+0.37)     &        &            &  \\
		$\Delta_{\mathrm{CT}}$ & 5.82      & 5.60        & 3.20      & 5.07     & 4.64      & 4.43        & 4.39   & 4.7        &  \\
		                       & ($-$0.01) & (+0.01)     & (+0.00)   &          & (+0.00)   & (+0.00)     &        &            &  \\ \hline
	\end{tabular*}
\end{table*}

\begin{figure}[t!]
	\centering
	\includegraphics[width=\columnwidth]{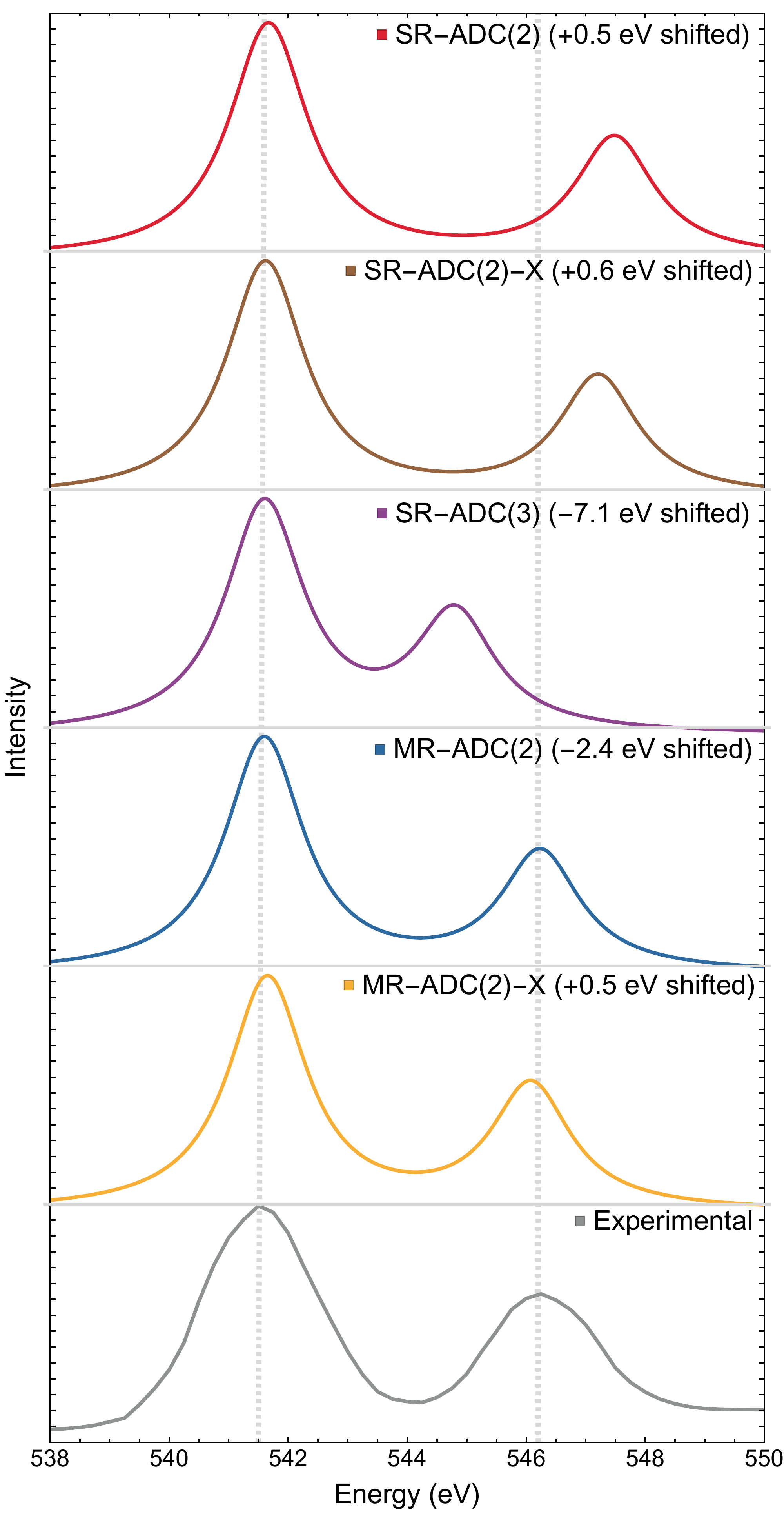}
	\caption{
		Oxygen K-edge photoelectron spectra of ozone computed using five ADC approximations compared to the experimental spectrum from Ref.\@ \citenum{Banna.1977}. 
		Simulated spectra used a 0.8 eV broadening parameter and were shifted to align with the first peak of the experimental spectrum. 
		All calculations were performed using the cc-pCVTZ-X2C basis set and the X2C scalar relativistic effects. 
		MR-ADC calculations used the CASSCF(12e,9o) wavefunction as a reference.}
	\label{fgr:spectra-ozone}
\end{figure}

Next, we consider the ozone molecule (\ce{O3}), which ground-state electronic structure has been shown to exhibit a multireference character.\cite{Hayes.1971,Hay.1975,Laidig.1981,Schmidt.1998,Kalemos.2008,Musial.2009,Oyedepo.2010,Nair.2012,Miliordos.2013,Miliordos.2014,Takeshita.2015} 
High-level calculations using multireference configuration interaction demonstrate that the $\mathrm{^1\mathrm{A}'}$ ground-state wavefunction of ozone has a significant ($\sim 18\%$) contribution from the open-shell singlet electronic configuration.\cite{Miliordos.2013,Miliordos.2014}
The singlet diradical character of \ce{O3} influences its reactivity\cite{Audran.2018} and must be properly accounted for in the calculations of excited states and spectra.

\autoref{fgr:natorbs-ozone} shows the natural orbitals and occupations of ozone computed using CASSCF with 12 electrons in 9 frontier active orbitals, corresponding to four electrons and three 2p atomic orbitals from each oxygen atom. 
The singlet diradical character of ozone can be noticed in the natural occupations of the $1a_2$ and $2b_2$ orbitals that significantly deviate from 2.0 and 0.0, respectively.

We now turn our attention to the oxygen K-edge ionization energies (\autoref{tbl:cvs-ozone}) and X-ray photoelectron spectra (XPS) of \ce{O3} (\autoref{fgr:spectra-ozone}) simulated using the CVS-MR-ADC and CVS-SR-ADC methods.
\autoref{tbl:cvs-ozone} also shows the core ionization energies computed using CVS-EOM-CCSD and MRCISD. 
The experimental gas-phase XPS spectrum\cite{Banna.1977} shown in \autoref{fgr:spectra-ozone} exhibits two peaks with 2:1 intensity ratio corresponding to the K-edge ionization in the terminal (\ce{O_T}) and central (\ce{O_C}) oxygen atoms, respectively. 
The relative ordering of these two peaks can be explained from the analysis of Mulliken atomic charges computed using CASSCF (Table S1 of ESI\dag) that are negative for the \ce{O_T} atoms ($-$0.15) and positive for the \ce{O_C} atom (0.3), in agreement with the Lewis resonance structures shown in \autoref{scheme:ozone}.
The excess electron density on the terminal oxygen atoms gives rise to a more efficient screening of the \ce{O_T} core hole relative to that for the \ce{O_C} atoms and a red shift of the corresponding peak in the XPS spectrum.
\begin{scheme}[ht]
\begin{center}
\includegraphics[width=0.7\columnwidth]{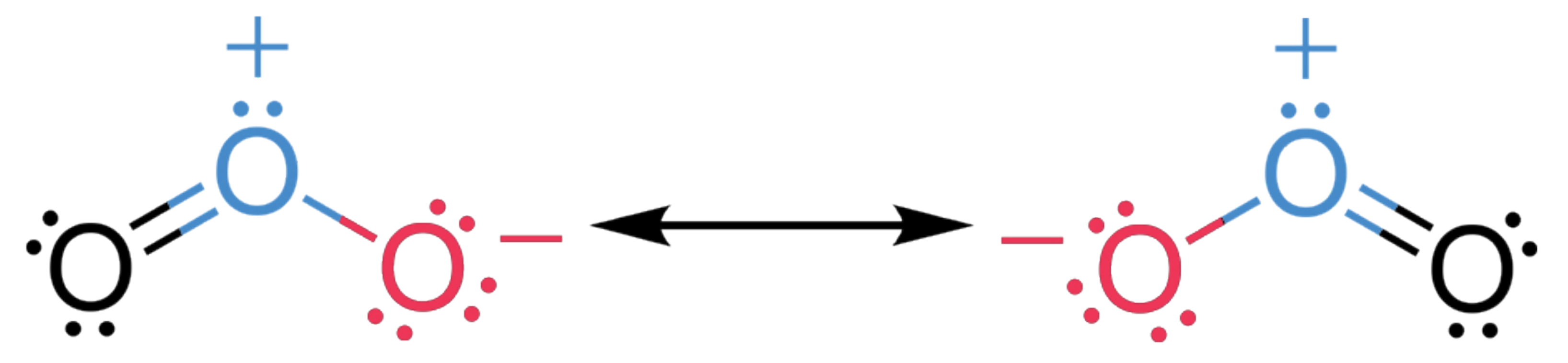}
\end{center}
\caption{Resonance structures of the ozone molecule.}
\label{scheme:ozone}
\end{scheme}

Since the computational methods considered in this work do not incorporate vibrational effects, when comparing the simulated \ce{O3} XPS spectra to the experiment we focus only on the relative intensities and energies of the two peaks.
All single- and multireference ADC methods correctly predict the ordering of \ce{O_T} and \ce{O_C} peaks, but differ in the description of their relative energy ($\Delta_{\mathrm{CT}}$).
The $\Delta_{\mathrm{CT}}$ computed using CVS-MR-ADC(2) (4.64 eV) and CVS-MR-ADC(2)-X (4.43 eV) are in excellent agreement with the peak spacing from experiment (4.7 eV) and MRCISD (4.39 eV, \autoref{tbl:cvs-ozone}).
The CVS-SR-ADC(2) and CVS-SR-ADC(2)-X methods overestimate $\Delta_{\mathrm{CT}}$ (5.82 and 5.60 eV, respectively), while it is significantly underestimated (3.20 eV) by the CVS-SR-ADC(3) approximation.

The large errors in $\Delta_{\mathrm{CT}}$ of the single-reference ADC approximations can be attributed to their inability to properly describe the singlet diradical character of ozone that reduces the electron density on \ce{O_T} while increasing it on \ce{O_C} (\autoref{fgr:natorbs-ozone}), affecting the screening and relative energies of the corresponding core holes in the simulated XPS spectrum.
The charge redistribution induced by the diradical character can be detected in the Mulliken atomic charges computed at the Hartree--Fock and CASSCF levels of theory (Table S1 of ESI\dag) that show significant differences for all oxygen atoms ($\sim$ 0.1 for \ce{O_C} and $\sim$ 0.05 for \ce{O_T}).
The unbalanced description of the \ce{O_T} and \ce{O_C} core-ionized states is also observed in the results of the single-reference CVS-EOM-CCSD method (\autoref{tbl:cvs-ozone}) that overestimates $\Delta_{\mathrm{CT}}$ by $\sim$ 0.7 eV relative to MRCISD.

\subsection{Simulating the X-Ray Photoelectron Spectra of Benzyne Diradicals}
\label{sec:benzynes}

\begin{figure}[t!]
	\begin{subfigure}[t]{1.00\columnwidth}
		\includegraphics[width=\columnwidth]{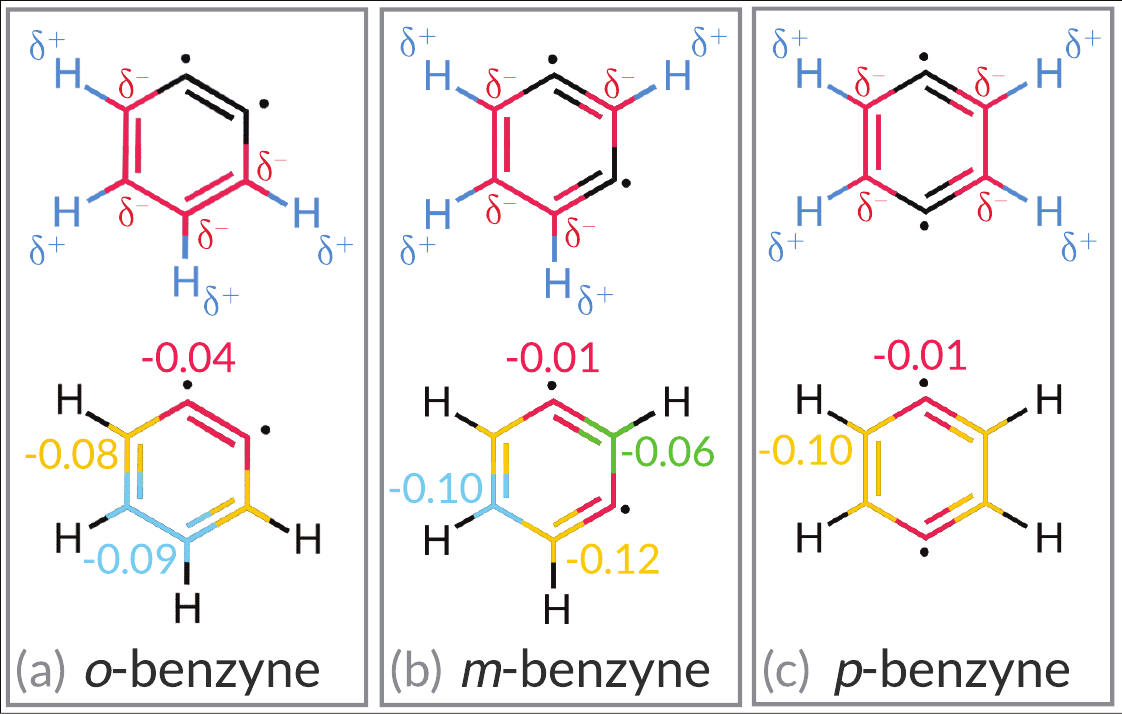}
		\phantomcaption
		\label{fgr:orto-benzyne}
	\end{subfigure}
	\begin{subfigure}[t]{0\columnwidth}
		\includegraphics[width=\columnwidth]{example-image-b}
		\phantomcaption
		\label{fgr:meta-benzyne}
	\end{subfigure}
	\begin{subfigure}[t]{0\columnwidth}
		\includegraphics[width=\columnwidth]{example-image-b}
		\phantomcaption
		\label{fgr:para-benzyne}
	\end{subfigure}
	\vspace*{-10mm}
	\caption{Molecular structures and Mulliken atomic charges of three benzyne isomers: (\subref{fgr:orto-benzyne}) \textit{o}-\ce{C6H4}, (\subref{fgr:meta-benzyne}) \textit{m}-\ce{C6H4}, and (\subref{fgr:para-benzyne}) \textit{p}-\ce{C6H4}. Calculations were performed using CASSCF(8e,8o) and the cc-pCVDZ basis set. Also shown are partial negative ($\delta^{-}$) and positive ($\delta^{+}$) charges due to the polarization of the C--H bonds.}
	\label{fgr:benzynes}
\end{figure}

\begin{figure}[t!]
\begin{subfigure}[t]{1.00\columnwidth}
    \includegraphics[width=\columnwidth]{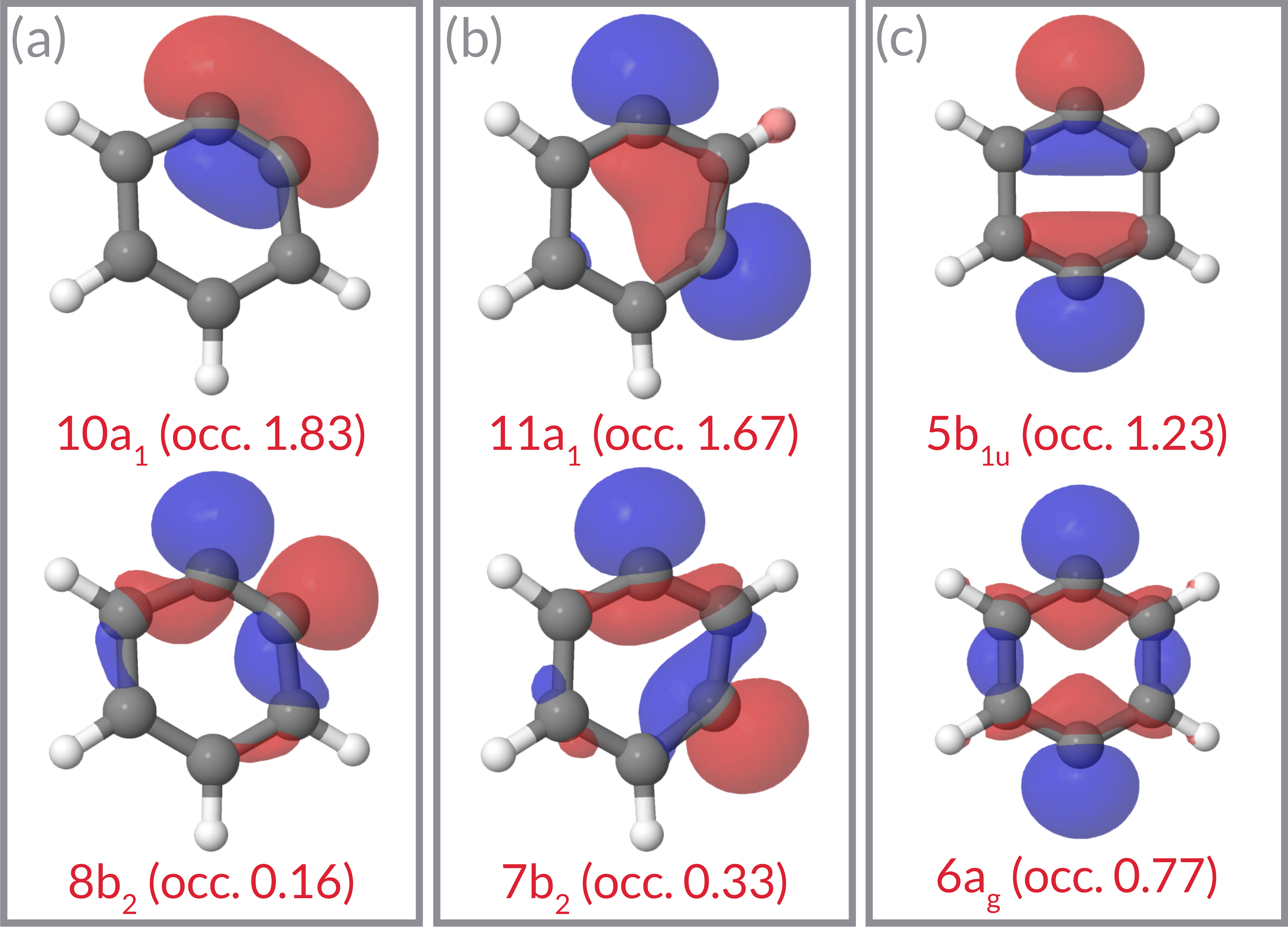}
    \phantomcaption
    \label{fgr:natorbs-orto-benzynes}
\end{subfigure}
\begin{subfigure}[t]{0\columnwidth}
    \includegraphics[width=\columnwidth]{example-image-b}
    \phantomcaption
    \label{fgr:natorbs-meta-benzynes}
\end{subfigure}
\begin{subfigure}[t]{0\columnwidth}
    \includegraphics[width=\columnwidth]{example-image-b}
    \phantomcaption
    \label{fgr:natorbs-para-benzynes}
\end{subfigure}
\vspace*{-10mm}
\caption{Frontier natural orbitals and occupations of three benzyne isomers: (\subref{fgr:natorbs-orto-benzynes}) \textit{o}-\ce{C6H4}, (\subref{fgr:natorbs-meta-benzynes}) \textit{m}-\ce{C6H4}, and (\subref{fgr:natorbs-para-benzynes}) \textit{p}-\ce{C6H4}. Calculations were performed using CASSCF(8e,8o) and the cc-pCVDZ basis set.}
\label{fgr:natorbs-benzynes}
\end{figure}

\begin{figure*}[t!]
	\centering
	\begin{subfigure}[t]{1.00\textwidth}
		\includegraphics[width=\textwidth]{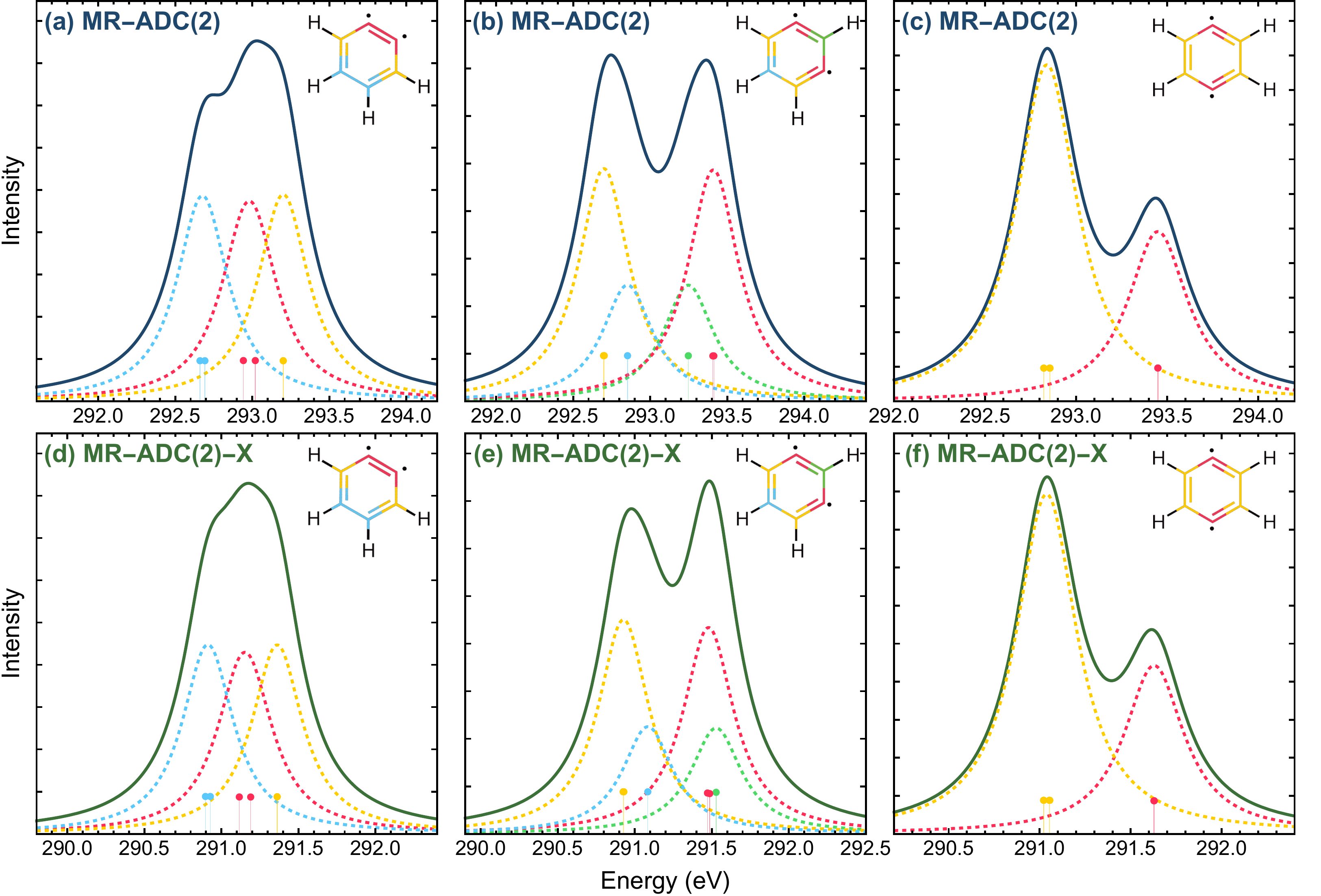}
		\phantomcaption
		\label{fgr:spectra-ortho-mradc2}
	\end{subfigure}
	\begin{subfigure}[t]{0\textwidth}
		\includegraphics[width=\textwidth]{example-image-b}
		\phantomcaption
		\label{fgr:spectra-meta-mradc2}
	\end{subfigure}
	\begin{subfigure}[t]{0\textwidth}
		\includegraphics[width=\textwidth]{example-image-b}
		\phantomcaption
		\label{fgr:spectra-para-mradc2}
	\end{subfigure}
	\begin{subfigure}[t]{0\textwidth}
		\includegraphics[width=\textwidth]{example-image-b}
		\phantomcaption
		\label{fgr:spectra-ortho-mradc2x}
	\end{subfigure}
	\begin{subfigure}[t]{0\textwidth}
		\includegraphics[width=\textwidth]{example-image-b}
		\phantomcaption
		\label{fgr:spectra-meta-mradc2x}
	\end{subfigure}
	\begin{subfigure}[t]{0\textwidth}
		\includegraphics[width=\textwidth]{example-image-b}
		\phantomcaption
		\label{fgr:spectra-para-mradc2x}
	\end{subfigure}
	\vspace*{-10mm}
	\caption{Carbon K-edge photoelectron spectra of \textit{ortho}-, \textit{meta}-, and \textit{para}-benzyne molecules computed using CVS-MR-ADC(2) (\subref{fgr:spectra-ortho-mradc2} -- \subref{fgr:spectra-para-mradc2}) and CVS-MR-ADC(2)-X (\subref{fgr:spectra-ortho-mradc2x} -- \subref{fgr:spectra-para-mradc2x}) methods, respectively. Solid lines show XPS spectra calculated using the 0.2 eV broadening. Dashed lines show spectral contributions from symmetry-equivalent carbon sites, color-coded as shown in each molecular structure. Calculations used the CASSCF(8e,8o) reference, cc-pCVDZ-X2C basis set, and the X2C description of scalar relativistic effects.}
	\label{fgr:spectra-benzynes-mradc}
\end{figure*}

\begin{figure*}[t!]
	\centering
	\begin{subfigure}[t]{1.00\textwidth}
		\includegraphics[width=\textwidth]{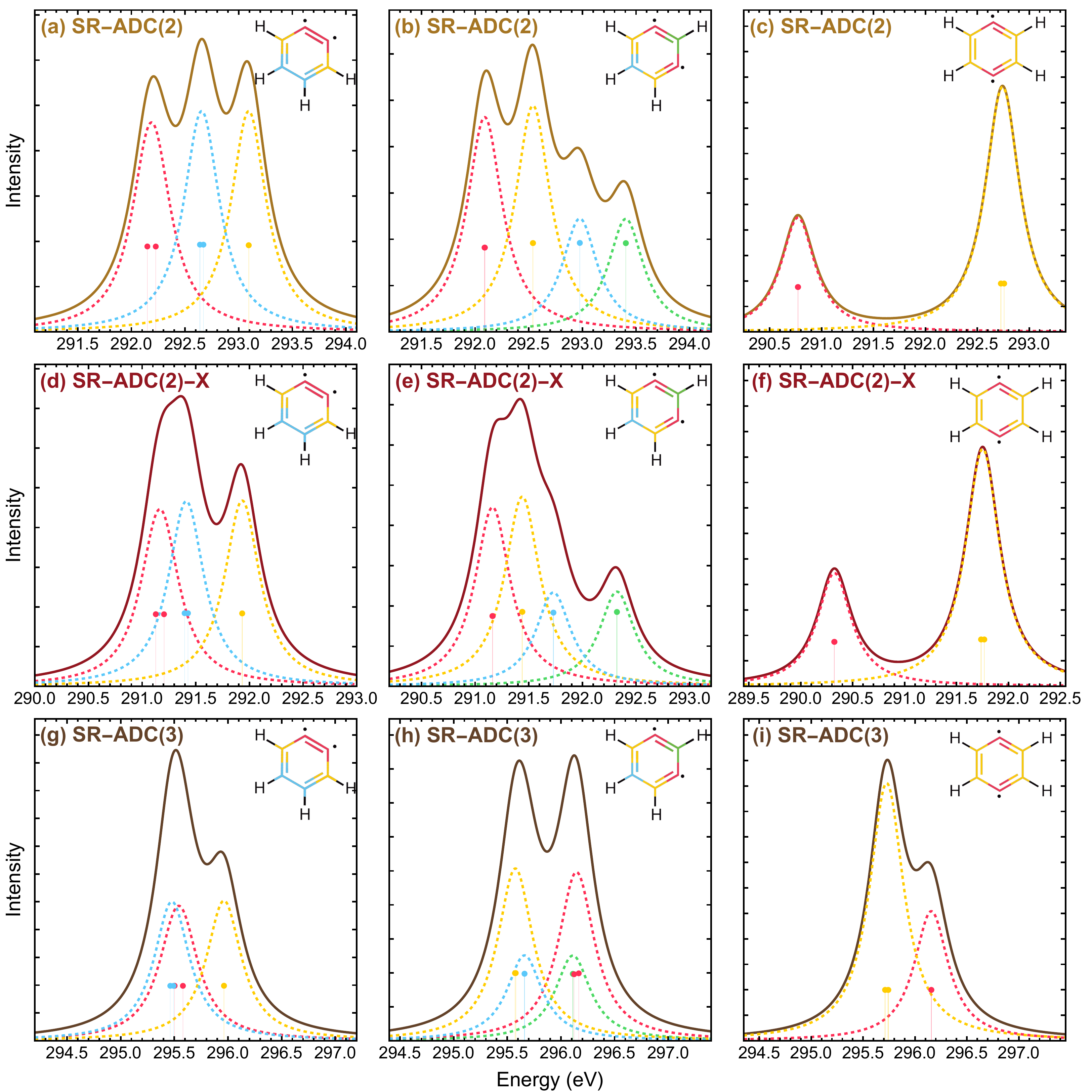}
		\phantomcaption
		\label{fgr:spectra-ortho-sradc2}
	\end{subfigure}
	\begin{subfigure}[t]{0\textwidth}
		\includegraphics[width=\textwidth]{example-image-b}
		\phantomcaption
		\label{fgr:spectra-meta-sradc2}
	\end{subfigure}
	\begin{subfigure}[t]{0\textwidth}
		\includegraphics[width=\textwidth]{example-image-b}
		\phantomcaption
		\label{fgr:spectra-para-sradc2}
	\end{subfigure}
	\begin{subfigure}[t]{0\textwidth}
		\includegraphics[width=\textwidth]{example-image-b}
		\phantomcaption
		\label{fgr:spectra-ortho-sradc2x}
	\end{subfigure}
	\begin{subfigure}[t]{0\textwidth}
		\includegraphics[width=\textwidth]{example-image-b}
		\phantomcaption
		\label{fgr:spectra-meta-sradc2x}
	\end{subfigure}
	\begin{subfigure}[t]{0\textwidth}
		\includegraphics[width=\textwidth]{example-image-b}
		\phantomcaption
		\label{fgr:spectra-para-sradc2x}
	\end{subfigure}
	\begin{subfigure}[t]{0\textwidth}
		\includegraphics[width=\textwidth]{example-image-b}
		\phantomcaption
		\label{fgr:spectra-ortho-sradc3}
	\end{subfigure}
	\begin{subfigure}[t]{0\textwidth}
		\includegraphics[width=\textwidth]{example-image-b}
		\phantomcaption
		\label{fgr:spectra-meta-sradc3}
	\end{subfigure}
	\begin{subfigure}[t]{0\textwidth}
		\includegraphics[width=\textwidth]{example-image-b}
		\phantomcaption
		\label{fgr:spectra-para-sradc3}
	\end{subfigure}
	\vspace*{-10mm}
	\caption{Carbon K-edge photoelectron spectra of \textit{ortho}-, \textit{meta}-, and \textit{para}-benzyne molecules computed using CVS-SR-ADC(2) (\subref{fgr:spectra-ortho-sradc2} -- \subref{fgr:spectra-para-sradc2}), CVS-SR-ADC(2)-X (\subref{fgr:spectra-ortho-sradc2x} -- \subref{fgr:spectra-para-sradc2x}), and CVS-SR-ADC(3) (\subref{fgr:spectra-ortho-sradc3} -- \subref{fgr:spectra-para-sradc3}), respectively. Solid lines show XPS spectra calculated using the 0.2 eV broadening. Dashed lines show spectral contributions from symmetry-equivalent carbon sites, color-coded as shown in each molecular structure. Calculations used the cc-pCVDZ-X2C basis set and the X2C description of scalar relativistic effects.}
	\label{fgr:spectra-benzynes-sradc}
\end{figure*}

Finally, we apply the CVS-MR-ADC methods to investigate the carbon K-edge XPS spectra of three benzyne diradicals (\textit{ortho}-, \textit{meta}-, and \textit{para}-isomers) shown in \autoref{fgr:benzynes}. 
Benzynes are highly reactive intermediates that are commonly formed in organic and combustion reactions\cite{Wenk.2003,Sato.2010,Wenthold.2010,Wentrup.2010,Winkler.2010} and can act as precursors in the formation of polycyclic aromatic hydrocarbons.\cite{Tranter.2010,Shukla.2011,Comandini.2011,Matsugi.2012,Monluc.2021}
Due to their open-shell singlet character, the electronic structure and properties of benzynes have been studied using a variety of quantum chemical methods.\cite{Evangelista:2007p024102,Li.2008,Evangelista:2009p4728,Hanauer.2012,Jagau.2012,Jagau.2012c,Jagau.2012b,Samanta.2014,Hannon.2016,ChenuangLi.2018,ChenuangLi.2018err,Ray.2019,NanHe.2020,JunShen.2021,Mullinax.2015,Nakano.201790p,Krylov.2018,Kleinpeter.2019}
All three molecules have the singlet ground electronic state with a significant diradical character that increases from \textit{ortho}- to \textit{para}-benzyne, \cite{Abe.2013,Hoffmann.2019} along with decreasing singlet--triplet gap.\cite{Leopold.1986,Wenthold.1998}
\autoref{fgr:natorbs-benzynes} shows the frontier natural orbitals of each isomer computed using CASSCF(8e,8o).
As the diradical character increases from \textit{ortho}- to \textit{para}-benzyne, the populations of two natural orbitals become increasingly similar.

\autoref{fgr:benzynes} depicts the distribution of Mulliken atomic charges for the carbon atoms in each benzyne isomer computed using CASSCF(8e,8o).
For all molecules, the carbon atoms bonded directly to the hydrogen atoms carry a higher negative charge compared to that of the carbon radical centers, as expected from formal considerations of charge distribution based on atomic electronegativities. 
This analysis has implications for understanding the relative energetics of core-ionized states in the carbon K-edge XPS spectra of benzynes, suggesting that the core holes created on the hydrogenated carbon atoms {\it will have lower energy} than those on the radical centers due to an increased core-hole screening by the excess electron density.

The carbon K-edge XPS spectra of benzynes simulated using CVS-MR-ADC(2) and CVS-MR-ADC(2)-X are shown in \autoref{fgr:spectra-benzynes-mradc}.
In agreement with our analysis based on core-hole screening, the lowest-energy transition in each spectrum corresponds to the K-edge ionization of the hydrogenated carbon atoms with the highest negative Mulliken charge in \autoref{fgr:benzynes}.
CVS-MR-ADC(2) and CVS-MR-ADC(2)-X show very similar spectra predicting that the core-ionized states localized on the radical carbon centers are significantly blue-shifted relative to the first peak in the XPS spectrum of each molecule.
The smallest blue shift is observed in \textit{ortho}-benzyne, which dehydrogenated carbon atoms carry a significant negative charge ($-$0.04).

To understand the importance of multireference effects in the simulations of benzyne core-ionized states, we consider the carbon K-edge XPS spectra simulated using CVS-SR-ADC (\autoref{fgr:spectra-benzynes-sradc}).
The CVS-SR-ADC(2) and CVS-SR-ADC(2)-X XPS spectra are {\it qualitatively different} from the CVS-MR-ADC spectra (\autoref{fgr:spectra-benzynes-mradc}) with the lowest-energy transition corresponding to the core ionization of carbon radical centers. 
The most significant difference between the CVS-SR-ADC and CVS-MR-ADC spectra is observed for \textit{para}-benzyne with the largest degree of multireference character, where including the strong correlation effects at the ADC(2) level changes the relative spacing between the two peaks in the XPS spectrum by $\sim$ 2.6 eV inverting their order ({\it cf.} \autoref{fgr:spectra-para-mradc2} and \autoref{fgr:spectra-para-sradc2}).
(Calculated core ionization energies and transition probabilities can be found in ESI.\dag)
In contrast to our benchmark for weakly-correlated systems (\autoref{subsec:CVS-benchmark}) where CVS-SR-ADC(2)-X and CVS-MR-ADC(2)-X showed similar performance, the results of these two methods are significantly different for all benzynes, especially for the \textit{para}-isomer.

The role of multireference effects can be rationalized by comparing the Mulliken charges computed using Hartree--Fock (SCF) and CASSCF(8e,8o) (Table S4 in ESI.\dag).
For all molecules, neglecting the strong correlation effects in SCF increases the negative charges on the dehydrogenated carbon atoms while making the hydrogen-bonded carbon centers more positively charged.
These differences in charge distribution between SCF and CASSCF can be traced to the inability of the former method to describe a non-zero population of the lowest-unoccupied molecular orbital (\autoref{fgr:natorbs-benzynes}) that is partially localized on the hydrogen-bonded carbon atoms. 
The largest difference between the SCF and CASSCF Mulliken charges is observed for \textit{para}-benzyne ($\sim$ 0.05), in agreement with the highest degree of diradical character in this molecule among all benzynes.

Interestingly, we find that the XPS spectra simulated using CVS-SR-ADC(3) (Figures \ref{fgr:spectra-ortho-sradc3} -- \ref{fgr:spectra-para-sradc3}) qualitatively agree with the results from CVS-MR-ADC, although the single-reference method significantly underestimates the blue shift of core-hole states localized at the radical centers relative to the first peak in the XPS spectrum.
Considering the poor performance of CVS-SR-ADC(3) for the dissociation of \ce{N2} (\autoref{subsec:dissociation-n2}), apparent lack of convergence of the CVS-SR-ADC simulated spectra in \autoref{fgr:spectra-benzynes-sradc} with increasing level of theory, and the fact that SR-ADC(3) has been shown to produce large errors in ionization energies of systems with strong multireference character,\cite{Chatterjee.2019,Chatterjee.2020} we believe that this result stems from fortuitous error cancellation rather than the higher-order description of electron correlation effects.

\section{Conclusions}\label{sec:Conclusions}

We presented implementation, benchmark, and applications of multireference algebraic diagrammatic construction theory with core-valence separation (CVS-MR-ADC) for calculations of core ionization energies and X-ray photoelectron spectra (XPS).
In contrast to conventional multireference methods, the CVS-MR-ADC approach does not require incorporating core orbitals in the active space and can simulate a large number transitions in the XPS spectra by starting with a single complete active-space self-consistent field (CASSCF) wavefunction computed for the ground electronic state. 

We benchmarked the accuracy of CVS-MR-ADC for the K-edge ionization energies of 16 small weakly-correlated molecules against the accurate results from equation-of-motion coupled cluster theory with single, double, and triple excitations.\cite{Liu.2019} 
For this benchmark set, the performance of CVS-MR-ADC methods is similar to that of the single-reference ADC approximations (CVS-SR-ADC), with CVS-MR-ADC(2)-X and CVS-SR-ADC(2)-X showing the smallest mean absolute errors of $\sim$ 0.4 eV.
Additionally, we investigated the dependence of the CVS-MR-ADC results on the choice of active space.
CVS-MR-ADC(2)-X showed much weaker dependence on reference CASSCF orbitals compared to CVS-MR-ADC(2), which is consistent with the higher-order description of orbital relaxation effects in the former method. 

To demonstrate the performance of CVS-MR-ADC for multireference systems, we used this approach to compute the potential energy curves (PEC's) of core-ionized nitrogen molecule and to simulate the XPS spectra of ozone and three benzyne singlet diradicals (\textit{ortho}-, \textit{meta}-, and \textit{para}-isomers). 
The PEC's computed using the CVS-MR-ADC methods were found to be in a good agreement with the reference PEC from multireference configuration interaction with single and double excitations (MRCISD), while the CVS-SR-ADC curves diverged with increasing N--N bond length.
For ozone, our results demonstrate that including multireference effects is crucial to accurately predict the energy spacing between the core-ionized states localized on the terminal and central oxygen atoms. 
The relative energies of these two states predicted by CVS-MR-ADC(2) and CVS-MR-ADC(2)-X are in an excellent agreement with MRCISD and experiment,\cite{Banna.1977} while the single-reference ADC and equation-of-motion coupled cluster theories show large deviations.

When applied to benzyne diradicals, CVS-MR-ADC(2) and CVS-MR-ADC(2)-X predict that the first peak in the carbon K-edge XPS spectra of all three molecules corresponds to the core ionization of hydrogen-bonded carbon atoms as opposed to carbon radical centers, in agreement with a formal analysis of core-hole screening effects. 
In contrast, the single-reference CVS-SR-ADC(2) and CVS-SR-ADC(2)-X methods make qualitatively different predictions, favoring the carbon radical centers to be the lowest-energy ionization sites.
We attribute this to the CVS-SR-ADC inability to describe the singlet diradical character that influences the charge distribution and core-hole screening.
Our calculations also demonstrate that the CVS-SR-ADC(3) XPS spectra agree qualitatively with those from CVS-MR-ADC, which we attribute to fortuitous error cancellation.

The results presented in this work demonstrate the importance of strong correlation effects for accurate predictions of potential energy surfaces of core-ionized molecules, as well as peak spacing and relative order in the XPS spectra of multireference systems.
Our results also provide evidence that CVS-MR-ADC is a promising approach for the XPS simulations of molecules with significant multireference effects.
Future extensions of this method will include a more efficient implementation to treat larger molecular systems, adding ability to simulate XPS spectra of open-shell systems, and incorporating spin-orbit coupling effects for molecules with heavy elements. 
Work along these directions is currently under way in our group.

\section*{Conflicts of interest}
There are no conflicts to declare.

\section*{Acknowledgements}

This work was supported by the National Science Foundation, under Grant No. CHE-2044648. Computations were performed at the Ohio Supercomputer Center under Project Nos. PAS1583 and PAS1963.\cite{OhioSupercomputerCenter1987}



\balance



\begin{mcitethebibliography}{175}
\providecommand*{\natexlab}[1]{#1}
\providecommand*{\mciteSetBstSublistMode}[1]{}
\providecommand*{\mciteSetBstMaxWidthForm}[2]{}
\providecommand*{\mciteBstWouldAddEndPuncttrue}
  {\def\EndOfBibitem{\unskip.}}
\providecommand*{\mciteBstWouldAddEndPunctfalse}
  {\let\EndOfBibitem\relax}
\providecommand*{\mciteSetBstMidEndSepPunct}[3]{}
\providecommand*{\mciteSetBstSublistLabelBeginEnd}[3]{}
\providecommand*{\EndOfBibitem}{}
\mciteSetBstSublistMode{f}
\mciteSetBstMaxWidthForm{subitem}
{(\emph{\alph{mcitesubitemcount}})}
\mciteSetBstSublistLabelBeginEnd{\mcitemaxwidthsubitemform\space}
{\relax}{\relax}

\bibitem[Lin \emph{et~al.}(2017)Lin, Liu, Yu, Cheng, Singer, Shpyrko, Xin,
  Tamura, Tian, Weng, Yang, Meng, Nordlund, Yang, and Doeff]{Lin.2017}
F.~Lin, Y.~Liu, X.~Yu, L.~Cheng, A.~Singer, O.~G. Shpyrko, H.~L. Xin,
  N.~Tamura, C.~Tian, T.-C. Weng, X.-Q. Yang, Y.~S. Meng, D.~Nordlund, W.~Yang
  and M.~M. Doeff, \emph{Chem. Rev.}, 2017, \textbf{117}, 13123--13186\relax
\mciteBstWouldAddEndPuncttrue
\mciteSetBstMidEndSepPunct{\mcitedefaultmidpunct}
{\mcitedefaultendpunct}{\mcitedefaultseppunct}\relax
\EndOfBibitem
\bibitem[Chergui and Collet(2017)]{Chergui.2017}
M.~Chergui and E.~Collet, \emph{Chem. Rev.}, 2017, \textbf{117},
  11025--11065\relax
\mciteBstWouldAddEndPuncttrue
\mciteSetBstMidEndSepPunct{\mcitedefaultmidpunct}
{\mcitedefaultendpunct}{\mcitedefaultseppunct}\relax
\EndOfBibitem
\bibitem[Kraus \emph{et~al.}(2018)Kraus, Z{\"u}rch, Cushing, Neumark, and
  Leone]{Kraus.2018}
P.~M. Kraus, M.~Z{\"u}rch, S.~K. Cushing, D.~M. Neumark and S.~R. Leone,
  \emph{Nat. Rev. Chem.}, 2018, \textbf{2}, 82--94\relax
\mciteBstWouldAddEndPuncttrue
\mciteSetBstMidEndSepPunct{\mcitedefaultmidpunct}
{\mcitedefaultendpunct}{\mcitedefaultseppunct}\relax
\EndOfBibitem
\bibitem[Norman and Dreuw(2018)]{Norman.2018}
P.~Norman and A.~Dreuw, \emph{Chem. Rev.}, 2018, \textbf{118}, 7208--7248\relax
\mciteBstWouldAddEndPuncttrue
\mciteSetBstMidEndSepPunct{\mcitedefaultmidpunct}
{\mcitedefaultendpunct}{\mcitedefaultseppunct}\relax
\EndOfBibitem
\bibitem[Pellegrini \emph{et~al.}(2016)Pellegrini, Marinelli, and
  Reiche]{Pellegrini.2016}
C.~Pellegrini, A.~Marinelli and S.~Reiche, \emph{Rev. Mod. Phys.}, 2016,
  \textbf{88}, 015006\relax
\mciteBstWouldAddEndPuncttrue
\mciteSetBstMidEndSepPunct{\mcitedefaultmidpunct}
{\mcitedefaultendpunct}{\mcitedefaultseppunct}\relax
\EndOfBibitem
\bibitem[Geloni \emph{et~al.}(2017)Geloni, Huang, and Pellegrini]{Geloni.2017}
G.~Geloni, Z.~Huang and C.~Pellegrini, \emph{Energy and Environment Series},
  2017,  1--44\relax
\mciteBstWouldAddEndPuncttrue
\mciteSetBstMidEndSepPunct{\mcitedefaultmidpunct}
{\mcitedefaultendpunct}{\mcitedefaultseppunct}\relax
\EndOfBibitem
\bibitem[Young \emph{et~al.}(2018)Young, Ueda, G{\"u}hr, Bucksbaum, Simon,
  Mukamel, Rohringer, Prince, Masciovecchio, Meyer, Rudenko, Rolles, Bostedt,
  Fuchs, Reis, Santra, Kapteyn, Murnane, Ibrahim, L\'{e}gar\'{e}, Vrakking,
  Isinger, Kroon, Gisselbrecht, L{\textquoteright}Huillier, W\"{o}rner, and
  Leone]{Young.2018}
L.~Young, K.~Ueda, M.~G{\"u}hr, P.~H. Bucksbaum, M.~Simon, S.~Mukamel,
  N.~Rohringer, K.~C. Prince, C.~Masciovecchio, M.~Meyer, A.~Rudenko,
  D.~Rolles, C.~Bostedt, M.~Fuchs, D.~A. Reis, R.~Santra, H.~Kapteyn,
  M.~Murnane, H.~Ibrahim, F.~L\'{e}gar\'{e}, M.~Vrakking, M.~Isinger, D.~Kroon,
  M.~Gisselbrecht, A.~L{\textquoteright}Huillier, H.~J. W\"{o}rner and S.~R.
  Leone, \emph{J. Phys. B}, 2018, \textbf{51}, 032003\relax
\mciteBstWouldAddEndPuncttrue
\mciteSetBstMidEndSepPunct{\mcitedefaultmidpunct}
{\mcitedefaultendpunct}{\mcitedefaultseppunct}\relax
\EndOfBibitem
\bibitem[Piancastelli \emph{et~al.}(2019)Piancastelli, Marchenko, Guillemin,
  Journel, Travnikova, Ismail, and Simon]{Piancastelli.2019}
M.~N. Piancastelli, T.~Marchenko, R.~Guillemin, L.~Journel, O.~Travnikova,
  I.~Ismail and M.~Simon, \emph{Rep. Prog. Phys.}, 2019, \textbf{83},
  016401\relax
\mciteBstWouldAddEndPuncttrue
\mciteSetBstMidEndSepPunct{\mcitedefaultmidpunct}
{\mcitedefaultendpunct}{\mcitedefaultseppunct}\relax
\EndOfBibitem
\bibitem[Seres \emph{et~al.}(2004)Seres, Seres, Krausz, and
  Spielmann]{Seres.2004}
E.~Seres, J.~Seres, F.~Krausz and C.~Spielmann, \emph{Phys. Rev. Lett.}, 2004,
  \textbf{92}, 163002\relax
\mciteBstWouldAddEndPuncttrue
\mciteSetBstMidEndSepPunct{\mcitedefaultmidpunct}
{\mcitedefaultendpunct}{\mcitedefaultseppunct}\relax
\EndOfBibitem
\bibitem[Popmintchev \emph{et~al.}(2010)Popmintchev, Chen, Arpin, Murnane, and
  Kapteyn]{Popmintchev.2010}
T.~Popmintchev, M.-C. Chen, P.~Arpin, M.~M. Murnane and H.~C. Kapteyn,
  \emph{Nat. Photonics}, 2010, \textbf{4}, 822--832\relax
\mciteBstWouldAddEndPuncttrue
\mciteSetBstMidEndSepPunct{\mcitedefaultmidpunct}
{\mcitedefaultendpunct}{\mcitedefaultseppunct}\relax
\EndOfBibitem
\bibitem[Li \emph{et~al.}(2017)Li, Ren, Yin, Zhao, Chew, Cheng, Cunningham,
  Wang, Hu, Wu, Chini, and Chang]{Li.2017}
J.~Li, X.~Ren, Y.~Yin, K.~Zhao, A.~Chew, Y.~Cheng, E.~Cunningham, Y.~Wang,
  S.~Hu, Y.~Wu, M.~Chini and Z.~Chang, \emph{Nat. Commun.}, 2017, \textbf{8},
  186\relax
\mciteBstWouldAddEndPuncttrue
\mciteSetBstMidEndSepPunct{\mcitedefaultmidpunct}
{\mcitedefaultendpunct}{\mcitedefaultseppunct}\relax
\EndOfBibitem
\bibitem[Barreau \emph{et~al.}(2020)Barreau, Ross, Garg, Kraus, Neumark, and
  Leone]{Barreau.2020}
L.~Barreau, A.~D. Ross, S.~Garg, P.~M. Kraus, D.~M. Neumark and S.~R. Leone,
  \emph{Sci. Rep.}, 2020, \textbf{10}, 5773\relax
\mciteBstWouldAddEndPuncttrue
\mciteSetBstMidEndSepPunct{\mcitedefaultmidpunct}
{\mcitedefaultendpunct}{\mcitedefaultseppunct}\relax
\EndOfBibitem
\bibitem[Greczynski and Hultman(2019)]{Greczynski.2019}
G.~Greczynski and L.~Hultman, \emph{Prog. Mater. Sci.}, 2019, \textbf{107},
  100591\relax
\mciteBstWouldAddEndPuncttrue
\mciteSetBstMidEndSepPunct{\mcitedefaultmidpunct}
{\mcitedefaultendpunct}{\mcitedefaultseppunct}\relax
\EndOfBibitem
\bibitem[Stevie and Donley(2020)]{Stevie.2020}
F.~A. Stevie and C.~L. Donley, \emph{J. Vac. Sci. Technol. A}, 2020,
  \textbf{38}, 063204\relax
\mciteBstWouldAddEndPuncttrue
\mciteSetBstMidEndSepPunct{\mcitedefaultmidpunct}
{\mcitedefaultendpunct}{\mcitedefaultseppunct}\relax
\EndOfBibitem
\bibitem[Briggs and Grant(2003)]{briggs2003surface}
D.~Briggs and J.~Grant, \emph{{Surface Analysis by Auger and X-ray
  Photoelectron Spectroscopy}}, SurfaceSpectra, 2003\relax
\mciteBstWouldAddEndPuncttrue
\mciteSetBstMidEndSepPunct{\mcitedefaultmidpunct}
{\mcitedefaultendpunct}{\mcitedefaultseppunct}\relax
\EndOfBibitem
\bibitem[Watts and Wolstenholme(2019)]{watts2019introduction}
J.~Watts and J.~Wolstenholme, \emph{{An Introduction to Surface Analysis by XPS
  and AES}}, Wiley, 2019\relax
\mciteBstWouldAddEndPuncttrue
\mciteSetBstMidEndSepPunct{\mcitedefaultmidpunct}
{\mcitedefaultendpunct}{\mcitedefaultseppunct}\relax
\EndOfBibitem
\bibitem[Winter \emph{et~al.}(2007)Winter, Aziz, Hergenhahn, Faubel, and
  Hertel]{Winter.2007}
B.~Winter, E.~F. Aziz, U.~Hergenhahn, M.~Faubel and I.~V. Hertel, \emph{J.
  Chem. Phys.}, 2007, \textbf{126}, 124504\relax
\mciteBstWouldAddEndPuncttrue
\mciteSetBstMidEndSepPunct{\mcitedefaultmidpunct}
{\mcitedefaultendpunct}{\mcitedefaultseppunct}\relax
\EndOfBibitem
\bibitem[Nishizawa \emph{et~al.}(2010)Nishizawa, Kurahashi, Sekiguchi, Mizuno,
  Ogi, Horio, Oura, Kosugi, and Suzuki]{Nishizawa.2010}
K.~Nishizawa, N.~Kurahashi, K.~Sekiguchi, T.~Mizuno, Y.~Ogi, T.~Horio, M.~Oura,
  N.~Kosugi and T.~Suzuki, \emph{Phys. Chem. Chem. Phys.}, 2010, \textbf{13},
  413--417\relax
\mciteBstWouldAddEndPuncttrue
\mciteSetBstMidEndSepPunct{\mcitedefaultmidpunct}
{\mcitedefaultendpunct}{\mcitedefaultseppunct}\relax
\EndOfBibitem
\bibitem[Th{\"u}rmer \emph{et~al.}(2021)Th{\"u}rmer, Malerz, Trinter,
  Hergenhahn, Lee, Neumark, Meijer, Winter, and Wilkinson]{Thurmer.2021}
S.~Th{\"u}rmer, S.~Malerz, F.~Trinter, U.~Hergenhahn, C.~Lee, D.~M. Neumark,
  G.~Meijer, B.~Winter and I.~Wilkinson, \emph{Chem. Sci.}, 2021, \textbf{12},
  10558--10582\relax
\mciteBstWouldAddEndPuncttrue
\mciteSetBstMidEndSepPunct{\mcitedefaultmidpunct}
{\mcitedefaultendpunct}{\mcitedefaultseppunct}\relax
\EndOfBibitem
\bibitem[Banna(2006)]{Banna.2006}
M.~S. Banna, \emph{Contemp. Phys.}, 2006, \textbf{25}, 159--176\relax
\mciteBstWouldAddEndPuncttrue
\mciteSetBstMidEndSepPunct{\mcitedefaultmidpunct}
{\mcitedefaultendpunct}{\mcitedefaultseppunct}\relax
\EndOfBibitem
\bibitem[Kraus \emph{et~al.}(2014)Kraus, Reichelt, G{\"u}nther, Gregoratti,
  Amati, Kiskinova, Yulaev, Vlassiouk, and Kolmakov]{Kraus.2014}
J.~Kraus, R.~Reichelt, S.~G{\"u}nther, L.~Gregoratti, M.~Amati, M.~Kiskinova,
  A.~Yulaev, I.~Vlassiouk and A.~Kolmakov, \emph{Nanoscale}, 2014, \textbf{6},
  14394--14403\relax
\mciteBstWouldAddEndPuncttrue
\mciteSetBstMidEndSepPunct{\mcitedefaultmidpunct}
{\mcitedefaultendpunct}{\mcitedefaultseppunct}\relax
\EndOfBibitem
\bibitem[Tao and Nguyen(2018)]{Tao.2018}
F.~F. Tao and L.~Nguyen, \emph{Phys. Chem. Chem. Phys.}, 2018, \textbf{20},
  9812--9823\relax
\mciteBstWouldAddEndPuncttrue
\mciteSetBstMidEndSepPunct{\mcitedefaultmidpunct}
{\mcitedefaultendpunct}{\mcitedefaultseppunct}\relax
\EndOfBibitem
\bibitem[Bokarev and K{\"u}hn(2020)]{Bokarev.2020}
S.~I. Bokarev and O.~K{\"u}hn, \emph{Wiley Interdiscip. Rev. Comput. Mol.
  Sci.}, 2020, \textbf{10}, e1433\relax
\mciteBstWouldAddEndPuncttrue
\mciteSetBstMidEndSepPunct{\mcitedefaultmidpunct}
{\mcitedefaultendpunct}{\mcitedefaultseppunct}\relax
\EndOfBibitem
\bibitem[Kasper \emph{et~al.}(2020)Kasper, Stetina, Jenkins, and
  Li]{Kasper.2020}
J.~M. Kasper, T.~F. Stetina, A.~J. Jenkins and X.~Li, \emph{Chem. Phys. Rev.},
  2020, \textbf{1}, 011304\relax
\mciteBstWouldAddEndPuncttrue
\mciteSetBstMidEndSepPunct{\mcitedefaultmidpunct}
{\mcitedefaultendpunct}{\mcitedefaultseppunct}\relax
\EndOfBibitem
\bibitem[Besley(2020)]{Besley.2020}
N.~A. Besley, \emph{Acc. Chem. Res.}, 2020, \textbf{53}, 1306--1315\relax
\mciteBstWouldAddEndPuncttrue
\mciteSetBstMidEndSepPunct{\mcitedefaultmidpunct}
{\mcitedefaultendpunct}{\mcitedefaultseppunct}\relax
\EndOfBibitem
\bibitem[Besley(2021)]{Besley.2021}
N.~A. Besley, \emph{Wiley Interdiscip. Rev. Comput. Mol. Sci.}, 2021,
  \textbf{11}, e1527\relax
\mciteBstWouldAddEndPuncttrue
\mciteSetBstMidEndSepPunct{\mcitedefaultmidpunct}
{\mcitedefaultendpunct}{\mcitedefaultseppunct}\relax
\EndOfBibitem
\bibitem[Rankine and Penfold(2021)]{Rankine.2021}
C.~D. Rankine and T.~J. Penfold, \emph{J. Phys. Chem. A}, 2021, \textbf{125},
  4276--4293\relax
\mciteBstWouldAddEndPuncttrue
\mciteSetBstMidEndSepPunct{\mcitedefaultmidpunct}
{\mcitedefaultendpunct}{\mcitedefaultseppunct}\relax
\EndOfBibitem
\bibitem[Salpeter and Bethe(1951)]{Salpeter:1951p1232}
E.~E. Salpeter and H.~A. Bethe, \emph{Phys. Rev.}, 1951, \textbf{84},
  1232--1242\relax
\mciteBstWouldAddEndPuncttrue
\mciteSetBstMidEndSepPunct{\mcitedefaultmidpunct}
{\mcitedefaultendpunct}{\mcitedefaultseppunct}\relax
\EndOfBibitem
\bibitem[Hedin(1965)]{Hedin:1965p796}
L.~Hedin, \emph{Phys. Rev. A}, 1965, \textbf{139}, 796--823\relax
\mciteBstWouldAddEndPuncttrue
\mciteSetBstMidEndSepPunct{\mcitedefaultmidpunct}
{\mcitedefaultendpunct}{\mcitedefaultseppunct}\relax
\EndOfBibitem
\bibitem[van Schilfgaarde \emph{et~al.}(2006)van Schilfgaarde, Kotani, and
  Faleev]{vanSchilfgaarde:2006p226402}
M.~van Schilfgaarde, T.~Kotani and S.~V. Faleev, \emph{Phys. Rev. Lett.}, 2006,
  \textbf{96}, 226402\relax
\mciteBstWouldAddEndPuncttrue
\mciteSetBstMidEndSepPunct{\mcitedefaultmidpunct}
{\mcitedefaultendpunct}{\mcitedefaultseppunct}\relax
\EndOfBibitem
\bibitem[van Setten \emph{et~al.}(2013)van Setten, Weigend, and
  Evers]{vanSetten:2013p232}
M.~J. van Setten, F.~Weigend and F.~Evers, \emph{J. Chem. Theory Comput.},
  2013, \textbf{9}, 232--246\relax
\mciteBstWouldAddEndPuncttrue
\mciteSetBstMidEndSepPunct{\mcitedefaultmidpunct}
{\mcitedefaultendpunct}{\mcitedefaultseppunct}\relax
\EndOfBibitem
\bibitem[Stener \emph{et~al.}(2003)Stener, Fronzoni, and
  de~Simone]{Stener:2003p115}
M.~Stener, G.~Fronzoni and M.~de~Simone, \emph{Chem. Phys. Lett.}, 2003,
  \textbf{373}, 115--123\relax
\mciteBstWouldAddEndPuncttrue
\mciteSetBstMidEndSepPunct{\mcitedefaultmidpunct}
{\mcitedefaultendpunct}{\mcitedefaultseppunct}\relax
\EndOfBibitem
\bibitem[Besley and Asmuruf(2010)]{Besley:2010p12024}
N.~A. Besley and F.~A. Asmuruf, \emph{Phys. Chem. Chem. Phys.}, 2010,
  \textbf{12}, 12024--12039\relax
\mciteBstWouldAddEndPuncttrue
\mciteSetBstMidEndSepPunct{\mcitedefaultmidpunct}
{\mcitedefaultendpunct}{\mcitedefaultseppunct}\relax
\EndOfBibitem
\bibitem[Peng \emph{et~al.}(2015)Peng, Lestrange, Goings, Caricato, and
  Li]{Peng:2015p4146}
B.~Peng, P.~J. Lestrange, J.~J. Goings, M.~Caricato and X.~Li, \emph{J. Chem.
  Theory Comput.}, 2015, \textbf{11}, 4146--4153\relax
\mciteBstWouldAddEndPuncttrue
\mciteSetBstMidEndSepPunct{\mcitedefaultmidpunct}
{\mcitedefaultendpunct}{\mcitedefaultseppunct}\relax
\EndOfBibitem
\bibitem[Barth \emph{et~al.}(1980)Barth, Buenker, Peyerimhoff, and
  Butscher]{Barth:1980p149}
A.~Barth, R.~J. Buenker, S.~D. Peyerimhoff and W.~Butscher, \emph{Chem. Phys.},
  1980, \textbf{46}, 149--164\relax
\mciteBstWouldAddEndPuncttrue
\mciteSetBstMidEndSepPunct{\mcitedefaultmidpunct}
{\mcitedefaultendpunct}{\mcitedefaultseppunct}\relax
\EndOfBibitem
\bibitem[Nooijen and Bartlett(1995)]{Nooijen:1995p6735}
M.~Nooijen and R.~J. Bartlett, \emph{J. Chem. Phys.}, 1995, \textbf{102},
  6735--6756\relax
\mciteBstWouldAddEndPuncttrue
\mciteSetBstMidEndSepPunct{\mcitedefaultmidpunct}
{\mcitedefaultendpunct}{\mcitedefaultseppunct}\relax
\EndOfBibitem
\bibitem[Sen \emph{et~al.}(2013)Sen, Shee, and Mukherjee]{Sen:2013p2625}
S.~Sen, A.~Shee and D.~Mukherjee, \emph{Mol. Phys.}, 2013, \textbf{111},
  2625--2639\relax
\mciteBstWouldAddEndPuncttrue
\mciteSetBstMidEndSepPunct{\mcitedefaultmidpunct}
{\mcitedefaultendpunct}{\mcitedefaultseppunct}\relax
\EndOfBibitem
\bibitem[Dutta \emph{et~al.}(2014)Dutta, Gupta, Vaval, and
  Pal]{Dutta:2014p3656}
A.~K. Dutta, J.~Gupta, N.~Vaval and S.~Pal, \emph{J. Chem. Theory Comput.},
  2014, \textbf{10}, 3656--3668\relax
\mciteBstWouldAddEndPuncttrue
\mciteSetBstMidEndSepPunct{\mcitedefaultmidpunct}
{\mcitedefaultendpunct}{\mcitedefaultseppunct}\relax
\EndOfBibitem
\bibitem[Coriani and Koch(2015)]{Coriani:2015p181103}
S.~Coriani and H.~Koch, \emph{J. Chem. Phys.}, 2015, \textbf{143}, 181103\relax
\mciteBstWouldAddEndPuncttrue
\mciteSetBstMidEndSepPunct{\mcitedefaultmidpunct}
{\mcitedefaultendpunct}{\mcitedefaultseppunct}\relax
\EndOfBibitem
\bibitem[Nascimento and A~Eugene~DePrince(2017)]{Nascimento:2017p2951}
D.~R. Nascimento and I.~A~Eugene~DePrince, \emph{J. Phys. Chem. Lett.}, 2017,
  2951--2957\relax
\mciteBstWouldAddEndPuncttrue
\mciteSetBstMidEndSepPunct{\mcitedefaultmidpunct}
{\mcitedefaultendpunct}{\mcitedefaultseppunct}\relax
\EndOfBibitem
\bibitem[Liu \emph{et~al.}(2019)Liu, Matthews, Coriani, and Cheng]{Liu.2019}
J.~Liu, D.~Matthews, S.~Coriani and L.~Cheng, \emph{J. Chem. Theory Comput.},
  2019, \textbf{15}, 1642--1651\relax
\mciteBstWouldAddEndPuncttrue
\mciteSetBstMidEndSepPunct{\mcitedefaultmidpunct}
{\mcitedefaultendpunct}{\mcitedefaultseppunct}\relax
\EndOfBibitem
\bibitem[Barth and Schirmer(1985)]{Barth:1985p867}
A.~Barth and J.~Schirmer, \emph{J. Phys. B: At. Mol. Phys.}, 1985, \textbf{18},
  867--885\relax
\mciteBstWouldAddEndPuncttrue
\mciteSetBstMidEndSepPunct{\mcitedefaultmidpunct}
{\mcitedefaultendpunct}{\mcitedefaultseppunct}\relax
\EndOfBibitem
\bibitem[Wenzel \emph{et~al.}(2014)Wenzel, Wormit, and Dreuw]{Wenzel.2014}
J.~Wenzel, M.~Wormit and A.~Dreuw, \emph{J. Comput. Chem.}, 2014, \textbf{35},
  1900--1915\relax
\mciteBstWouldAddEndPuncttrue
\mciteSetBstMidEndSepPunct{\mcitedefaultmidpunct}
{\mcitedefaultendpunct}{\mcitedefaultseppunct}\relax
\EndOfBibitem
\bibitem[Schirmer and Thiel(2001)]{Schirmer.2001}
J.~Schirmer and A.~Thiel, \emph{J. Chem. Phys.}, 2001, \textbf{115},
  10621--10635\relax
\mciteBstWouldAddEndPuncttrue
\mciteSetBstMidEndSepPunct{\mcitedefaultmidpunct}
{\mcitedefaultendpunct}{\mcitedefaultseppunct}\relax
\EndOfBibitem
\bibitem[Thiel \emph{et~al.}(2003)Thiel, Schirmer, and K\"{o}ppel]{Thiel.2003}
A.~Thiel, J.~Schirmer and H.~K\"{o}ppel, \emph{J. Chem. Phys.}, 2003,
  \textbf{119}, 2088--2101\relax
\mciteBstWouldAddEndPuncttrue
\mciteSetBstMidEndSepPunct{\mcitedefaultmidpunct}
{\mcitedefaultendpunct}{\mcitedefaultseppunct}\relax
\EndOfBibitem
\bibitem[Lischka \emph{et~al.}(2018)Lischka, Nachtigallov\'{a}, Aquino, Szalay,
  Plasser, Machado, and Barbatti]{Lischka.2018}
H.~Lischka, D.~Nachtigallov\'{a}, A.~J.~A. Aquino, P.~G. Szalay, F.~Plasser,
  F.~B.~C. Machado and M.~Barbatti, \emph{Chem. Rev.}, 2018, \textbf{118},
  7293--7361\relax
\mciteBstWouldAddEndPuncttrue
\mciteSetBstMidEndSepPunct{\mcitedefaultmidpunct}
{\mcitedefaultendpunct}{\mcitedefaultseppunct}\relax
\EndOfBibitem
\bibitem[Park \emph{et~al.}(2020)Park, Al-Saadon, MacLeod, Shiozaki, and
  Vlaisavljevich]{Park.2020}
J.~W. Park, R.~Al-Saadon, M.~K. MacLeod, T.~Shiozaki and B.~Vlaisavljevich,
  \emph{Chem. Rev.}, 2020, \textbf{120}, 5878--5909\relax
\mciteBstWouldAddEndPuncttrue
\mciteSetBstMidEndSepPunct{\mcitedefaultmidpunct}
{\mcitedefaultendpunct}{\mcitedefaultseppunct}\relax
\EndOfBibitem
\bibitem[Khedkar and Roemelt(2021)]{Khedkar.2021}
A.~Khedkar and M.~Roemelt, \emph{Phys. Chem. Chem. Phys.}, 2021, \textbf{23},
  17097--17112\relax
\mciteBstWouldAddEndPuncttrue
\mciteSetBstMidEndSepPunct{\mcitedefaultmidpunct}
{\mcitedefaultendpunct}{\mcitedefaultseppunct}\relax
\EndOfBibitem
\bibitem[Rocha(2011)]{Rocha.2011}
A.~B. Rocha, \emph{J. Chem. Phys.}, 2011, \textbf{134}, 024107\relax
\mciteBstWouldAddEndPuncttrue
\mciteSetBstMidEndSepPunct{\mcitedefaultmidpunct}
{\mcitedefaultendpunct}{\mcitedefaultseppunct}\relax
\EndOfBibitem
\bibitem[Rocha and de~Moura(2011)]{Rocha.2011vjw}
A.~B. Rocha and C.~E.~V. de~Moura, \emph{J. Chem. Phys.}, 2011, \textbf{135},
  224112\relax
\mciteBstWouldAddEndPuncttrue
\mciteSetBstMidEndSepPunct{\mcitedefaultmidpunct}
{\mcitedefaultendpunct}{\mcitedefaultseppunct}\relax
\EndOfBibitem
\bibitem[de~Moura \emph{et~al.}(2013)de~Moura, Oliveira, and Rocha]{Moura.2013}
C.~E.~V. de~Moura, R.~R. Oliveira and A.~B. Rocha, \emph{J. Mol. Model.}, 2013,
  \textbf{19}, 2027--2033\relax
\mciteBstWouldAddEndPuncttrue
\mciteSetBstMidEndSepPunct{\mcitedefaultmidpunct}
{\mcitedefaultendpunct}{\mcitedefaultseppunct}\relax
\EndOfBibitem
\bibitem[Corral \emph{et~al.}(2017)Corral, Gonz\'{a}lez-V\'{a}zquez, and
  Mart\'{i}n]{Corral.201794l}
I.~Corral, J.~Gonz\'{a}lez-V\'{a}zquez and F.~Mart\'{i}n, \emph{J. Chem. Theory
  Comput.}, 2017, \textbf{13}, 1723--1736\relax
\mciteBstWouldAddEndPuncttrue
\mciteSetBstMidEndSepPunct{\mcitedefaultmidpunct}
{\mcitedefaultendpunct}{\mcitedefaultseppunct}\relax
\EndOfBibitem
\bibitem[Bhattacharya \emph{et~al.}(2021)Bhattacharya, Shamasundar, and
  Emmanouilidou]{Battacharya.2021}
D.~Bhattacharya, K.~R. Shamasundar and A.~Emmanouilidou, \emph{J. Phys. Chem.
  A}, 2021, \textbf{125}, 7778--7787\relax
\mciteBstWouldAddEndPuncttrue
\mciteSetBstMidEndSepPunct{\mcitedefaultmidpunct}
{\mcitedefaultendpunct}{\mcitedefaultseppunct}\relax
\EndOfBibitem
\bibitem[{\AA}gren and Jensen(1993)]{Agren:1993p45}
H.~{\AA}gren and H.~J.~A. Jensen, \emph{Chem. Phys.}, 1993, \textbf{172},
  45--57\relax
\mciteBstWouldAddEndPuncttrue
\mciteSetBstMidEndSepPunct{\mcitedefaultmidpunct}
{\mcitedefaultendpunct}{\mcitedefaultseppunct}\relax
\EndOfBibitem
\bibitem[Josefsson \emph{et~al.}(2012)Josefsson, Kunnus, Schreck, F\"{o}hlisch,
  Groot, Wernet, and Odelius]{Josefsson.2012}
I.~Josefsson, K.~Kunnus, S.~Schreck, A.~F\"{o}hlisch, F.~d. Groot, P.~Wernet
  and M.~Odelius, \emph{J. Phys. Chem. Lett.}, 2012, \textbf{3},
  3565--3570\relax
\mciteBstWouldAddEndPuncttrue
\mciteSetBstMidEndSepPunct{\mcitedefaultmidpunct}
{\mcitedefaultendpunct}{\mcitedefaultseppunct}\relax
\EndOfBibitem
\bibitem[Pinjari \emph{et~al.}(2016)Pinjari, Delcey, Guo, Odelius, and
  Lundberg]{Pinjari:2016p477}
R.~V. Pinjari, M.~G. Delcey, M.~Guo, M.~Odelius and M.~Lundberg, \emph{J.
  Comput. Chem.}, 2016, \textbf{37}, 477--486\relax
\mciteBstWouldAddEndPuncttrue
\mciteSetBstMidEndSepPunct{\mcitedefaultmidpunct}
{\mcitedefaultendpunct}{\mcitedefaultseppunct}\relax
\EndOfBibitem
\bibitem[Guo \emph{et~al.}(2016)Guo, S{\o}rensen, Delcey, Pinjari, and
  Lundberg]{Guo:2016p3250}
M.~Guo, L.~K. S{\o}rensen, M.~G. Delcey, R.~V. Pinjari and M.~Lundberg,
  \emph{Phys. Chem. Chem. Phys.}, 2016, \textbf{18}, 3250--3259\relax
\mciteBstWouldAddEndPuncttrue
\mciteSetBstMidEndSepPunct{\mcitedefaultmidpunct}
{\mcitedefaultendpunct}{\mcitedefaultseppunct}\relax
\EndOfBibitem
\bibitem[Yeager and J{\o}rgensen(1979)]{Yeager:1979p77}
D.~L. Yeager and P.~J{\o}rgensen, \emph{Chem. Phys. Lett.}, 1979, \textbf{65},
  77--80\relax
\mciteBstWouldAddEndPuncttrue
\mciteSetBstMidEndSepPunct{\mcitedefaultmidpunct}
{\mcitedefaultendpunct}{\mcitedefaultseppunct}\relax
\EndOfBibitem
\bibitem[Graham and Yeager(1991)]{Graham:1991p2884}
R.~L. Graham and D.~L. Yeager, \emph{J. Chem. Phys.}, 1991, \textbf{94},
  2884--2893\relax
\mciteBstWouldAddEndPuncttrue
\mciteSetBstMidEndSepPunct{\mcitedefaultmidpunct}
{\mcitedefaultendpunct}{\mcitedefaultseppunct}\relax
\EndOfBibitem
\bibitem[Yeager(1992)]{Yeager:1992p133}
D.~L. Yeager, \emph{Applied Many-Body Methods in Spectroscopy and Electronic
  Structure}, Springer, Boston, MA, Boston, MA, 1992, pp. 133--161\relax
\mciteBstWouldAddEndPuncttrue
\mciteSetBstMidEndSepPunct{\mcitedefaultmidpunct}
{\mcitedefaultendpunct}{\mcitedefaultseppunct}\relax
\EndOfBibitem
\bibitem[Nichols \emph{et~al.}(1984)Nichols, Yeager, and
  J{\o}rgensen]{Nichols:1998p293}
J.~A. Nichols, D.~L. Yeager and P.~J{\o}rgensen, \emph{J. Chem. Phys.}, 1984,
  \textbf{80}, 293--314\relax
\mciteBstWouldAddEndPuncttrue
\mciteSetBstMidEndSepPunct{\mcitedefaultmidpunct}
{\mcitedefaultendpunct}{\mcitedefaultseppunct}\relax
\EndOfBibitem
\bibitem[Helmich-Paris(2019)]{HelmichParis:2019p174121}
B.~Helmich-Paris, \emph{J. Chem. Phys.}, 2019, \textbf{150}, 174121\relax
\mciteBstWouldAddEndPuncttrue
\mciteSetBstMidEndSepPunct{\mcitedefaultmidpunct}
{\mcitedefaultendpunct}{\mcitedefaultseppunct}\relax
\EndOfBibitem
\bibitem[Helmich-Paris(2021)]{Paris.2021}
B.~Helmich-Paris, \emph{Int. J. Quantum Chem.}, 2021, \textbf{121},
  e26559\relax
\mciteBstWouldAddEndPuncttrue
\mciteSetBstMidEndSepPunct{\mcitedefaultmidpunct}
{\mcitedefaultendpunct}{\mcitedefaultseppunct}\relax
\EndOfBibitem
\bibitem[K{\"o}hn and Bargholz(2019)]{Kohn:2019p041106}
A.~K{\"o}hn and A.~Bargholz, \emph{J. Chem. Phys.}, 2019, \textbf{151},
  041106\relax
\mciteBstWouldAddEndPuncttrue
\mciteSetBstMidEndSepPunct{\mcitedefaultmidpunct}
{\mcitedefaultendpunct}{\mcitedefaultseppunct}\relax
\EndOfBibitem
\bibitem[Datta and Nooijen(2012)]{Datta:2012p204107}
D.~Datta and M.~Nooijen, \emph{J. Chem. Phys.}, 2012, \textbf{137},
  204107\relax
\mciteBstWouldAddEndPuncttrue
\mciteSetBstMidEndSepPunct{\mcitedefaultmidpunct}
{\mcitedefaultendpunct}{\mcitedefaultseppunct}\relax
\EndOfBibitem
\bibitem[Maganas \emph{et~al.}(2019)Maganas, Kowalska, Nooijen, DeBeer, and
  Neese]{Maganas.2019}
D.~Maganas, J.~K. Kowalska, M.~Nooijen, S.~DeBeer and F.~Neese, \emph{J. Chem.
  Phys.}, 2019, \textbf{150}, 104106\relax
\mciteBstWouldAddEndPuncttrue
\mciteSetBstMidEndSepPunct{\mcitedefaultmidpunct}
{\mcitedefaultendpunct}{\mcitedefaultseppunct}\relax
\EndOfBibitem
\bibitem[Sokolov(2018)]{Sokolov.2018}
A.~Y. Sokolov, \emph{J. Chem. Phys.}, 2018, \textbf{149}, 204113\relax
\mciteBstWouldAddEndPuncttrue
\mciteSetBstMidEndSepPunct{\mcitedefaultmidpunct}
{\mcitedefaultendpunct}{\mcitedefaultseppunct}\relax
\EndOfBibitem
\bibitem[Chatterjee and Sokolov(2019)]{Chatterjee.2019}
K.~Chatterjee and A.~Y. Sokolov, \emph{J. Chem. Theory Comput.}, 2019,
  \textbf{15}, 5908--5924\relax
\mciteBstWouldAddEndPuncttrue
\mciteSetBstMidEndSepPunct{\mcitedefaultmidpunct}
{\mcitedefaultendpunct}{\mcitedefaultseppunct}\relax
\EndOfBibitem
\bibitem[Chatterjee and Sokolov(2020)]{Chatterjee.2020}
K.~Chatterjee and A.~Y. Sokolov, \emph{J. Chem. Theory Comput.}, 2020,
  \textbf{16}, 6343--6357\relax
\mciteBstWouldAddEndPuncttrue
\mciteSetBstMidEndSepPunct{\mcitedefaultmidpunct}
{\mcitedefaultendpunct}{\mcitedefaultseppunct}\relax
\EndOfBibitem
\bibitem[Mazin and Sokolov(2021)]{Mazin.2021}
I.~M. Mazin and A.~Y. Sokolov, \emph{J. Chem. Theory Comput.}, 2021,
  \textbf{17}, 6152--6165\relax
\mciteBstWouldAddEndPuncttrue
\mciteSetBstMidEndSepPunct{\mcitedefaultmidpunct}
{\mcitedefaultendpunct}{\mcitedefaultseppunct}\relax
\EndOfBibitem
\bibitem[Cederbaum \emph{et~al.}(1980)Cederbaum, Domcke, and
  Schirmer]{Cederbaum.1980}
L.~S. Cederbaum, W.~Domcke and J.~Schirmer, \emph{Phys. Rev. A}, 1980,
  \textbf{22}, 206--222\relax
\mciteBstWouldAddEndPuncttrue
\mciteSetBstMidEndSepPunct{\mcitedefaultmidpunct}
{\mcitedefaultendpunct}{\mcitedefaultseppunct}\relax
\EndOfBibitem
\bibitem[Barth and Cederbaum(1981)]{Barth.1981}
A.~Barth and L.~S. Cederbaum, \emph{Phys. Rev. A}, 1981, \textbf{23},
  1038--1061\relax
\mciteBstWouldAddEndPuncttrue
\mciteSetBstMidEndSepPunct{\mcitedefaultmidpunct}
{\mcitedefaultendpunct}{\mcitedefaultseppunct}\relax
\EndOfBibitem
\bibitem[Finley \emph{et~al.}(1998)Finley, Malmqvist, Roos, and
  Serrano-Andr{\'e}s]{Finley:1998p299}
J.~P. Finley, P.~{\AA}. Malmqvist, B.~O. Roos and L.~Serrano-Andr{\'e}s,
  \emph{Chem. Phys. Lett.}, 1998, \textbf{288}, 299--306\relax
\mciteBstWouldAddEndPuncttrue
\mciteSetBstMidEndSepPunct{\mcitedefaultmidpunct}
{\mcitedefaultendpunct}{\mcitedefaultseppunct}\relax
\EndOfBibitem
\bibitem[Andersson \emph{et~al.}(1990)Andersson, Malmqvist, Roos, Sadlej, and
  Wolinski]{Andersson:1990p5483}
K.~Andersson, P.~{\AA}. Malmqvist, B.~O. Roos, A.~J. Sadlej and K.~Wolinski,
  \emph{J. Phys. Chem.}, 1990, \textbf{94}, 5483--5488\relax
\mciteBstWouldAddEndPuncttrue
\mciteSetBstMidEndSepPunct{\mcitedefaultmidpunct}
{\mcitedefaultendpunct}{\mcitedefaultseppunct}\relax
\EndOfBibitem
\bibitem[Andersson \emph{et~al.}(1992)Andersson, Malmqvist, and
  Roos]{Andersson:1992p1218}
K.~Andersson, P.~{\AA}. Malmqvist and B.~O. Roos, \emph{J. Chem. Phys.}, 1992,
  \textbf{96}, 1218--1226\relax
\mciteBstWouldAddEndPuncttrue
\mciteSetBstMidEndSepPunct{\mcitedefaultmidpunct}
{\mcitedefaultendpunct}{\mcitedefaultseppunct}\relax
\EndOfBibitem
\bibitem[Angeli \emph{et~al.}(2001)Angeli, Cimiraglia, Evangelisti, Leininger,
  and Malrieu]{Angeli.2001b}
C.~Angeli, R.~Cimiraglia, S.~Evangelisti, T.~Leininger and J.-P. Malrieu,
  \emph{J. Chem. Phys.}, 2001, \textbf{114}, 10252--10264\relax
\mciteBstWouldAddEndPuncttrue
\mciteSetBstMidEndSepPunct{\mcitedefaultmidpunct}
{\mcitedefaultendpunct}{\mcitedefaultseppunct}\relax
\EndOfBibitem
\bibitem[Angeli \emph{et~al.}(2001)Angeli, Cimiraglia, and
  Malrieu]{Angeli.2001}
C.~Angeli, R.~Cimiraglia and J.-P. Malrieu, \emph{Chem. Phys. Lett.}, 2001,
  \textbf{350}, 297--305\relax
\mciteBstWouldAddEndPuncttrue
\mciteSetBstMidEndSepPunct{\mcitedefaultmidpunct}
{\mcitedefaultendpunct}{\mcitedefaultseppunct}\relax
\EndOfBibitem
\bibitem[Angeli \emph{et~al.}(2004)Angeli, Borini, Cestari, and
  Cimiraglia]{Angeli.2004}
C.~Angeli, S.~Borini, M.~Cestari and R.~Cimiraglia, \emph{J. Chem. Phys.},
  2004, \textbf{121}, 4043--4049\relax
\mciteBstWouldAddEndPuncttrue
\mciteSetBstMidEndSepPunct{\mcitedefaultmidpunct}
{\mcitedefaultendpunct}{\mcitedefaultseppunct}\relax
\EndOfBibitem
\bibitem[Coriani and Koch(2015)]{Coriani.2015c1e}
S.~Coriani and H.~Koch, \emph{J. Chem. Phys.}, 2015, \textbf{143}, 181103\relax
\mciteBstWouldAddEndPuncttrue
\mciteSetBstMidEndSepPunct{\mcitedefaultmidpunct}
{\mcitedefaultendpunct}{\mcitedefaultseppunct}\relax
\EndOfBibitem
\bibitem[Vidal \emph{et~al.}(2019)Vidal, Feng, Epifanovsky, Krylov, and
  Coriani]{Vidal.2019}
M.~L. Vidal, X.~Feng, E.~Epifanovsky, A.~I. Krylov and S.~Coriani, \emph{J.
  Chem. Theory Comput.}, 2019, \textbf{15}, 3117--3133\relax
\mciteBstWouldAddEndPuncttrue
\mciteSetBstMidEndSepPunct{\mcitedefaultmidpunct}
{\mcitedefaultendpunct}{\mcitedefaultseppunct}\relax
\EndOfBibitem
\bibitem[Vidal \emph{et~al.}(2019)Vidal, Krylov, and Coriani]{Vidal.2019b}
M.~L. Vidal, A.~I. Krylov and S.~Coriani, \emph{Phys. Chem. Chem. Phys.}, 2019,
  \textbf{22}, 2693--2703\relax
\mciteBstWouldAddEndPuncttrue
\mciteSetBstMidEndSepPunct{\mcitedefaultmidpunct}
{\mcitedefaultendpunct}{\mcitedefaultseppunct}\relax
\EndOfBibitem
\bibitem[Vidal \emph{et~al.}(2020)Vidal, Pokhilko, Krylov, and
  Coriani]{Vidal.2020}
M.~L. Vidal, P.~Pokhilko, A.~I. Krylov and S.~Coriani, \emph{J. Phys. Chem.
  Lett.}, 2020,  8314--8321\relax
\mciteBstWouldAddEndPuncttrue
\mciteSetBstMidEndSepPunct{\mcitedefaultmidpunct}
{\mcitedefaultendpunct}{\mcitedefaultseppunct}\relax
\EndOfBibitem
\bibitem[Thielen \emph{et~al.}(2021)Thielen, Hodecker, Piazolo, Rehn, and
  Dreuw]{Thielen.2021}
S.~M. Thielen, M.~Hodecker, J.~Piazolo, D.~R. Rehn and A.~Dreuw, \emph{J. Chem.
  Phys.}, 2021, \textbf{154}, 154108\relax
\mciteBstWouldAddEndPuncttrue
\mciteSetBstMidEndSepPunct{\mcitedefaultmidpunct}
{\mcitedefaultendpunct}{\mcitedefaultseppunct}\relax
\EndOfBibitem
\bibitem[K\"{o}ppel \emph{et~al.}(1997)K\"{o}ppel, Gadea, Klatt, Schirmer, and
  Cederbaum]{Koppel.1997}
H.~K\"{o}ppel, F.~X. Gadea, G.~Klatt, J.~Schirmer and L.~S. Cederbaum, \emph{J.
  Chem. Phys.}, 1997, \textbf{106}, 4415--4429\relax
\mciteBstWouldAddEndPuncttrue
\mciteSetBstMidEndSepPunct{\mcitedefaultmidpunct}
{\mcitedefaultendpunct}{\mcitedefaultseppunct}\relax
\EndOfBibitem
\bibitem[Barth and Schirmer(1999)]{Barth.1999}
A.~Barth and J.~Schirmer, \emph{J. Phys. B}, 1999, \textbf{18}, 867\relax
\mciteBstWouldAddEndPuncttrue
\mciteSetBstMidEndSepPunct{\mcitedefaultmidpunct}
{\mcitedefaultendpunct}{\mcitedefaultseppunct}\relax
\EndOfBibitem
\bibitem[Trofimov \emph{et~al.}(2000)Trofimov, Moskovskaya, Gromov,
  Vitkovskaya, and Schirmer]{Trofimov.2000}
A.~B. Trofimov, T.~E. Moskovskaya, E.~V. Gromov, N.~M. Vitkovskaya and
  J.~Schirmer, \emph{J. Struct. Chem.}, 2000, \textbf{41}, 483--494\relax
\mciteBstWouldAddEndPuncttrue
\mciteSetBstMidEndSepPunct{\mcitedefaultmidpunct}
{\mcitedefaultendpunct}{\mcitedefaultseppunct}\relax
\EndOfBibitem
\bibitem[Wenzel \emph{et~al.}(2015)Wenzel, Holzer, Wormit, and
  Dreuw]{Wenzel.2015}
J.~Wenzel, A.~Holzer, M.~Wormit and A.~Dreuw, \emph{J. Chem. Phys.}, 2015,
  \textbf{142}, 214104\relax
\mciteBstWouldAddEndPuncttrue
\mciteSetBstMidEndSepPunct{\mcitedefaultmidpunct}
{\mcitedefaultendpunct}{\mcitedefaultseppunct}\relax
\EndOfBibitem
\bibitem[Zheng and Cheng(2019)]{Zheng.2019}
X.~Zheng and L.~Cheng, \emph{J. Chem. Theory Comput.}, 2019, \textbf{15},
  4945--4955\relax
\mciteBstWouldAddEndPuncttrue
\mciteSetBstMidEndSepPunct{\mcitedefaultmidpunct}
{\mcitedefaultendpunct}{\mcitedefaultseppunct}\relax
\EndOfBibitem
\bibitem[Peng \emph{et~al.}(2019)Peng, Copan, and Sokolov]{Peng.2019}
R.~Peng, A.~V. Copan and A.~Y. Sokolov, \emph{J. Phys. Chem. A}, 2019,
  \textbf{123}, 1840--1850\relax
\mciteBstWouldAddEndPuncttrue
\mciteSetBstMidEndSepPunct{\mcitedefaultmidpunct}
{\mcitedefaultendpunct}{\mcitedefaultseppunct}\relax
\EndOfBibitem
\bibitem[Garner and Neuscamman(2020)]{Garner.2020}
S.~M. Garner and E.~Neuscamman, \emph{J. Chem. Phys.}, 2020, \textbf{153},
  154102\relax
\mciteBstWouldAddEndPuncttrue
\mciteSetBstMidEndSepPunct{\mcitedefaultmidpunct}
{\mcitedefaultendpunct}{\mcitedefaultseppunct}\relax
\EndOfBibitem
\bibitem[Seidu \emph{et~al.}(2019)Seidu, Neville, Kleinschmidt, Heil, Marian,
  and Schuurman]{Seidu.2019}
I.~Seidu, S.~P. Neville, M.~Kleinschmidt, A.~Heil, C.~M. Marian and M.~S.
  Schuurman, \emph{J. Chem. Phys.}, 2019, \textbf{151}, 144104\relax
\mciteBstWouldAddEndPuncttrue
\mciteSetBstMidEndSepPunct{\mcitedefaultmidpunct}
{\mcitedefaultendpunct}{\mcitedefaultseppunct}\relax
\EndOfBibitem
\bibitem[Dickhoff and Neck(2008)]{dickhoff2008many}
W.~Dickhoff and D.~V. Neck, \emph{{Many-body Theory Exposed! Propagator
  Description Of Quantum Mechanics In Many-body Systems (2nd Edition)}}, World
  Scientific Publishing Company, 2008\relax
\mciteBstWouldAddEndPuncttrue
\mciteSetBstMidEndSepPunct{\mcitedefaultmidpunct}
{\mcitedefaultendpunct}{\mcitedefaultseppunct}\relax
\EndOfBibitem
\bibitem[Fetter and Walecka(2012)]{fetter2012quantum}
A.~Fetter and J.~Walecka, \emph{{Quantum Theory of Many-Particle Systems}},
  Dover Publications, 2012\relax
\mciteBstWouldAddEndPuncttrue
\mciteSetBstMidEndSepPunct{\mcitedefaultmidpunct}
{\mcitedefaultendpunct}{\mcitedefaultseppunct}\relax
\EndOfBibitem
\bibitem[Dirac(1927)]{Dirac.1927}
P.~A.~M. Dirac, \emph{Proc. R. Soc. Lond. A}, 1927, \textbf{114},
  243--265\relax
\mciteBstWouldAddEndPuncttrue
\mciteSetBstMidEndSepPunct{\mcitedefaultmidpunct}
{\mcitedefaultendpunct}{\mcitedefaultseppunct}\relax
\EndOfBibitem
\bibitem[Feynman(1998)]{feynman1998statistical}
R.~Feynman, \emph{{Statistical Mechanics: A Set Of Lectures}}, Avalon
  Publishing, 1998\relax
\mciteBstWouldAddEndPuncttrue
\mciteSetBstMidEndSepPunct{\mcitedefaultmidpunct}
{\mcitedefaultendpunct}{\mcitedefaultseppunct}\relax
\EndOfBibitem
\bibitem[Werner and Meyer(1980)]{Werner.1980}
H.~Werner and W.~Meyer, \emph{J. Chem. Phys.}, 1980, \textbf{73},
  2342--2356\relax
\mciteBstWouldAddEndPuncttrue
\mciteSetBstMidEndSepPunct{\mcitedefaultmidpunct}
{\mcitedefaultendpunct}{\mcitedefaultseppunct}\relax
\EndOfBibitem
\bibitem[Werner and Meyer(1981)]{Werner.1981}
H.~Werner and W.~Meyer, \emph{J. Chem. Phys.}, 1981, \textbf{74},
  5794--5801\relax
\mciteBstWouldAddEndPuncttrue
\mciteSetBstMidEndSepPunct{\mcitedefaultmidpunct}
{\mcitedefaultendpunct}{\mcitedefaultseppunct}\relax
\EndOfBibitem
\bibitem[Knowles and Werner(1985)]{Knowles.1985}
P.~J. Knowles and H.-J. Werner, \emph{Chem. Phys. Lett.}, 1985, \textbf{115},
  259--267\relax
\mciteBstWouldAddEndPuncttrue
\mciteSetBstMidEndSepPunct{\mcitedefaultmidpunct}
{\mcitedefaultendpunct}{\mcitedefaultseppunct}\relax
\EndOfBibitem
\bibitem[Dyall(1995)]{Dyall.1995}
K.~G. Dyall, \emph{J. Chem. Phys.}, 1995, \textbf{102}, 4909--4918\relax
\mciteBstWouldAddEndPuncttrue
\mciteSetBstMidEndSepPunct{\mcitedefaultmidpunct}
{\mcitedefaultendpunct}{\mcitedefaultseppunct}\relax
\EndOfBibitem
\bibitem[Schirmer(1982)]{Schirmer:1982p2395}
J.~Schirmer, \emph{Phys. Rev. A}, 1982, \textbf{26}, 2395--2416\relax
\mciteBstWouldAddEndPuncttrue
\mciteSetBstMidEndSepPunct{\mcitedefaultmidpunct}
{\mcitedefaultendpunct}{\mcitedefaultseppunct}\relax
\EndOfBibitem
\bibitem[Schirmer \emph{et~al.}(1983)Schirmer, Cederbaum, and
  Walter]{Schirmer:1983p1237}
J.~Schirmer, L.~S. Cederbaum and O.~Walter, \emph{Phys. Rev. A}, 1983,
  \textbf{28}, 1237--1259\relax
\mciteBstWouldAddEndPuncttrue
\mciteSetBstMidEndSepPunct{\mcitedefaultmidpunct}
{\mcitedefaultendpunct}{\mcitedefaultseppunct}\relax
\EndOfBibitem
\bibitem[Davidson(1975)]{Davidson.1975}
E.~R. Davidson, \emph{J. Comput. Phys.}, 1975, \textbf{17}, 87--94\relax
\mciteBstWouldAddEndPuncttrue
\mciteSetBstMidEndSepPunct{\mcitedefaultmidpunct}
{\mcitedefaultendpunct}{\mcitedefaultseppunct}\relax
\EndOfBibitem
\bibitem[Liu(1978)]{Liu.1978}
B.~Liu, \emph{Numerical Algorithms in Chemistry: Algebraic Methods}, 1978,
  49--53\relax
\mciteBstWouldAddEndPuncttrue
\mciteSetBstMidEndSepPunct{\mcitedefaultmidpunct}
{\mcitedefaultendpunct}{\mcitedefaultseppunct}\relax
\EndOfBibitem
\bibitem[Banerjee and Sokolov(2021)]{Banerjee.2021}
S.~Banerjee and A.~Y. Sokolov, \emph{J. Chem. Phys.}, 2021, \textbf{154},
  074105\relax
\mciteBstWouldAddEndPuncttrue
\mciteSetBstMidEndSepPunct{\mcitedefaultmidpunct}
{\mcitedefaultendpunct}{\mcitedefaultseppunct}\relax
\EndOfBibitem
\bibitem[Asmuruf and Besley(2008)]{Asmuruf.2008}
F.~A. Asmuruf and N.~A. Besley, \emph{Chem. Phys. Lett.}, 2008, \textbf{463},
  267--271\relax
\mciteBstWouldAddEndPuncttrue
\mciteSetBstMidEndSepPunct{\mcitedefaultmidpunct}
{\mcitedefaultendpunct}{\mcitedefaultseppunct}\relax
\EndOfBibitem
\bibitem[Sun \emph{et~al.}(2020)Sun, Zhang, Banerjee, Bao, Barbry, Blunt,
  Bogdanov, Booth, Chen, Cui, Eriksen, Gao, Guo, Hermann, Hermes, Koh, Koval,
  Lehtola, Li, Liu, Mardirossian, McClain, Motta, Mussard, Pham, Pulkin,
  Purwanto, Robinson, Ronca, Sayfutyarova, Scheurer, Schurkus, Smith, Sun, Sun,
  Upadhyay, Wagner, Wang, White, Whitfield, Williamson, Wouters, Yang, Yu, Zhu,
  Berkelbach, Sharma, Sokolov, and Chan]{Sun.2020}
Q.~Sun, X.~Zhang, S.~Banerjee, P.~Bao, M.~Barbry, N.~S. Blunt, N.~A. Bogdanov,
  G.~H. Booth, J.~Chen, Z.-H. Cui, J.~J. Eriksen, Y.~Gao, S.~Guo, J.~Hermann,
  M.~R. Hermes, K.~Koh, P.~Koval, S.~Lehtola, Z.~Li, J.~Liu, N.~Mardirossian,
  J.~D. McClain, M.~Motta, B.~Mussard, H.~Q. Pham, A.~Pulkin, W.~Purwanto,
  P.~J. Robinson, E.~Ronca, E.~R. Sayfutyarova, M.~Scheurer, H.~F. Schurkus,
  J.~E.~T. Smith, C.~Sun, S.-N. Sun, S.~Upadhyay, L.~K. Wagner, X.~Wang,
  A.~White, J.~D. Whitfield, M.~J. Williamson, S.~Wouters, J.~Yang, J.~M. Yu,
  T.~Zhu, T.~C. Berkelbach, S.~Sharma, A.~Y. Sokolov and G.~K.-L. Chan,
  \emph{J. Chem. Phys.}, 2020, \textbf{153}, 024109\relax
\mciteBstWouldAddEndPuncttrue
\mciteSetBstMidEndSepPunct{\mcitedefaultmidpunct}
{\mcitedefaultendpunct}{\mcitedefaultseppunct}\relax
\EndOfBibitem
\bibitem[Banerjee and Sokolov(2019)]{Banerjee.2019}
S.~Banerjee and A.~Y. Sokolov, \emph{J. Chem. Phys.}, 2019, \textbf{151},
  224112\relax
\mciteBstWouldAddEndPuncttrue
\mciteSetBstMidEndSepPunct{\mcitedefaultmidpunct}
{\mcitedefaultendpunct}{\mcitedefaultseppunct}\relax
\EndOfBibitem
\bibitem[Neese(2012)]{Neese.2012}
F.~Neese, \emph{Wiley Interdiscip. Rev. Comput. Mol. Sci.}, 2012, \textbf{2},
  73--78\relax
\mciteBstWouldAddEndPuncttrue
\mciteSetBstMidEndSepPunct{\mcitedefaultmidpunct}
{\mcitedefaultendpunct}{\mcitedefaultseppunct}\relax
\EndOfBibitem
\bibitem[Neese \emph{et~al.}(2020)Neese, Wennmohs, Becker, and
  Riplinger]{Neese.2020}
F.~Neese, F.~Wennmohs, U.~Becker and C.~Riplinger, \emph{J. Chem. Phys.}, 2020,
  \textbf{152}, 224108\relax
\mciteBstWouldAddEndPuncttrue
\mciteSetBstMidEndSepPunct{\mcitedefaultmidpunct}
{\mcitedefaultendpunct}{\mcitedefaultseppunct}\relax
\EndOfBibitem
\bibitem[Stanton \emph{et~al.}()Stanton, Gauss, Cheng, Harding, Matthews, and
  Szalay]{CFOUR.page}
J.~Stanton, J.~Gauss, L.~Cheng, M.~Harding, D.~Matthews and P.~Szalay,
  \emph{{CFOUR Recontracted Correlation-consistent Basis Functions}},
  \url{http://www.cfour.de/}\relax
\mciteBstWouldAddEndPuncttrue
\mciteSetBstMidEndSepPunct{\mcitedefaultmidpunct}
{\mcitedefaultendpunct}{\mcitedefaultseppunct}\relax
\EndOfBibitem
\bibitem[Dyall(2001)]{Dyall.2001.4}
K.~G. Dyall, \emph{J. Chem. Phys.}, 2001, \textbf{115}, 9136--9143\relax
\mciteBstWouldAddEndPuncttrue
\mciteSetBstMidEndSepPunct{\mcitedefaultmidpunct}
{\mcitedefaultendpunct}{\mcitedefaultseppunct}\relax
\EndOfBibitem
\bibitem[Liu and Peng(2009)]{Liu.2009}
W.~Liu and D.~Peng, \emph{J. Chem. Phys.}, 2009, \textbf{131}, 031104\relax
\mciteBstWouldAddEndPuncttrue
\mciteSetBstMidEndSepPunct{\mcitedefaultmidpunct}
{\mcitedefaultendpunct}{\mcitedefaultseppunct}\relax
\EndOfBibitem
\bibitem[M{\o}ller and Plesset(1934)]{Moller.1934}
C.~M{\o}ller and M.~S. Plesset, \emph{Phys. Rev.}, 1934, \textbf{46},
  618--622\relax
\mciteBstWouldAddEndPuncttrue
\mciteSetBstMidEndSepPunct{\mcitedefaultmidpunct}
{\mcitedefaultendpunct}{\mcitedefaultseppunct}\relax
\EndOfBibitem
\bibitem[Dunning(1989)]{Dunning.1989.1}
T.~H. Dunning, \emph{J. Chem. Phys.}, 1989, \textbf{90}, 1007--1023\relax
\mciteBstWouldAddEndPuncttrue
\mciteSetBstMidEndSepPunct{\mcitedefaultmidpunct}
{\mcitedefaultendpunct}{\mcitedefaultseppunct}\relax
\EndOfBibitem
\bibitem[Kendall \emph{et~al.}(1992)Kendall, Dunning, and
  Harrison]{Kendall.1992}
R.~A. Kendall, T.~H. Dunning and R.~J. Harrison, \emph{J. Chem. Phys.}, 1992,
  \textbf{96}, 6796--6806\relax
\mciteBstWouldAddEndPuncttrue
\mciteSetBstMidEndSepPunct{\mcitedefaultmidpunct}
{\mcitedefaultendpunct}{\mcitedefaultseppunct}\relax
\EndOfBibitem
\bibitem[Woon and Dunning(1995)]{Woon.1995}
D.~E. Woon and T.~H. Dunning, \emph{J. Chem. Phys.}, 1995, \textbf{103},
  4572--4585\relax
\mciteBstWouldAddEndPuncttrue
\mciteSetBstMidEndSepPunct{\mcitedefaultmidpunct}
{\mcitedefaultendpunct}{\mcitedefaultseppunct}\relax
\EndOfBibitem
\bibitem[Werner \emph{et~al.}(2012)Werner, Knowles, Knizia, Manby, and
  Sch{\"u}tz]{Werner.2012}
H.~Werner, P.~J. Knowles, G.~Knizia, F.~R. Manby and M.~Sch{\"u}tz, \emph{Wiley
  Interdiscip. Rev. Comput. Mol. Sci.}, 2012, \textbf{2}, 242--253\relax
\mciteBstWouldAddEndPuncttrue
\mciteSetBstMidEndSepPunct{\mcitedefaultmidpunct}
{\mcitedefaultendpunct}{\mcitedefaultseppunct}\relax
\EndOfBibitem
\bibitem[Werner \emph{et~al.}()Werner, Knowles, Knizia, Manby, Sch{\"u}tz,
  Celani, Gy\"{o}rffy, Kats, Korona, Lindh, Mitrushenkov, Rauhut, Shamasundar,
  Adler, Amos, Bennie, Bernhardsson, Berning, Cooper, Deegan, Dobbyn, Eckert,
  Goll, Hampel, Hesselmann, Hetzer, Hrenar, Jansen, K\"{o}ppl, Lee, Liu, Lloyd,
  Ma, Mata, May, McNicholas, Meyer, III, Mura, Nicklass, O'Neill, Palmieri,
  Peng, Pfluger, Pitzer, Reiher, Shiozaki, Stoll, Stone, Tarroni,
  Thorsteinsson, Wang, and Welborn]{MOLPRO}
H.-J. Werner, P.~J. Knowles, G.~Knizia, F.~R. Manby, M.~Sch{\"u}tz, P.~Celani,
  W.~Gy\"{o}rffy, D.~Kats, T.~Korona, R.~Lindh, A.~Mitrushenkov, G.~Rauhut,
  K.~R. Shamasundar, T.~B. Adler, R.~D. Amos, S.~J. Bennie, A.~Bernhardsson,
  A.~Berning, D.~L. Cooper, M.~J.~O. Deegan, A.~J. Dobbyn, F.~Eckert, E.~Goll,
  C.~Hampel, A.~Hesselmann, G.~Hetzer, T.~Hrenar, G.~Jansen, C.~K\"{o}ppl,
  S.~J.~R. Lee, Y.~Liu, A.~W. Lloyd, Q.~Ma, R.~A. Mata, A.~J. May, S.~J.
  McNicholas, W.~Meyer, T.~F.~M. III, M.~E. Mura, A.~Nicklass, D.~P. O'Neill,
  P.~Palmieri, D.~Peng, K.~Pfluger, R.~Pitzer, M.~Reiher, T.~Shiozaki,
  H.~Stoll, A.~J. Stone, R.~Tarroni, T.~Thorsteinsson, M.~Wang and M.~Welborn,
  \emph{{MOLPRO, version , a package of ab initio programs}}, see
  https://www.molpro.net\relax
\mciteBstWouldAddEndPuncttrue
\mciteSetBstMidEndSepPunct{\mcitedefaultmidpunct}
{\mcitedefaultendpunct}{\mcitedefaultseppunct}\relax
\EndOfBibitem
\bibitem[Werner \emph{et~al.}(2020)Werner, Knowles, Manby, Black, Doll,
  He{\ss}elmann, Kats, K\"{o}hn, Korona, Kreplin, Ma, Miller, Mitrushchenkov,
  Peterson, Polyak, Rauhut, and Sibaev]{Werner.2020}
H.-J. Werner, P.~J. Knowles, F.~R. Manby, J.~A. Black, K.~Doll,
  A.~He{\ss}elmann, D.~Kats, A.~K\"{o}hn, T.~Korona, D.~A. Kreplin, Q.~Ma,
  T.~F. Miller, A.~Mitrushchenkov, K.~A. Peterson, I.~Polyak, G.~Rauhut and
  M.~Sibaev, \emph{J. Chem. Phys.}, 2020, \textbf{152}, 144107\relax
\mciteBstWouldAddEndPuncttrue
\mciteSetBstMidEndSepPunct{\mcitedefaultmidpunct}
{\mcitedefaultendpunct}{\mcitedefaultseppunct}\relax
\EndOfBibitem
\bibitem[Buenker and Peyerimhoff(1974)]{Buenker.1974}
R.~J. Buenker and S.~D. Peyerimhoff, \emph{Theor. Chim. Acta}, 1974,
  \textbf{35}, 33--58\relax
\mciteBstWouldAddEndPuncttrue
\mciteSetBstMidEndSepPunct{\mcitedefaultmidpunct}
{\mcitedefaultendpunct}{\mcitedefaultseppunct}\relax
\EndOfBibitem
\bibitem[Shepard \emph{et~al.}(1992)Shepard, Lischka, Szalay, Kovar, and
  Ernzerhof]{Shepard.1992}
R.~Shepard, H.~Lischka, P.~G. Szalay, T.~Kovar and M.~Ernzerhof, \emph{J. Chem.
  Phys.}, 1992, \textbf{96}, 2085--2098\relax
\mciteBstWouldAddEndPuncttrue
\mciteSetBstMidEndSepPunct{\mcitedefaultmidpunct}
{\mcitedefaultendpunct}{\mcitedefaultseppunct}\relax
\EndOfBibitem
\bibitem[Ivanic(2003)]{Ivanic.2003d2e}
J.~Ivanic, \emph{J. Chem. Phys.}, 2003, \textbf{119}, 9364--9376\relax
\mciteBstWouldAddEndPuncttrue
\mciteSetBstMidEndSepPunct{\mcitedefaultmidpunct}
{\mcitedefaultendpunct}{\mcitedefaultseppunct}\relax
\EndOfBibitem
\bibitem[Ivanic(2003)]{Ivanic.2003}
J.~Ivanic, \emph{J. Chem. Phys.}, 2003, \textbf{119}, 9377--9385\relax
\mciteBstWouldAddEndPuncttrue
\mciteSetBstMidEndSepPunct{\mcitedefaultmidpunct}
{\mcitedefaultendpunct}{\mcitedefaultseppunct}\relax
\EndOfBibitem
\bibitem[Barca \emph{et~al.}(2020)Barca, Bertoni, Carrington, Datta, Silva,
  Deustua, Fedorov, Gour, Gunina, Guidez, Harville, Irle, Ivanic, Kowalski,
  Leang, Li, Li, Lutz, Magoulas, Mato, Mironov, Nakata, Pham, Piecuch, Poole,
  Pruitt, Rendell, Roskop, Ruedenberg, Sattasathuchana, Schmidt, Shen,
  Slipchenko, Sosonkina, Sundriyal, Tiwari, Vallejo, Westheimer, W\l{}och, Xu,
  Zahariev, and Gordon]{GAMESS.2020}
G.~M.~J. Barca, C.~Bertoni, L.~Carrington, D.~Datta, N.~D. Silva, J.~E.
  Deustua, D.~G. Fedorov, J.~R. Gour, A.~O. Gunina, E.~Guidez, T.~Harville,
  S.~Irle, J.~Ivanic, K.~Kowalski, S.~S. Leang, H.~Li, W.~Li, J.~J. Lutz,
  I.~Magoulas, J.~Mato, V.~Mironov, H.~Nakata, B.~Q. Pham, P.~Piecuch,
  D.~Poole, S.~R. Pruitt, A.~P. Rendell, L.~B. Roskop, K.~Ruedenberg,
  T.~Sattasathuchana, M.~W. Schmidt, J.~Shen, L.~Slipchenko, M.~Sosonkina,
  V.~Sundriyal, A.~Tiwari, J.~L.~G. Vallejo, B.~Westheimer, M.~W\l{}och, P.~Xu,
  F.~Zahariev and M.~S. Gordon, \emph{J. Chem. Phys.}, 2020, \textbf{152},
  154102\relax
\mciteBstWouldAddEndPuncttrue
\mciteSetBstMidEndSepPunct{\mcitedefaultmidpunct}
{\mcitedefaultendpunct}{\mcitedefaultseppunct}\relax
\EndOfBibitem
\bibitem[de~Moura(2021)]{isGAMESS}
C.~E.~V. de~Moura, \emph{{isGAMESS: a Python interface for multiconfigurational
  Inner-Shell Excited States calculations using GAMESS}}, 2021,
  \url{https://github.com/carlosevmoura/isGAMESS}\relax
\mciteBstWouldAddEndPuncttrue
\mciteSetBstMidEndSepPunct{\mcitedefaultmidpunct}
{\mcitedefaultendpunct}{\mcitedefaultseppunct}\relax
\EndOfBibitem
\bibitem[Jolly \emph{et~al.}(1984)Jolly, Bomben, and Eyermann]{Jolly.1984}
W.~Jolly, K.~Bomben and C.~Eyermann, \emph{At. Data Nucl. Data Tables}, 1984,
  \textbf{31}, 433--493\relax
\mciteBstWouldAddEndPuncttrue
\mciteSetBstMidEndSepPunct{\mcitedefaultmidpunct}
{\mcitedefaultendpunct}{\mcitedefaultseppunct}\relax
\EndOfBibitem
\bibitem[Beach \emph{et~al.}(1984)Beach, Eyermann, Smit, Xiang, and
  Jolly]{Beach.1984}
D.~B. Beach, C.~J. Eyermann, S.~P. Smit, S.~F. Xiang and W.~L. Jolly, \emph{J.
  Am. Chem. Soc.}, 1984, \textbf{106}, 536--539\relax
\mciteBstWouldAddEndPuncttrue
\mciteSetBstMidEndSepPunct{\mcitedefaultmidpunct}
{\mcitedefaultendpunct}{\mcitedefaultseppunct}\relax
\EndOfBibitem
\bibitem[Nakayama \emph{et~al.}(1998)Nakayama, Nakano, and
  Hirao]{Nakayama.1998}
K.~Nakayama, H.~Nakano and K.~Hirao, \emph{Int. J. Quantum Chem.}, 1998,
  \textbf{66}, 157--175\relax
\mciteBstWouldAddEndPuncttrue
\mciteSetBstMidEndSepPunct{\mcitedefaultmidpunct}
{\mcitedefaultendpunct}{\mcitedefaultseppunct}\relax
\EndOfBibitem
\bibitem[Andrzejak and Witek(2011)]{Andrzejak.2011}
M.~Andrzejak and H.~A. Witek, \emph{Theor. Chem. Acc.}, 2011, \textbf{129},
  161\relax
\mciteBstWouldAddEndPuncttrue
\mciteSetBstMidEndSepPunct{\mcitedefaultmidpunct}
{\mcitedefaultendpunct}{\mcitedefaultseppunct}\relax
\EndOfBibitem
\bibitem[Banna \emph{et~al.}(1977)Banna, Frost, McDowell, Noodleman, and
  Wallbank]{Banna.1977}
M.~Banna, D.~C. Frost, C.~A. McDowell, L.~Noodleman and B.~Wallbank,
  \emph{Chem. Phys. Lett.}, 1977, \textbf{49}, 213--217\relax
\mciteBstWouldAddEndPuncttrue
\mciteSetBstMidEndSepPunct{\mcitedefaultmidpunct}
{\mcitedefaultendpunct}{\mcitedefaultseppunct}\relax
\EndOfBibitem
\bibitem[Hayes and Siu(1971)]{Hayes.1971}
E.~F. Hayes and A.~K.~Q. Siu, \emph{J. Am. Chem. Soc.}, 1971, \textbf{93},
  2090--2091\relax
\mciteBstWouldAddEndPuncttrue
\mciteSetBstMidEndSepPunct{\mcitedefaultmidpunct}
{\mcitedefaultendpunct}{\mcitedefaultseppunct}\relax
\EndOfBibitem
\bibitem[Hay \emph{et~al.}(1975)Hay, Dunning, and Goddard]{Hay.1975}
P.~J. Hay, T.~H. Dunning and W.~A. Goddard, \emph{J. Chem. Phys.}, 1975,
  \textbf{62}, 3912--3924\relax
\mciteBstWouldAddEndPuncttrue
\mciteSetBstMidEndSepPunct{\mcitedefaultmidpunct}
{\mcitedefaultendpunct}{\mcitedefaultseppunct}\relax
\EndOfBibitem
\bibitem[Laidig and Schaefer(1981)]{Laidig.1981}
W.~D. Laidig and H.~F. Schaefer, \emph{J. Chem. Phys.}, 1981, \textbf{74},
  3411--3414\relax
\mciteBstWouldAddEndPuncttrue
\mciteSetBstMidEndSepPunct{\mcitedefaultmidpunct}
{\mcitedefaultendpunct}{\mcitedefaultseppunct}\relax
\EndOfBibitem
\bibitem[Schmidt and Gordon(1998)]{Schmidt.1998}
M.~W. Schmidt and M.~S. Gordon, \emph{Annu. Rev. Phys. Chem.}, 1998,
  \textbf{49}, 233--266\relax
\mciteBstWouldAddEndPuncttrue
\mciteSetBstMidEndSepPunct{\mcitedefaultmidpunct}
{\mcitedefaultendpunct}{\mcitedefaultseppunct}\relax
\EndOfBibitem
\bibitem[Kalemos and Mavridis(2008)]{Kalemos.2008}
A.~Kalemos and A.~Mavridis, \emph{J. Chem. Phys.}, 2008, \textbf{129},
  054312\relax
\mciteBstWouldAddEndPuncttrue
\mciteSetBstMidEndSepPunct{\mcitedefaultmidpunct}
{\mcitedefaultendpunct}{\mcitedefaultseppunct}\relax
\EndOfBibitem
\bibitem[Musia\l{} \emph{et~al.}(2009)Musia\l{}, Kucharski, Zerzucha, Ku\'{s},
  and Bartlett]{Musial.2009}
M.~Musia\l{}, S.~A. Kucharski, P.~Zerzucha, T.~Ku\'{s} and R.~J. Bartlett,
  \emph{J. Chem. Phys.}, 2009, \textbf{131}, 194104\relax
\mciteBstWouldAddEndPuncttrue
\mciteSetBstMidEndSepPunct{\mcitedefaultmidpunct}
{\mcitedefaultendpunct}{\mcitedefaultseppunct}\relax
\EndOfBibitem
\bibitem[Oyedepo and Wilson(2010)]{Oyedepo.2010}
G.~A. Oyedepo and A.~K. Wilson, \emph{J. Phys. Chem. A}, 2010, \textbf{114},
  8806--8816\relax
\mciteBstWouldAddEndPuncttrue
\mciteSetBstMidEndSepPunct{\mcitedefaultmidpunct}
{\mcitedefaultendpunct}{\mcitedefaultseppunct}\relax
\EndOfBibitem
\bibitem[Bhaskaran-Nair and Kowalski(2012)]{Nair.2012}
K.~Bhaskaran-Nair and K.~Kowalski, \emph{J. Chem. Phys.}, 2012, \textbf{137},
  216101\relax
\mciteBstWouldAddEndPuncttrue
\mciteSetBstMidEndSepPunct{\mcitedefaultmidpunct}
{\mcitedefaultendpunct}{\mcitedefaultseppunct}\relax
\EndOfBibitem
\bibitem[Miliordos \emph{et~al.}(2013)Miliordos, Ruedenberg, and
  Xantheas]{Miliordos.2013}
E.~Miliordos, K.~Ruedenberg and S.~S. Xantheas, \emph{Angew. Chem. Int. Ed.},
  2013, \textbf{52}, 5736--5739\relax
\mciteBstWouldAddEndPuncttrue
\mciteSetBstMidEndSepPunct{\mcitedefaultmidpunct}
{\mcitedefaultendpunct}{\mcitedefaultseppunct}\relax
\EndOfBibitem
\bibitem[Miliordos and Xantheas(2014)]{Miliordos.2014}
E.~Miliordos and S.~S. Xantheas, \emph{J. Am. Chem. Soc.}, 2014, \textbf{136},
  2808--2817\relax
\mciteBstWouldAddEndPuncttrue
\mciteSetBstMidEndSepPunct{\mcitedefaultmidpunct}
{\mcitedefaultendpunct}{\mcitedefaultseppunct}\relax
\EndOfBibitem
\bibitem[Takeshita \emph{et~al.}(2015)Takeshita, Lindquist, and
  Dunning]{Takeshita.2015}
T.~Y. Takeshita, B.~A. Lindquist and T.~H. Dunning, \emph{J. Phys. Chem. A},
  2015, \textbf{119}, 7683--7694\relax
\mciteBstWouldAddEndPuncttrue
\mciteSetBstMidEndSepPunct{\mcitedefaultmidpunct}
{\mcitedefaultendpunct}{\mcitedefaultseppunct}\relax
\EndOfBibitem
\bibitem[Audran \emph{et~al.}(2018)Audran, Marque, and Santelli]{Audran.2018}
G.~Audran, S.~R. Marque and M.~Santelli, \emph{Tetrahedron}, 2018, \textbf{74},
  6221--6261\relax
\mciteBstWouldAddEndPuncttrue
\mciteSetBstMidEndSepPunct{\mcitedefaultmidpunct}
{\mcitedefaultendpunct}{\mcitedefaultseppunct}\relax
\EndOfBibitem
\bibitem[Wenk \emph{et~al.}(2003)Wenk, Winkler, and Sander]{Wenk.2003}
H.~H. Wenk, M.~Winkler and W.~Sander, \emph{Angew. Chem. Int. Ed.}, 2003,
  \textbf{42}, 502--528\relax
\mciteBstWouldAddEndPuncttrue
\mciteSetBstMidEndSepPunct{\mcitedefaultmidpunct}
{\mcitedefaultendpunct}{\mcitedefaultseppunct}\relax
\EndOfBibitem
\bibitem[Sato and Niino(2010)]{Sato.2010}
T.~Sato and H.~Niino, \emph{Aust. J. Chem.}, 2010, \textbf{63},
  1048--1060\relax
\mciteBstWouldAddEndPuncttrue
\mciteSetBstMidEndSepPunct{\mcitedefaultmidpunct}
{\mcitedefaultendpunct}{\mcitedefaultseppunct}\relax
\EndOfBibitem
\bibitem[Wenthold(2010)]{Wenthold.2010}
P.~G. Wenthold, \emph{Aust. J. Chem.}, 2010, \textbf{63}, 1091--1098\relax
\mciteBstWouldAddEndPuncttrue
\mciteSetBstMidEndSepPunct{\mcitedefaultmidpunct}
{\mcitedefaultendpunct}{\mcitedefaultseppunct}\relax
\EndOfBibitem
\bibitem[Wentrup(2010)]{Wentrup.2010}
C.~Wentrup, \emph{Aust. J. Chem.}, 2010, \textbf{63}, 979--986\relax
\mciteBstWouldAddEndPuncttrue
\mciteSetBstMidEndSepPunct{\mcitedefaultmidpunct}
{\mcitedefaultendpunct}{\mcitedefaultseppunct}\relax
\EndOfBibitem
\bibitem[Winkler and Sander(2010)]{Winkler.2010}
M.~Winkler and W.~Sander, \emph{Aust. J. Chem.}, 2010, \textbf{63},
  1013--1047\relax
\mciteBstWouldAddEndPuncttrue
\mciteSetBstMidEndSepPunct{\mcitedefaultmidpunct}
{\mcitedefaultendpunct}{\mcitedefaultseppunct}\relax
\EndOfBibitem
\bibitem[Tranter \emph{et~al.}(2010)Tranter, Klippenstein, Harding, Giri, Yang,
  and Kiefer]{Tranter.2010}
R.~S. Tranter, S.~J. Klippenstein, L.~B. Harding, B.~R. Giri, X.~Yang and J.~H.
  Kiefer, \emph{J. Phys. Chem. A}, 2010, \textbf{114}, 8240--8261\relax
\mciteBstWouldAddEndPuncttrue
\mciteSetBstMidEndSepPunct{\mcitedefaultmidpunct}
{\mcitedefaultendpunct}{\mcitedefaultseppunct}\relax
\EndOfBibitem
\bibitem[Shukla \emph{et~al.}(2011)Shukla, Tsuchiya, and Koshi]{Shukla.2011}
B.~Shukla, K.~Tsuchiya and M.~Koshi, \emph{J. Phys. Chem. A}, 2011,
  \textbf{115}, 5284--5293\relax
\mciteBstWouldAddEndPuncttrue
\mciteSetBstMidEndSepPunct{\mcitedefaultmidpunct}
{\mcitedefaultendpunct}{\mcitedefaultseppunct}\relax
\EndOfBibitem
\bibitem[Comandini and Brezinsky(2011)]{Comandini.2011}
A.~Comandini and K.~Brezinsky, \emph{J. Phys. Chem. A}, 2011, \textbf{115},
  5547--5559\relax
\mciteBstWouldAddEndPuncttrue
\mciteSetBstMidEndSepPunct{\mcitedefaultmidpunct}
{\mcitedefaultendpunct}{\mcitedefaultseppunct}\relax
\EndOfBibitem
\bibitem[Matsugi and Miyoshi(2012)]{Matsugi.2012}
A.~Matsugi and A.~Miyoshi, \emph{Phys. Chem. Chem. Phys.}, 2012, \textbf{14},
  9722--9728\relax
\mciteBstWouldAddEndPuncttrue
\mciteSetBstMidEndSepPunct{\mcitedefaultmidpunct}
{\mcitedefaultendpunct}{\mcitedefaultseppunct}\relax
\EndOfBibitem
\bibitem[Monluc \emph{et~al.}(2021)Monluc, Nikolayev, Medvedkov, Azyazov,
  Morozov, and Mebel]{Monluc.2021}
L.~Monluc, A.~A. Nikolayev, I.~A. Medvedkov, V.~N. Azyazov, A.~N. Morozov and
  A.~M. Mebel, \emph{ChemPhysChem}, 2021\relax
\mciteBstWouldAddEndPuncttrue
\mciteSetBstMidEndSepPunct{\mcitedefaultmidpunct}
{\mcitedefaultendpunct}{\mcitedefaultseppunct}\relax
\EndOfBibitem
\bibitem[Evangelista \emph{et~al.}(2007)Evangelista, Allen, and
  Schaefer]{Evangelista:2007p024102}
F.~A. Evangelista, W.~D. Allen and H.~F. Schaefer, \emph{J. Chem. Phys.}, 2007,
  \textbf{127}, 024102\relax
\mciteBstWouldAddEndPuncttrue
\mciteSetBstMidEndSepPunct{\mcitedefaultmidpunct}
{\mcitedefaultendpunct}{\mcitedefaultseppunct}\relax
\EndOfBibitem
\bibitem[Li and Paldus(2008)]{Li.2008}
X.~Li and J.~Paldus, \emph{J. Chem. Phys.}, 2008, \textbf{129}, 174101\relax
\mciteBstWouldAddEndPuncttrue
\mciteSetBstMidEndSepPunct{\mcitedefaultmidpunct}
{\mcitedefaultendpunct}{\mcitedefaultseppunct}\relax
\EndOfBibitem
\bibitem[Evangelista \emph{et~al.}(2009)Evangelista, Simmonett, Schaefer,
  Mukherjee, and Allen]{Evangelista:2009p4728}
F.~A. Evangelista, A.~C. Simmonett, H.~F. Schaefer, D.~Mukherjee and W.~D.
  Allen, \emph{Phys. Chem. Chem. Phys.}, 2009, \textbf{11}, 4728\relax
\mciteBstWouldAddEndPuncttrue
\mciteSetBstMidEndSepPunct{\mcitedefaultmidpunct}
{\mcitedefaultendpunct}{\mcitedefaultseppunct}\relax
\EndOfBibitem
\bibitem[Hanauer and K\"{o}hn(2012)]{Hanauer.2012}
M.~Hanauer and A.~K\"{o}hn, \emph{J. Chem. Phys.}, 2012, \textbf{136},
  204107\relax
\mciteBstWouldAddEndPuncttrue
\mciteSetBstMidEndSepPunct{\mcitedefaultmidpunct}
{\mcitedefaultendpunct}{\mcitedefaultseppunct}\relax
\EndOfBibitem
\bibitem[Jagau and Gauss(2012)]{Jagau.2012}
T.-C. Jagau and J.~Gauss, \emph{Chem. Phys.}, 2012, \textbf{401}, 73--87\relax
\mciteBstWouldAddEndPuncttrue
\mciteSetBstMidEndSepPunct{\mcitedefaultmidpunct}
{\mcitedefaultendpunct}{\mcitedefaultseppunct}\relax
\EndOfBibitem
\bibitem[Jagau and Gauss(2012)]{Jagau.2012c}
T.-C. Jagau and J.~Gauss, \emph{J. Chem. Phys.}, 2012, \textbf{137},
  044115\relax
\mciteBstWouldAddEndPuncttrue
\mciteSetBstMidEndSepPunct{\mcitedefaultmidpunct}
{\mcitedefaultendpunct}{\mcitedefaultseppunct}\relax
\EndOfBibitem
\bibitem[Jagau and Gauss(2012)]{Jagau.2012b}
T.-C. Jagau and J.~Gauss, \emph{J. Chem. Phys.}, 2012, \textbf{137},
  044116\relax
\mciteBstWouldAddEndPuncttrue
\mciteSetBstMidEndSepPunct{\mcitedefaultmidpunct}
{\mcitedefaultendpunct}{\mcitedefaultseppunct}\relax
\EndOfBibitem
\bibitem[Samanta \emph{et~al.}(2014)Samanta, Mukherjee, Hanauer, and
  K\"{o}hn]{Samanta.2014}
P.~K. Samanta, D.~Mukherjee, M.~Hanauer and A.~K\"{o}hn, \emph{J. Chem. Phys.},
  2014, \textbf{140}, 134108\relax
\mciteBstWouldAddEndPuncttrue
\mciteSetBstMidEndSepPunct{\mcitedefaultmidpunct}
{\mcitedefaultendpunct}{\mcitedefaultseppunct}\relax
\EndOfBibitem
\bibitem[Hannon \emph{et~al.}(2016)Hannon, Li, and Evangelista]{Hannon.2016}
K.~P. Hannon, C.~Li and F.~A. Evangelista, \emph{J. Chem. Phys.}, 2016,
  \textbf{144}, 204111\relax
\mciteBstWouldAddEndPuncttrue
\mciteSetBstMidEndSepPunct{\mcitedefaultmidpunct}
{\mcitedefaultendpunct}{\mcitedefaultseppunct}\relax
\EndOfBibitem
\bibitem[Li and Evangelista(2016)]{ChenuangLi.2018}
C.~Li and F.~A. Evangelista, \emph{J. Chem. Phys.}, 2016, \textbf{144},
  164114\relax
\mciteBstWouldAddEndPuncttrue
\mciteSetBstMidEndSepPunct{\mcitedefaultmidpunct}
{\mcitedefaultendpunct}{\mcitedefaultseppunct}\relax
\EndOfBibitem
\bibitem[Li and Evangelista(2018)]{ChenuangLi.2018err}
C.~Li and F.~A. Evangelista, \emph{J. Chem. Phys.}, 2018, \textbf{148},
  079903\relax
\mciteBstWouldAddEndPuncttrue
\mciteSetBstMidEndSepPunct{\mcitedefaultmidpunct}
{\mcitedefaultendpunct}{\mcitedefaultseppunct}\relax
\EndOfBibitem
\bibitem[Ray \emph{et~al.}(2019)Ray, Manna, Ghosh, Chaudhuri, and
  Chattopadhyay]{Ray.2019}
S.~S. Ray, S.~Manna, A.~Ghosh, R.~K. Chaudhuri and S.~Chattopadhyay, \emph{Int.
  J. Quantum Chem.}, 2019, \textbf{119}, e25776\relax
\mciteBstWouldAddEndPuncttrue
\mciteSetBstMidEndSepPunct{\mcitedefaultmidpunct}
{\mcitedefaultendpunct}{\mcitedefaultseppunct}\relax
\EndOfBibitem
\bibitem[He and Evangelista(2020)]{NanHe.2020}
N.~He and F.~A. Evangelista, \emph{J. Chem. Phys.}, 2020, \textbf{152},
  094107\relax
\mciteBstWouldAddEndPuncttrue
\mciteSetBstMidEndSepPunct{\mcitedefaultmidpunct}
{\mcitedefaultendpunct}{\mcitedefaultseppunct}\relax
\EndOfBibitem
\bibitem[Shen and Piecuch(2021)]{JunShen.2021}
J.~Shen and P.~Piecuch, \emph{Mol. Phys.}, 2021,  e1966534\relax
\mciteBstWouldAddEndPuncttrue
\mciteSetBstMidEndSepPunct{\mcitedefaultmidpunct}
{\mcitedefaultendpunct}{\mcitedefaultseppunct}\relax
\EndOfBibitem
\bibitem[Mullinax \emph{et~al.}(2015)Mullinax, Sokolov, and
  Schaefer]{Mullinax.2015}
J.~W. Mullinax, A.~Y. Sokolov and H.~F. Schaefer, \emph{J. Chem. Theory
  Comput.}, 2015, \textbf{11}, 2487--2495\relax
\mciteBstWouldAddEndPuncttrue
\mciteSetBstMidEndSepPunct{\mcitedefaultmidpunct}
{\mcitedefaultendpunct}{\mcitedefaultseppunct}\relax
\EndOfBibitem
\bibitem[Nakano(2017)]{Nakano.201790p}
M.~Nakano, \emph{Topics in Current Chemistry}, 2017, \textbf{375}, 47\relax
\mciteBstWouldAddEndPuncttrue
\mciteSetBstMidEndSepPunct{\mcitedefaultmidpunct}
{\mcitedefaultendpunct}{\mcitedefaultseppunct}\relax
\EndOfBibitem
\bibitem[Krylov(2018)]{Krylov.2018}
A.~I. Krylov, \emph{Reviews in Computational Chemistry}, 2018,  151--224\relax
\mciteBstWouldAddEndPuncttrue
\mciteSetBstMidEndSepPunct{\mcitedefaultmidpunct}
{\mcitedefaultendpunct}{\mcitedefaultseppunct}\relax
\EndOfBibitem
\bibitem[Kleinpeter and Koch(2019)]{Kleinpeter.2019}
E.~Kleinpeter and A.~Koch, \emph{Tetrahedron}, 2019, \textbf{75},
  4663--4668\relax
\mciteBstWouldAddEndPuncttrue
\mciteSetBstMidEndSepPunct{\mcitedefaultmidpunct}
{\mcitedefaultendpunct}{\mcitedefaultseppunct}\relax
\EndOfBibitem
\bibitem[Abe(2013)]{Abe.2013}
M.~Abe, \emph{Chem. Rev.}, 2013, \textbf{113}, 7011--7088\relax
\mciteBstWouldAddEndPuncttrue
\mciteSetBstMidEndSepPunct{\mcitedefaultmidpunct}
{\mcitedefaultendpunct}{\mcitedefaultseppunct}\relax
\EndOfBibitem
\bibitem[Stuyver \emph{et~al.}(2019)Stuyver, Chen, Zeng, Geerlings, Proft, and
  Hoffmann]{Hoffmann.2019}
T.~Stuyver, B.~Chen, T.~Zeng, P.~Geerlings, F.~D. Proft and R.~Hoffmann,
  \emph{Chem. Rev.}, 2019, \textbf{119}, 11291--11351\relax
\mciteBstWouldAddEndPuncttrue
\mciteSetBstMidEndSepPunct{\mcitedefaultmidpunct}
{\mcitedefaultendpunct}{\mcitedefaultseppunct}\relax
\EndOfBibitem
\bibitem[Leopold \emph{et~al.}(1986)Leopold, Miller, and
  Lineberger]{Leopold.1986}
D.~G. Leopold, A.~E.~S. Miller and W.~C. Lineberger, \emph{J. Am. Chem. Soc.},
  1986, \textbf{108}, 1379--1384\relax
\mciteBstWouldAddEndPuncttrue
\mciteSetBstMidEndSepPunct{\mcitedefaultmidpunct}
{\mcitedefaultendpunct}{\mcitedefaultseppunct}\relax
\EndOfBibitem
\bibitem[Wenthold \emph{et~al.}(1998)Wenthold, Squires, and
  Lineberger]{Wenthold.1998}
P.~G. Wenthold, R.~R. Squires and W.~C. Lineberger, \emph{J. Am. Chem. Soc.},
  1998, \textbf{120}, 5279--5290\relax
\mciteBstWouldAddEndPuncttrue
\mciteSetBstMidEndSepPunct{\mcitedefaultmidpunct}
{\mcitedefaultendpunct}{\mcitedefaultseppunct}\relax
\EndOfBibitem
\bibitem[Center()]{OhioSupercomputerCenter1987}
O.~S. Center, \emph{{Ohio Supercomputer Center}},
  \url{http://osc.edu/ark:/19495/f5s1ph73}\relax
\mciteBstWouldAddEndPuncttrue
\mciteSetBstMidEndSepPunct{\mcitedefaultmidpunct}
{\mcitedefaultendpunct}{\mcitedefaultseppunct}\relax
\EndOfBibitem
\end{mcitethebibliography}
\bibliographystyle{rsc} 

\providecommand*{\mcitethebibliography}{\thebibliography}
\csname @ifundefined\endcsname{endmcitethebibliography}
{\let\endmcitethebibliography\endthebibliography}{}

\end{document}